\documentclass[twocolumn,showpacs,amsmath,amssymb,pre,superscriptaddress,a4paper,floatfix,showkeys]{revtex4}
\usepackage{dcolumn}% Align table columns on decimal point
\usepackage{graphicx}
\usepackage{epsfig}

\begin{document}

\title{Event-by-event simulation of quantum phenomena:\\
Application to Einstein-Podolosky-Rosen-Bohm experiments\footnote{J. Comp. Theor. Nanosci. {\bf4}, 957 - 991, (2007), with minor corrections.}}

\author{H. De Raedt}
\email{h.a.de.raedt@rug.nl}
\affiliation{Department of Applied Physics, Zernike Institute for Advanced Materials,
University of Groningen, Nijenborgh 4, NL-9747 AG Groningen, The Netherlands}
\author{K. De Raedt}
\affiliation{Department of Computer Science,
University of Groningen, Blauwborgje 3, NL-9747 AC Groningen, The Netherlands}
\author{K. Michielsen}
\affiliation{EMBD, Vlasakker 21, B-2160 Wommelgem, Belgium}
\author{K. Keimpema}
\affiliation{Department of Applied Physics, Materials Science Centre,
University of Groningen, Nijenborgh 4, NL-9747 AG Groningen, The Netherlands}
\author{S. Miyashita}
\affiliation{Department of Physics, Graduate School of Science, University of
Tokyo, Bunkyo-ku,Tokyo 113-0033, Japan}
 % end of PACS codes
\date{\today}

% 03.65.Ud % Entanglement and quantum nonlocality (e.g. EPR paradox, Bell's inequalities, GHZ states, etc.)
% 03.65.Ta % Foundations of quantum mechanics; measurement theory

\begin{abstract}
We review the data gathering and analysis procedure used in real Einstein-Podolsky-Rosen-Bohm experiments with photons
and we illustrate the procedure by analyzing experimental data.
Based on this analysis, we construct event-based computer simulation models in which every
essential element in the experiment has a counterpart.
The data is analyzed by counting single-particle events and two-particle coincidences,
using the same procedure as in experiments.
The simulation models strictly satisfy Einstein's criteria of local causality,
do not rely on any concept of quantum theory or probability theory,
and reproduce the results of quantum theory for a quantum system of two $S=1/2$ particles.
We present a rigorous analytical treatment of these models and show that
they may yield results that are in exact agreement with quantum theory.
The apparent conflict with the folklore on Bell's theorem, stating that
such models are not supposed to exist, is resolved.
Finally, starting from the principles of probable inference, we derive the probability distributions of quantum theory
of the Einstein-Podolsky-Rosen-Bohm experiment without invoking concepts of quantum theory.
\end{abstract}

\maketitle
\tableofcontents
\def\sumprime{\mathop{{\sum}'}}

\section{Introduction}
\label{Introduction}

As nanofabrication technology is advancing from the stage of scientific
experiments to the stage of building nanoscopic systems that perform useful tasks,
it is important to have computational tools that allow the designer to assess,
with adequate reliability, how the system will behave~\cite{DREX06}.
Quantum theory provides the foundation for developing these tools.
However, just like any other theory, quantum theory has its own limitations.
If the successful operation of the device depends on individual events rather
than on the statistical properties of many events, quantum theory can no longer be used
to describe the behavior of the device.
Indeed, as is well-known from the early days in the development of quantum theory,
quantum theory has nothing to say about individual events~\cite{BOHM51,HOME97,BALL03}.
Reconciling the mathematical formalism that does not describe individual events
with the experimental fact that each observation yields a definite outcome
is referred to as the quantum measurement paradox and is
the most fundamental problem in the foundation of quantum theory~\cite{HOME97}.

Computer simulation is widely regarded as complementary to theory and experiment~\cite{LAND00}.
If computer simulation is indeed a third methodology,
it should be possible to simulate quantum phenomena on an event-by-event basis.
In view of the fundamental problem alluded to above, there
is little hope that we can find a simulation algorithm
within the framework of quantum theory.
However, if we think of quantum theory as a recipe to compute probability
distributions only, there is nothing that prevents us from stepping outside the framework
that quantum theory provides.

To head off possible misunderstandings, it may be important to rephrase what has been said.
Of course, we could simply use pseudo-random numbers to generate events according to the probability distribution
that is obtained by solving the time-independent Schr{\"o}dinger equation.
However, that is not what we mean when we say that within the framework of quantum theory,
there is little hope to find an algorithm that simulates the individual events
and reproduces the expectation values obtained from quantum theory.
The challenge is to find algorithms that simulate, event-by-event,
the experimental observations that, for instance, interference patterns appear only after a considerable
number of individual events have been recorded by the detector~\cite{GRAN86,TONO98}, without
first solving the Schr\"odinger equation.

In a number of recent papers~\cite{RAED05b,RAED05c,RAED05d,MICH05,RAED06z,MICH06z},
we have demonstrated that locally-connected networks of processing units
with a primitive learning capability can simulate event-by-event,
the single-photon beam splitter and Mach-Zehnder interferometer
experiments of Grangier \textit{et al}.~\cite{GRAN86}.
Furthermore, we have shown that this approach can be generalized to simulate
universal quantum computation by an event-by-event process~\cite{RAED05c,MICH05,MICH06z}.
Therefore, at least in principle, our approach can be used to simulate
all wave interference phenomena and many-body quantum systems using particle-like processes only.
This work suggests that we may have discovered a procedure to simulate quantum phenomena
using causal, Einstein-local, event-based processes.
Our approach is not an extension of quantum theory in any sense
nor is it a proposal for another interpretation of quantum mechanics.
The probability distributions of quantum theory are
generated by local, causal processes.

According to the folklore about Bell's theorem, a procedure
such as the one that we discovered should not exist.
Bell's theorem states that any local, hidden variable
model will produce results that are in conflict with the quantum theory of a system of two $S=1/2$ particles~\cite{BELL93}.
However, it is often overlooked that this statement can be proven for a (very) restricted class of probabilistic models only.
Indeed, minor modifications to the original model of Bell lead to the conclusion that there is no conflict~\cite{LARS04,SANT05,ZUKO06}.
In fact, Bell's theorem does not necessarily apply to the systems that we are interested in
as both simulation algorithms and actual data do not need to
satisfy the (hidden) conditions under which Bell's theorem hold~\cite{SICA99,HESS04,HESS05b}.
Furthermore, we have given analytical proofs that
two-particle correlations of the simulation models agree {\sl exactly} with
the quantum theoretical expression~\cite{RAED07z,RAED06c}.

\begin{figure}[t]
\mbox{
\includegraphics[width=8cm]{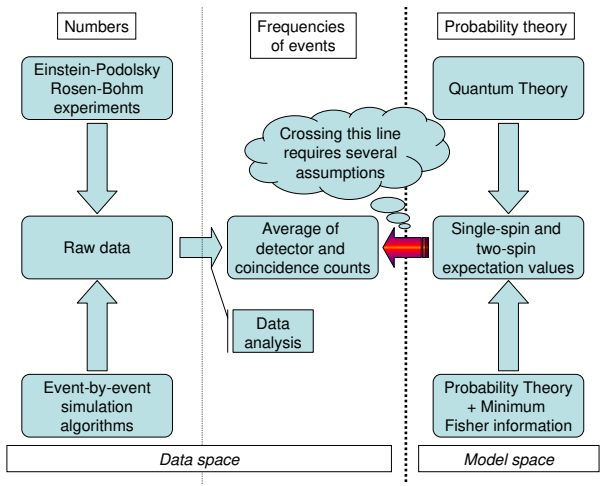}
}
\caption{Logical relationship between data and theory.
}
\label{fig0}
\end{figure}

\subsection{Aim of this work}

In this paper, we take the point of view that the fundamental problem, originating
from the work of Einstein, Podolsky, and Rosen (EPR)~\cite{EPR35}, reformulated by Bohm~\cite{BOHM51} and
studied in detail by Bell~\cite{BELL93}, is to explain how individual events,
registered by different detectors in such a way that a measurement on
one particle does not have a causal effect on the result of the measurement
on the other particle (Einstein's criterion of local causality), exhibit the
correlations that are characteristic for a quantum system in the entangled state.
We assume that:
\begin{itemize}
\item{The experimental data, including the (post) processing of it, constitutes the set of facts}
\item{The experimental facts are a faithful representation of the results of the ideal experiment}
\item{Quantum theory is compatible with these facts. In the quantum physics community, it is generally accepted that
the results of Einstein-Podolsky-Rosen-Bohm (EPRB)
experiments agree with the predictions of quantum theory~\cite{FREE72,ASPE82a,ASPE82b,TAPS94,TITT98,WEIH98,ROWE01,FATA04,SAKA06}
}
\end{itemize}

In this paper, we review constructive proofs that there exist (simple)
computer simulation algorithms that satisfy Einstein's criterion of
local causality and \textit{exactly} reproduce the results of the quantum
theoretical description of real EPRB experiments~\cite{RAED07z,RAED06c,RAED07a,RAED07b,ZHAO07a}.
These algorithms generate the same type of data as experiments and employ
the same procedure as used in experiments to analyze the data. In view of
the quantum measurement paradox~\cite{HOME97,BALL03}, the latter prohibits the use of algorithms
that rely on (concepts of) quantum theory.
In addition, for the reasons explained later, these simulation
algorithms do not rely on techniques of inductive inference (probability theory)
to draw conclusions from the data.
In this paper, we also discuss the apparent conflict with Bell's theorem.

To appreciate the fundamental issues that are involved, it is necessary to understand well
the logical relation between computer simulation, experiment and theory
on the one hand and data and theory on the other hand.
Therefore, we first elaborate on these relationships.

\subsection{Computer simulation versus experiment and theory}

In general, and in the analysis of real EPRB experiments~\cite{EPR35,BOHM51}
in particular, it is important to recognize that there are fundamental, conceptual
differences between the set of experimental facts, their interpretation in
terms of a mathematical model, and a computer simulation of the facts.

Obviously, because of limited precision of the instruments, any record of
experimental facts is just a set of integer numbers (floating point numbers
have a finite number of digits and can therefore be regarded as integer
numbers). Theories that describe Newtonian mechanics or electrodynamics
assign real numbers to experimentally observable quantities. The relation
between theory and experimental data is one-to-one: The experimental
accuracy determines the number of significant digits of the real numbers.
These theories have a deductive character.

Quantum theory assigns a probability, a real number between zero and one, for an event (=
experimental fact) to occur~\cite{BALL03,BELL93,JAYN89}. However, we can always
use an integer number to represent the event itself (in any real experiment
the number of events is necessary finite). By assigning probabilities to
events, we change the character of the theoretical description on a
fundamental level: Instead of deduction, we (have to) use inductive
inference to relate a theoretical description to the facts~\cite{BALL03,JAYN03}.

Although probability theory
provides a rigorous mathematical framework to make such inferences,
there are ample examples that illustrate how easy it is to
make the wrong inference, also for mundane, every-day problems~\cite{GRIM95,TRIB69,JAYN03,HONE02}
that are not related to quantum mechanics at all.
Subtle mistakes such as dropping some of the
conditions~\cite{BALL86}, or mixing up the meaning of physical independence and
logical independence, can give rise to all kinds of paradoxes~\cite{JAYN89,ACCA05,HESS00,HESS01,HESS04,HESS05,KRAC05,HESS05b}.

In general, a computer simulation approach does not need the machinery of
probability theory to relate simulation data to the experimental facts. A
digital computer can generate sets of integer numbers only.
We can compare these numbers to the experimental data
directly, without recourse to inductive inference. On the one hand, this
puts computer simulation in the luxury position that it cannot suffer
from mistakes of the kind alluded to earlier, simply because there is no
need to use inductive inference. On the other hand, using the computer, we
are strictly bound to the elementary rules of logic and arithmetic.
Therefore, it is not legitimate to use arguments such as ``in an experiment
it is impossible to repeat the experiment twice and get exactly the same
answer''. While this statement is correct with very high probability, when
we use a digital computer it is logically false because we can always
exactly repeat the same calculation (we exclude the possibility that the
computer is malfunctioning). Therefore, in a computer simulation,
it should be possible to explain the facts without invoking ``loopholes''
such as detection efficiency or counterfactual reasoning.

A graphical representation of the point of view taken in this paper is given
in Fig.~\ref{fig0}. On the left, we have processes that generate events.
Each event is represented by one or more numbers, which we call raw data.
Experience or a new idea provide inspiration to choose one or more methods to analyze the data.
Typically, this analysis maps the raw data
onto a few numbers (called averages and coincidence counts in Fig.~\ref{fig0}),
that is the raw data is being compressed.
On the right hand side, we have several candidate mathematical models, ``theories'',
that may ``explain'' the results of the data analysis.

But, how do we relate data to (quantum) theory?
It is essential to recognize that before we can address this question,
we have to make the hypothesis that there exists some process that gives rise to the observed data.
Otherwise, we cannot go beyond the description of merely giving the data as it is.
Furthermore, a useful theoretical model should give a description of the data that is considerably
more compact than the data itself.

Crossing the line that separates the model space
from the data space requires making the fundamental hypothesis
that the process that gives rise to the data can be described within
the framework of probability theory.
Only then, we are in the position that we can use probability theory to relate the
mathematical model to the observed frequencies.
Of course, this is consistent with the fact that quantum theory does not describe the
individual events themselves~\cite{HOME97,BALL03}.

In this paper, the rules of probability theory are mainly used as a
tool to reason in a logically consistent manner~\cite{COX61,JAYN03}, to make logical inferences
about the frequencies that we can compute from the observed data~\cite{TRIB69,JAYN03}.
These inferences concern logical relations which may or may not correspond to causal
physical influences~\cite{JAYN03}.
As we will see later, much of the mysticism
surrounding Bell's theorem can be traced back to the failure to
recognize that probability theory is not defined through frequencies.

To avoid misunderstandings of what we are aiming to accomplish here, it may be
useful to draw an analogy with methods for simulating classical statistical mechanics~\cite{LAND00}.
According to the theory of equilibrium statistical mechanics,
the probability that a system is in the state with label $n$ is given by
\begin{equation}
\label{b3}
p_n=\frac{e^{-\beta E_n}}{\sum_{n=1}^N e^{-\beta E_n}},
\end{equation}
where
$N$ is the number of different states of the system, which usually is very large,
$E_n$ is the energy of the state, and $\beta=1/k_B T$ where $k_B$
is Boltzmann's constant and $T$ is the temperature.
Disregarding exceptional cases such as the two-dimensional
Ising model, for a nontrivial many-body system the partition function $Z={\sum_{n=1}^N e^{-\beta E_n}}$
is unknown. Hence, $p_n$ is not known.

Can we construct a simulation algorithm that
generates states according to the unknown probability distribution $(p_1,\ldots,p_N)$?
An affirmative answer to this question was given by Metropolis \textit{et al}.~\cite{METR53,HAMM64,LAND00}.
The basic idea is to design an artificial dynamical system, a Markov chain or master equation
that samples the space of $N$ states such that in the long run, the frequency with which this system visits the state $n$
approaches $p_n$ with probability one~\cite{HAMM64,LAND00}.

Looking back at Fig.~\ref{fig0}, if we replace ``event-by-event simulation algorithm(s)''
by ``Metropolis Monte Carlo Method'', ``Average ... counts'' by ``Average energy ...'',
and ``Quantum theory'' by ``Equilibrium Statistical Mechanics'',
the status of simulation algorithms and theoretical models
in these two different fields of physics is the same.

Although in applications to statistical mechanics, the Markov chain dynamics is of considerable interest in itself,
there obviously is no relation to the Newtonian dynamics of the particles involved~\cite{LAND00}.
The same holds for the dynamical processes that reproduce the results of quantum theory:
If an event-by-event simulation algorithm generates the same type
of raw data as the experiment does and the data analysis yields results that agree with quantum theory
we should be pleased with this achievement and not ask for this dynamics to be ``unique''.
In fact, in our earlier work we have already shown that
there exist both deterministic and pseudo-random processes that reproduce equally well the
probability distributions obtained from quantum theory and experiments~\cite{RAED05b,RAED05c,RAED05d,MICH05,RAED06z,MICH06z}.

\subsection{Disclaimer}

The work reviewed here is not concerned with the interpretation or extension of quantum theory.
The fact that there exist simulation algorithms that reproduce the results of quantum theory
has no direct implications to the foundations of quantum theory:  The algorithm describes the process of generating events
on a level of detail about which quantum theory has nothing to say (quantum measurement paradox)~\cite{HOME97,BALL03}.
The average properties of the data may be in perfect agreement with quantum theory but
the algorithms that generate such data are outside of the scope of what quantum theory can describe.
This may sound a little strange but it is not if one recognizes that probability theory
does not contain nor provides an algorithm to generate the values of the random variables either,
which in a sense, is at the heart of the quantum measurement paradox.

\subsection{Structure of the paper}

The paper is organized as follows.
In Section~\ref{EPRBexperiment}, we review the EPRB gedanken experiment
with magnetic particles and its experimental realization using the
photon polarization as a two-state system.
We elaborate on the data gathering and analysis procedures.
An essential ingredient of the data analysis procedure
is the time window that is used to identify coincidences.
In constrast to textbook treatments of EPRB experiments
in which the window is implicitly assumed to be infinite,
in real experiments the time window is made as small as possible.
We illustrate the importance of the choice of the time window
by analyzing a data set of a real EPRB experiment with photons~\cite{WEIH98}.

Section~\ref{Quantumtheory} briefly recalls the essentials
of the quantum theoretical description of the EPRB experiment
in terms of a system of two $S=1/2$ particles.

Section~\ref{theorydata} addresses the problem of relating quantum theory and
real data.
In Section~\ref{quantumdata}, we discuss how to generate
individual events from the solution of the quantum theoretical problem
and how to relate the quantum theoretical expectation values to
the actual data.
Section~\ref{dataquantum} deals with the inverse problem:
How do we relate data to (quantum) theory?
We elaborate on the fundamental difference between probabilities
(quantities that appear in the mathematical theory)
and frequencies (numbers obtained by counting events).

Section~\ref{SimulationModel} introduces deterministic and pseudo-random
event-based computer simulation models that satisfy
Einstein's criteria of local causality and reproduce
the results of the quantum theory of two $S=1/2$ particles.
We also prove that these models can exhibit correlations
that are stronger than those obtained from the quantum theory of two $S=1/2$ particles.

In Section~\ref{Discussion}, we resolve the apparent conflict between
the fact that there exist event-based simulation models
that satisfy Einstein's criteria of local causality and reproduce
the results of the quantum theory of two $S=1/2$ particles
and the folklore about Bell's theorem, stating
that such models are not supposed to exist.
We show that Bell's extension of Einstein's concept of locality implicitly assumes
that the absence of a causal influence implies logical independence~\cite{JAYN89},
an assumption which, in general, leads to logical inconsistencies~\cite{JAYN89,JAYN03}.

In Section~\ref{Kolmogorov}, we use standard Kolmogorov
probability calculus to analyze the probabilistic version
of our simulation models. We give a rigorous proof that these models can reproduce
{\sl exactly} the results of the quantum theory of two $S=1/2$ particles.

In Section~\ref{ProbabilisticModel}, we propose a
principle to derive the probability distributions of quantum theory of the EPRB experiment
by using the algebra of probable inference~\cite{COX61,JAYN03}, that is the axioms of probability theory,
without making recourse to quantum theory.
Our conclusions are summarized in Section~\ref{Conclusions}.

\section{EPRB experiments}
\label{EPRBexperiment}

\begin{figure}[t]
\mbox{
\includegraphics[width=8cm]{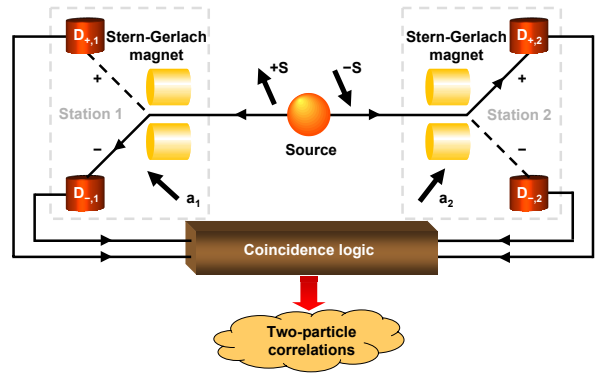}
}
\caption{Schematic diagram of an EPRB experiment with magnetic particles~\cite{BOHM51}.
}
\label{fig1}
\end{figure}

\subsection{Spin 1/2 particles}

Many experimental realizations and quantum theoretical descriptions
of the EPR gedanken experiment~\cite{EPR35}
adopt the model proposed by Bohm~\cite{BOHM51}.
A schematic diagram of the EPRB experiment is shown in Fig.~\ref{fig1}.
A source emits charge-neutral pairs of particles with opposite magnetic moments.
The two particles separate spatially and propagate in free space to an
observation station in which they are detected.
As the particle arrives at station $i=1,2$, it passes through a Stern-Gerlach magnet~\cite{GERL22}.
The magnetic moment of a particle interacts with the inhomogeneous magnetic field of a Stern-Gerlach magnet.
The Stern-Gerlach magnet deflects the particle, depending on the orientation of the magnet and
the magnetic moment of the particle.
The Stern-Gerlach magnet divides the beam of particles in two, spatially well-separated parts~\cite{GERL22}.
The observation that the beam splits into two, and not in a continuum of beams
is interpreted as evidence that the particles carry a magnetic moment that
can take two discrete values; it is quantized~\cite{GERL22}.
In quantum theory, we describe such a magnetic moment using $S=1/2$ operators.
By changing the orientation of the Stern-Gerlach magnet, we change the direction of the
plane that divides the two beams of particles.
In quantum theory language, we say that the quantization axis is determined
by the orientation of the Stern-Gerlach magnet.
As the particle leaves the Stern-Gerlach magnet, it generates a signal in one of the
two detectors. The firing of a detector corresponds to a detection event.

Charge-neutral, magnetic particles that pass through a Stern-Gerlach magnet
not only change their direction of motion but also experience a time-delay,
depending on the direction of their magnetic moment, relative to the direction
of the field in the Stern-Gerlach magnet.
The time-delays in Stern-Gerlach magnets are used to perform spectroscopy of atomic size
magnetic clusters~\cite{HEER89} and atomic interferometry~\cite{CHOR93}.

Real experiments require a criterion to decide which
events, registered in stations 1 and 2, correspond to the detection of particles belonging to a pair
(a single two-particle system).
In EPRB experiments, this criterion is the coincidence in time of the
events~\cite{CLAU74,WEIH98,SAKA06}, as is most clearly illustrated by the EPRB experiments
that use the photon polarization as a two-state system~\cite{FREE72,ASPE82a,ASPE82b,TAPS94,TITT98,WEIH98,ROWE01,FATA04}.

\subsection{Photon polarization}

\begin{figure}[t]
\begin{center}
\includegraphics[width=8cm]{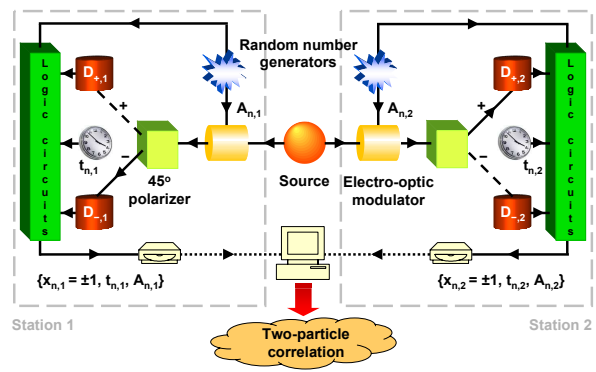}
\caption{Schematic diagram of an EPRB experiment with photons~\cite{WEIH98}.
}
\label{fig1a}
\end{center}
\end{figure}

In Fig.~\ref{fig1a}, we show a schematic diagram of an EPRB experiment
with photons (see also Fig.~2 in~\cite{WEIH98}).
Here, a source emits pairs of photons with opposite polarization.
Each photon of a pair propagates to an observation station
in which it is manipulated and detected.
The two stations are separated spatially and temporally~\cite{WEIH98}.
This arrangement prevents the observation at
station 1 (2) to have a causal effect on the
data registered at station $2$ (1)~\cite{WEIH98}.
As the photon arrives at station $i=1,2$, it passes through an electro-optic
modulator that rotates the polarization of the photon by an angle depending
on the voltage applied to the modulator.
These voltages are controlled by two independent binary random number generators.
As the photon leaves the polarizer, it generates a signal in one of the
two detectors.
The station's clock assigns a time-tag to each generated signal.
Effectively, this procedure discretizes time in intervals of a width that is
determined by the time-tag resolution $\tau$~\cite{WEIH98}.
In the experiment, the firing of a detector is regarded as an event.

As light is supposed to consist of non-interacting photons,
it is not unreasonable to assume that the individual photons
experience a time delay as they pass through the electro-optic modulators or polarizers.
Indeed, according to Maxwell's equation, in the optically anisotropic materials used to fabricate
these devices, plane waves with different polarization propagate
with different velocity and are refracted differently~\cite{BORN64}.

It is clear that, at least conceptually, the EPRB experiments with
photons or massive $S=1/2$ particles are very similar.

\subsection{Idealized experiments}

As it is one of the goals of this paper to demonstrate
that it is possible to reproduce the results of quantum theory (which implicitly assumes idealized conditions)
for the EPRB gedanken experiment by an event-based simulation algorithm,
it would be logically inconsistent to ``recover'' the results of the former
by simulating nonideal experiments.
Therefore, in this paper, we consider ideal experiments only,
meaning that we assume that detectors operate with 100\% efficiency,
clocks remain synchronized forever, the ``fair sampling'' assumption is satisfied~\cite{ADEN07}, and so on.
We assume that the two stations are separated spatially and temporally
such that the manipulation and observation at station 1 (2) cannot have a causal effect on the
data registered at station $2$ (1).
Furthermore, to realize the EPRB gedanken experiment on the computer,
we assume that the orientation of each Stern-Gerlach magnet or electro-optic modulator
can be changed at will, at any time.
Although these conditions are very difficult to satisfy in real experiments,
they are trivially realized in computer experiments.

\subsection{Particle source}

In general, on logical grounds (without counterfactual reasoning),
it is impossible to make a statement about
the directions of the spin (or polarization) of particles emitted by
the source unless we have performed an experiment to determine these directions.
Of course, in a computer experiment we have perfect control and we
can select any direction that we like.
Conceptually, we should distinguish between two extreme cases.
In the first case, we assume that we know nothing about the direction
of the spin (or polarization). We mimic this situation by using
pseudo-random numbers to select the directions.
This is the case that is typical for an EPRB experiment and
we will refer to it as Case I.
In the second case, refered to as Case II, we assume that we know that
the directions of both spins (or polarizations) are fixed (but
not necessarily the same).
A simulation algorithm that aims to reproduce the results
of quantum theory of two $S=1/2$ particles should be able
to reproduce these results for both Case I and II,
{\sl without any change to the simulation algorithm except for the part that
simulates the source}.

\subsection{Data gathered in an EPRB experiment}

Here and in the sequel, we use the EPRB experiment with
$S=1/2$ particles as the primary example.
The case of EPRB experiments that use the photon polarization
can be treated in exactly the same manner, replacing three-dimensional unit vectors by
two-dimensional ones and so on.

In the experiment, the firing of a detector is regarded as an event.
At the $n$th event, the data recorded on a hard disk at station $i=1,2$
consists of $x_{n,i}=\pm 1$, specifying which of the two detectors fired,
the time tag $t_{n,i}$ indicating the time at which a detector fired,
and the unit vector ${\bf a}_{n,i}$ that specifies the direction
of the magnetic field in the Stern-Gerlach magnet.
Hence, the set of data collected at station $i=1,2$ during a run of $N$ events
may be written as
\begin{eqnarray}
\label{Ups}
\Upsilon_i=\left\{ {x_{n,i} =\pm 1,t_{n,i},{\bf a}_{n,i} \vert n =1,\ldots ,N } \right\}
.
\end{eqnarray}

In the (computer) experiment, the data $\{\Upsilon_1,\Upsilon_2\}$ may be analyzed
long after the data has been collected~\cite{WEIH98}.
Coincidences are identified by comparing the time differences
$\{ t_{n,1}-t_{n,2} \vert n =1,\ldots ,N \}$ with a time window $W$~\cite{WEIH98}. %(typically a few ns~\cite{WEIH98}).
Introducing the symbol $\sum'$ to indicate that the sum
has to be taken over all events that satisfy
$\mathbf{a}_i=\mathbf{a}_{n,i}$ for $i=1,2$,
for each pair of directions $\mathbf{a}_1$ and $\mathbf{a}_2$ of the Stern-Gerlach magnets,
the number of coincidences $C_{xy}\equiv C_{xy}(\mathbf{a}_1,\mathbf{a}_2)$ between detectors $D_{x,1}$ ($x =\pm 1$) at station
1 and detectors $D_{y,2}$ ($y =\pm1 $) at station 2 is given by
\begin{eqnarray}
\label{Cxy}
C_{xy}&=&\sumprime_{n=1}^N\delta_{x,x_{n ,1}} \delta_{y,x_{n ,2}}
\Theta(W-\vert t_{n,1} -t_{n ,2}\vert)
,
\end{eqnarray}
where $\Theta (t)$ is the Heaviside step function.
We emphasize that we count
all events that, according to the same criterion as the one employed in experiment,
correspond to the detection of pairs.

The average single-particle counts are defined by
\begin{eqnarray}
\label{Ex}
E_1(\mathbf{a}_1,\mathbf{a}_2)&=&
\frac{\sum_{x,y=\pm1} xC_{xy}}{\sum_{x,y=\pm1} C_{xy}}
,
\nonumber \\
\noalign{and}
E_2(\mathbf{a}_1,\mathbf{a}_2)&=&\frac{\sum_{x,y=\pm1} yC_{xy}}{\sum_{x,y=\pm1} C_{xy}}
,
\end{eqnarray}
where the denominator is the sum of all coincidences.
According to standard terminology, the correlation between $x=\pm1$ and $y=\pm1$ events is defined by~\cite{GRIM95}
\medskip
\begin{widetext}
\begin{eqnarray}
\label{rhoxy}
\rho(\mathbf{a}_1,\mathbf{a}_2)&=&
\frac{
\frac{\sum_{x,y} xyC_{xy}}{\sum_{x,y} C_{xy}}
-\frac{\sum_{x,y} xC_{xy}}{\sum_{x,y} C_{xy}}
\frac{\sum_{x,y} yC_{xy}}{\sum_{x,y} C_{xy}}
}{
\sqrt{
\left(
\frac{ \sum_{x,y} x^2C_{xy} }{\sum_{x,y} C_{xy}}-
(\frac{\sum_{x,y} xC_{xy} }{\sum_{x,y} C_{xy} })^2
\right)%^{1/2}
\left(
\frac{\sum_{x,y} y^2C_{xy}}{\sum_{x,y} C_{xy}}-
(\frac{(\sum_{x,y} yC_{xy}}{\sum_{x,y} C_{xy}})^2
\right)%^{1/2}
}
}
.
\end{eqnarray}
\medskip
\end{widetext}
The correlation $\rho(\mathbf{a}_1,\mathbf{a}_2)$
is $+1$ ($-1$) in the case that $x=y$ ($x=-y$) with certainty.
If the values of $x$ and $y$ are independent, the correlation $\rho(\mathbf{a}_1,\mathbf{a}_2)$
is zero, but the converse is not necessarily true.

In the case of dichotomic variables $x$ and $y$, the correlation $\rho(\mathbf{a}_1,\mathbf{a}_2)$
is entirely determined by the average single-particle counts Eq.~(\ref{Ex})
and the two-particle average
\begin{eqnarray}
\label{Exy}
E(\mathbf{a}_1,\mathbf{a}_2)&=&
\frac{\sum_{x,y} xyC_{xy}}{\sum_{x,y} C_{xy}}
\nonumber \\
&=&\frac{C_{++}+C_{--}-C_{+-}-C_{-+}}{C_{++}+C_{--}+C_{+-}+C_{-+}}
.
\end{eqnarray}
For later use, it is expedient to introduce the function
\begin{equation}
\label{Sab}
S(\mathbf{a},\mathbf{b},\mathbf{c},\mathbf{d})=
E(\mathbf{a},\mathbf{c})-E(\mathbf{a},\mathbf{d})
+
E(\mathbf{b},\mathbf{c})+E(\mathbf{b},\mathbf{d})
,
\end{equation}
and its maximum
\begin{eqnarray}
\label{Smax}
S_{max}&\equiv&\max_{\mathbf{a},\mathbf{b},\mathbf{c},\mathbf{d}} S(\mathbf{a},\mathbf{b},\mathbf{c},\mathbf{d})
.
\end{eqnarray}

In general, the values for the average
single-particle counts $E_1(\mathbf{a}_1,\mathbf{a}_2)$ and $E_2(\mathbf{a}_1,\mathbf{a}_2)$
the coincidences $C_{xy}(\mathbf{a}_1,\mathbf{a}_2)$,
the two-particle averages $E(\mathbf{a}_1,\mathbf{a}_2)$,
$S(\mathbf{a},\mathbf{b},\mathbf{c},\mathbf{d})$, and $S_{max}$
not only depend on the directions $\mathbf{a}_1$ and $\mathbf{a}_2$ but also
on the time-tag resolution $\tau$
and the time window $W$ used to identify the coincidences.

\subsection{Role of the time window}

Most theoretical treatments of the EPRB experiment assume that the correlation,
as measured in the experiment, is given by~\cite{BELL93}
\begin{eqnarray}
\label{CxyBell}
C_{xy}^{(\infty)}&=&\sumprime_{n=1}^N\delta_{x,x_{n ,1}} \delta_{y,x_{n ,2}}
,
\end{eqnarray}
which we obtain from Eq.~(\ref{Cxy}) by taking the limit $W\rightarrow\infty$.
Although this limit defines a valid theoretical model, there is no reason
why this model should have any bearing on the real experiments, in particular
because experiments pay considerable attention to the choice of $W$.
A rational argument that might justify taking this limit is
the hypothesis that for ideal experiments, the value of $W$ should
not matter. However, in experiments a lot of effort is made to reduce (not increase) $W$~\cite{WEIH98,WEIH00}.

As we will see later, using our model it is relatively easy to reproduce the experimental facts
and the results of quantum theory if we neglect contributions that are ${\cal O}(W^2)$.
Furthermore, keeping $W$ arbitrary does not render the mathematics more complicated
so there really is no point of studying the simplified model defined by Eq.~(\ref{CxyBell}):
We may always consider the limiting case $W\rightarrow\infty$ afterwards.

\begin{figure}[t]
\begin{center}
\includegraphics[width=8cm]{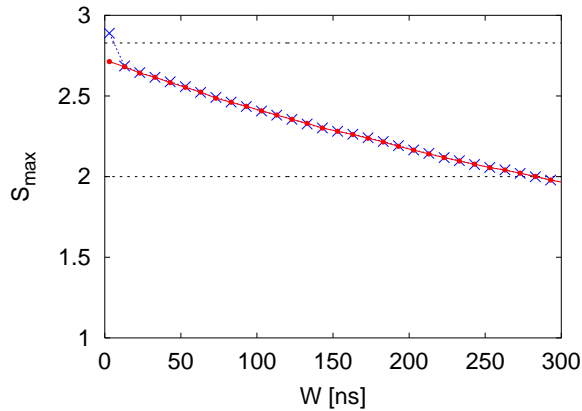}
\caption{$S_{max}$ as a function of the time window $W$,
computed from the data sets contained in the archives
Alice.zip and Bob.zip that can be downloaded from Ref.~\onlinecite{WEIHdownload}.
Bullets (red): Data obtained by
using the relative time shift $\Delta=4$ ns that maximizes the
number of coincidences.
Crosses (blue): Raw data ($\Delta=0$).
Dashed line at $2\sqrt 2 $: $S_{max}$ if the system is described by quantum theory (see Section~\ref{Quantumtheory}).
Dashed line at $2$: $S_{max}$ if the system is described by the
class of models introduced by Bell~\cite{BELL93}.
}
\label{exp1}
\end{center}
\end{figure}

\subsection{Case study: Analysis of experimental EPRB data}
\label{IIG}

It is remarkable that all textbook treatments of the EPRB
experiment assume that the experimental data
is obtained by using Eq.~(\ref{CxyBell}).
This is definitely not the case~\cite{WEIH98,WEIH00}.
We illustrate the importance of the choice of the time window
$W$ by analyzing a data set (the archives
Alice.zip and Bob.zip) of an EPRB experiment with photons
that is publically available~\cite{WEIHdownload}.

In the real experiment, the number of events detected at station 1 is unlikely
to be the same as the number of events detected at station 2.
In fact, the data sets of Ref.~\onlinecite{WEIHdownload} show that
station 1 (Alice.zip) recorded 388455 events while
station 2 (Bob.zip) recorded  302271 events.
Furthermore, in the real EPRB experiment, there may be an
unknown shift $\Delta$ (assumed to be constant during the experiment)
between the times $t_{n,1}$ gathered at station 1 and
the times $t_{n,2}$ recorded at station 2.
Therefore, there is some extra ambiguity in
matching the data of station 1 to the data of station 2.

A simple data processing procedure that resolves this
ambiguity consists of two steps~\cite{WEIH00}.
First, we make a histogram of the time differences
$t_{n,1}-t_{m,2}$ with a small but reasonable resolution
(we used $0.5$ ns).
Then, we fix the value of the time-shift $\Delta$
by searching for the time difference for which
the histogram reaches its maximum, that is we maximize
the number of coincidences by a suitable choice of $\Delta$.
For the case at hand, we find $\Delta=4$ ns.
Finally, we compute the coincidences,
the two-particle average, and $S_{max}$ using
the expressions given earlier.
The average times between two detection events is $2.5$ ms and $3.3$ ms
for Alice and Bob, respectively.
The number of coincidences (with double counts removed) is
13975 and 2899 for ($\Delta=4$ ns, $W=2$ ns) and
($\Delta=0$ , $W=3$ ns) respectively.

\begin{figure}[t]
\begin{center}
\includegraphics[width=8cm]{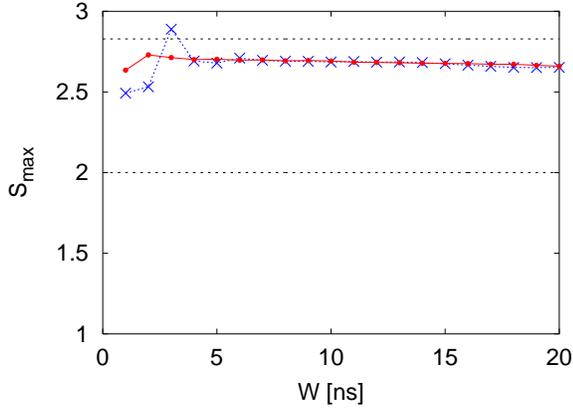}
\caption{Same as Fig.~\ref{exp1} except for the range of $W$.
Bullets (red): Data obtained by
using the relative time shift $\Delta=4$ ns that maximizes the
number of coincidences.
The maximum value of $S_{max}\approx2.73$ is found at $W=2$ ns.
Crosses (blue): Raw data $\Delta=0$.
The maximum value of $S_{max}\approx2.89$ is found at $W=3$ ns.
}
\label{exp2}
\end{center}
\end{figure}

In Figs.~\ref{exp1} and \ref{exp2} we present the results
for $S_{max}$ as a function of the time window $W$.
First, it is clear that $S_{max}$ decreases significantly
as $W$ increases but it is also clear that as $W\rightarrow0$,
$S_{max}$ is not very sensitive to the choice of $W$~\cite{WEIH00}.
Second, the procedure of maximizing the coincidence count
by varying $\Delta$ reduces the maximum value of $S_{max}$ from
a value 2.89 that considerably exceeds the maximum
for the quantum system ($2\sqrt{2}$, see Section~\ref{Quantumtheory}) to a value 2.73 that
violates the Bell inequality and is less than
the maximum for the quantum system.

The fact that the ``uncorrected'' data ($\Delta=0$) violate the rigorous bound for the quantum system
should not been taken as evidence that quantum theory is ``wrong'': As we explain
later, it merely indicates that the way in which the data of the two stations
has been grouped in two-particle events is not optimal.
Put more bluntly, there is no reason why a correlation between
similar but otherwise unrelated data should be described by quantum theory.
In any case, the analysis of the experimental data shows beyond doubt
that a model which aims to describe real EPRB experiments
should include the time window $W$ and that the interesting regime
is $W\rightarrow0$, not $W\rightarrow\infty$ as is assumed
in all textbook treatments of the EPRB experiment.
In Sections~\ref{SimulationModel} and \ref{Kolmogorov}, we show that
our simulation models reproduce the salient features of Figs.~\ref{exp1} and \ref{exp2} quite well
if contributions that are ${\cal O}(W^2)$ can be neglected.

\section{Quantum theory}
\label{Quantumtheory}

In this section we briefly review  some well-known results for the quantum theory of a system of two $S=1/2$ particles and
we give a brief account of the quantum theoretical description of Case I and Case II.
In quantum theory, the state of a system of two $S=1/2$ objects is described by a $4\times4$
density matrix $\rho$~\cite{BALL03}.
The average value of a dynamical variable, represented by
the $4\times4$ matrix $X$ is $\langle X \rangle=\hbox{\bf Tr}\rho X$~\cite{BALL03}.
According to the axioms of quantum theory~\cite{BALL03},
repeated measurements on the two-particle system described by the density matrix $\rho$
yield statistical estimates
for the single-particle expectation values
\begin{eqnarray}
\label{Ei}
\widehat
E_i(\mathbf{a})&=&\langle \mathbf{\sigma}_i\cdot\mathbf{a}\rangle
,
\end{eqnarray}
for $i=1,2$ and the two-particle correlations
\begin{eqnarray}
\label{Eab}
\widehat E(\mathbf{a},\mathbf{b})&=&\langle \mathbf{\sigma}_1\cdot\mathbf{a}\; \mathbf{\sigma}_2\cdot\mathbf{b} \rangle
,
\end{eqnarray}
where $\sigma_i=(\sigma_i^x ,\sigma_i^y ,\sigma_i^z )$
are the Pauli spin-1/2 matrices describing the spin of particle $i=1,2$~\cite{BALL03},
and $\mathbf{a}$ and $\mathbf{b}$ are unit vectors.
We introduce the notation
$\widehat{\phantom{E}}$
to distinguish the quantum theoretical results from the results
obtained by analysis of the data $\{\Upsilon_1,\Upsilon_2\}$.

If the density matrix of the quantum system factorizes, $\rho=\rho_1\otimes\rho_2$,
where $\rho_i$ is the $2\times2$ density matrix of particle $i$.
Then $\widehat E({\bf a},{\bf b})=\widehat E_1({\bf a})\widehat E_2({\bf b})$
and the correlation $\widehat\rho(\mathbf{a}_1,\mathbf{a}_2)=\widehat E({\bf a},{\bf b})-\widehat E_1({\bf a})\widehat E_2({\bf b})=0$.
Hence, $\rho=\rho_1\otimes\rho_2$ is called the uncorrelated quantum state.

Let us denote
$a=\widehat E_1({\bf a})$, $b=\widehat E_1({\bf b})$, and
$c=\widehat E_2({\bf c})$, $d=\widehat E_2({\bf d})$,
such that $S(\mathbf{a},\mathbf{b},\mathbf{c},\mathbf{d})=ac-ad+bc+bd$.
Clearly, $a,b,c,d\in[-1,1]$. For any $a,b,c,d\in[-1,1]$ we have~\cite{ACCA05}
\begin{eqnarray}
\label{eq30a}
|ac-ad+bc+bd|&\le&|ac-ad|+|bc+bd|
\nonumber \\
&\le&|a||c-d|+|b||c+d|
\nonumber \\
&\le&|c-d|+|c+d|
\nonumber \\
&\le&1-cd+1+cd
\nonumber \\
&\le&2.
\end{eqnarray}
Thus, we conclude that if the quantum system is in the uncorrelated state
we must have
\begin{eqnarray}
\label{Smax1}
\widehat S_{max}&\equiv&\max_{\mathbf{a},\mathbf{b},\mathbf{c},\mathbf{d}}
\widehat S(\mathbf{a},\mathbf{b},\mathbf{c},\mathbf{d})\le2
.
\end{eqnarray}

If the density matrix $\rho$ does not factorize, the upperbound to $S_{max}$ can be found as follows~\cite{CIRE80}.
Using the algebraic properties of the Pauli-spin matrices, a simple calculation yields,
\medskip
\begin{widetext}
\begin{eqnarray}
\label{CIRC}
\left(
\sigma_1\cdot\mathbf{a}\;\sigma_2\cdot\mathbf{c}
-
\sigma_1\cdot\mathbf{a}\;\sigma_2\cdot\mathbf{d}
+
\sigma_1\cdot\mathbf{b}\;\sigma_2\cdot\mathbf{c}
+
\sigma_1\cdot\mathbf{b}\;\sigma_2\cdot\mathbf{d}
\right)^2
&=&
4+4\sigma_1\cdot(\mathbf{a}\times\mathbf{b})\; \sigma_2\cdot(\mathbf{c}\times\mathbf{d})
.
\end{eqnarray}
Noting that $\hbox{\bf Tr}\rho X^\dagger Y$ defines an inner product on
the vector space of $4\times4$ matrices $X$ and $Y$, making
use of the fact that $\rho$ is positive semi-definite and that
$\hbox{\bf Tr}\rho=1$, we have
\begin{eqnarray}
\label{Schwarz}
\left| \hbox{\bf Tr}\rho X \right|^2
&=&
\left| \hbox{\bf Tr}\rho^{1/2}\rho^{1/2} X \right|^2
%\nonumber \\
%&\le&
\le
\hbox{\bf Tr}(\rho^{1/2})^\dagger\rho^{1/2}
\hbox{\bf Tr}(\rho^{1/2}X)^\dagger\rho^{1/2} X
%\nonumber \\
%&\le&
=
\hbox{\bf Tr}\rho X^\dagger X
.
\end{eqnarray}
For $X=X^\dagger=
\sigma_1\cdot\mathbf{a}\;\sigma_2\cdot\mathbf{c}
-
\sigma_1\cdot\mathbf{a}\;\sigma_2\cdot\mathbf{d}
+
\sigma_1\cdot\mathbf{b}\;\sigma_2\cdot\mathbf{c}
+
\sigma_1\cdot\mathbf{b}\;\sigma_2\cdot\mathbf{d}
$,
Eq.~(\ref{Schwarz}) becomes
\begin{eqnarray}
\label{CIRC1}
[
\widehat E(\mathbf{a},\mathbf{c})-\widehat E(\mathbf{a},\mathbf{d})
+
\widehat E(\mathbf{b},\mathbf{c})+\widehat E(\mathbf{b},\mathbf{d})
]^2
&\le&
4+4\widehat E(\mathbf{a}\times\mathbf{b},\mathbf{c}\times\mathbf{d})
.
\end{eqnarray}
\medskip
\end{widetext}
As the eigenvalues of $\sigma_1\cdot\mathbf{a}\; \sigma_2\cdot\mathbf{b}$
are $\pm\,\mathbf{a}\cdot\mathbf{b}$, and since $\mathbf{a}$, $\mathbf{b}$,
$\mathbf{c}$, and $\mathbf{d}$ are unit vectors,
we have $|E(\mathbf{a}\times\mathbf{b},\mathbf{c}\times\mathbf{d})|\le1$.
Hence~\cite{CIRE80}
\begin{eqnarray}
\label{Smax2}
|
\widehat S(\mathbf{a},\mathbf{b},\mathbf{c},\mathbf{d})|
&\le&2\sqrt{2}
,
\end{eqnarray}
independent of the quantum state $\rho$.
According to Eqs.~(\ref{Smax1}) and ~(\ref{Smax2}),
if $2<\widehat S_{max}\le2\sqrt{2}$
the quantum system is in a correlated state,
that is $\rho\not=\rho_1\otimes\rho_2$.
For pure states ($\hbox{\bf Tr}\rho^2=1$),
the converse is also true~\cite{GISI91} but, for general states
$\rho$ it is not~\cite{WERN89,HORO95,ARAV95}.
If, in an experiment or simulation, we would find
that $S_{max}>2\sqrt{2}$, the results of this
experiment or simulation cannot be described
by the quantum theory of a system of two $S=1/2$ particles.

We now examine the examples of a maximally correlated (entangled)
quantum state (called Case I) and the uncorrelated quantum state (called Case II)
in more detail.

\subsection{Case I: Singlet state}

The quantum theoretical description of the EPRB experiment assumes that the state of the two spin-1/2 particles
is described by the singlet state $\rho=|\Psi\rangle\langle\Psi|$ where
$|\Psi\rangle=\left(| \uparrow \downarrow \rangle -| \downarrow\uparrow\rangle\right)/\sqrt 2 $
and $|\uparrow\rangle$ ($|\downarrow\rangle$) is the eigenstate
of $\sigma^z$ with eigenvalue $+1$ ($-1$).
For the singlet state, the single-particle expectation values and the two-particle correlations are given by
\begin{eqnarray}
\label{Eis}
\widehat E_i(\mathbf{a}_i)&=&
\langle\Psi| \mathbf{\sigma}_i\cdot\mathbf{a}_i |\Psi\rangle=0\quad;\quad i=1,2
,
\end{eqnarray}
and
\begin{eqnarray}
\label{Eabs}
 \widehat E(\mathbf{a}_1,\mathbf{a}_2)&=&\langle\Psi| \mathbf{\sigma}_1\cdot\mathbf{a}_1\; \mathbf{\sigma}_2\cdot\mathbf{a}_2|\Psi \rangle
 =-\mathbf{a}_1\cdot\mathbf{a}_2
,
\end{eqnarray}
respectively.
A simple calculation shows that $S_{max}=2\sqrt{2}$, in other words, the singlet state
satisfies Eq.~(\ref{Smax2}) with equality.

For the singlet state,
the probability $P(x,y|{\bf a}_1,{\bf a}_2)$
that we observe a pair of events $x,y=\pm1$
under the (fixed) condition $({\bf a}_1,{\bf a}_2)$
is given by
\begin{eqnarray}
\label{Pxy}
P(x,y|{\bf a}_1,{\bf a}_2)&=&
\frac{1-xy{\bf a}_1\cdot{\bf a}_2}{4}
,
\end{eqnarray}
from which it follows that
\begin{eqnarray}
P(x|{\bf a}_1,{\bf a}_2)&=&\sum_{y=\pm1}P(x,y|{\bf a}_1,{\bf a}_2)=1/2,
\nonumber\\
P(y|{\bf a}_1,{\bf a}_2)&=&\sum_{x=\pm1}P(x,y|{\bf a}_1,{\bf a}_2)=1/2,
\end{eqnarray}
and
\begin{eqnarray}
\sum_{x,y=\pm1}P(x,y|{\bf a}_1,{\bf a}_2)&=&1,
\nonumber\\
\sum_{x,y=\pm1}xP(x,y|{\bf a}_1,{\bf a}_2)&=&0,
\nonumber\\
\sum_{x,y=\pm1}yP(x,y|{\bf a}_1,{\bf a}_2)&=&0,
\nonumber\\
\sum_{x,y=\pm1}xyP(x,y|{\bf a}_1,{\bf a}_2)&=&-{\bf a}_1\cdot{\bf a}_2,
\end{eqnarray}
in agreement with the second column of Table~\ref{tab1}.

In the quantum theoretical description,
the state of the two spin-1/2 particles may be correlated
($\rho(\mathbf{a}_1,\mathbf{a}_2)=\widehat E(\mathbf{a}_1,\mathbf{a}_2)$),
even though the particles are
spatially and temporally separated and do not necessarily interact.

\subsection{Case II: Spin-polarized state}

In Case II, $\rho=\rho_1\otimes\rho_2$
where $\rho_j=|\theta_j\phi_j\rangle\langle\theta_j\phi_j|$
and $|\theta_j\phi_j\rangle=\cos(\theta_j/2)|\uparrow\rangle+e^{i\phi_j}\sin(\theta_j/2)|\downarrow\rangle$
for $j=1,2$.
A straightforward calculation shows that
\begin{eqnarray}
\label{Eip}
\widehat E_i(\mathbf{a}_i)&=&\mathbf{a}_i\cdot\mathbf{S}_i;\quad i=1,2,
\\
\widehat E(\mathbf{a}_1,\mathbf{a}_2)&=&\widehat E_1(\mathbf{a}_1)\widehat E_2(\mathbf{a}_2)
,
\end{eqnarray}
where ${\bf S}_{i}=(\cos\phi_i\sin\theta_i,\sin\phi_{i}\sin\theta_i,\cos\theta_i)$.

For the product state,
the probability $P(x,y|{\bf a}_1,{\bf a}_2,{\bf S}_1,{\bf S}_2)$
that we observe a pair of events $x,y=\pm1$
under the (fixed) condition $({\bf a}_1,{\bf a}_2,{\bf S}_1,{\bf S}_2)$
is given by
\begin{equation}
\label{Pxyp}
P(x,y|{\bf a}_1,{\bf a}_2,{\bf S}_1,{\bf S}_2)=
\frac{1+x{\bf a}_1\cdot{\bf S}_1}{2}
\frac{1+y{\bf a}_2\cdot{\bf S}_2}{2},
\end{equation}
and yields expectation values that are
in agreement with the third column of Table~\ref{tab1}.
Obviously, for the spin-polarized state
$\rho(\mathbf{a}_1,\mathbf{a}_2)=
\widehat E(\mathbf{a}_1,\mathbf{a}_2)-\widehat E_1(\mathbf{a}_1)\widehat E_2(\mathbf{a}_2)=0$,
hence there is no correlation in this case.

\begin{table}
\begin{center}
\caption{Quantum system of two $S=1/2$ objects:
The expectation values in the singlet state (Case I)
and in the product state (Case II).
}
\label{tab1}       % Give a unique label
\begin{ruledtabular}
\begin{tabular}{lcc}
%\hline\noalign{\smallskip}
& Case I & Case II  \\
\noalign{\smallskip}\hline\noalign{\smallskip}
$\widehat E_1({\bf a}_1)$         & $0$ & ${\bf a}_1\cdot{\bf S}_1$\\
$\widehat E_2({\bf a}_2)$         & $0$ & ${\bf a}_2\cdot{\bf S}_2$\\
$\widehat E({\bf a}_1,{\bf a}_2)$ & $-{\bf a}_1\cdot{\bf a}_2$& $({\bf a}_1\cdot{\bf S}_1)({\bf a}_2\cdot{\bf S}_2)$\\
%\noalign{\smallskip}\hline
\end{tabular}
\end{ruledtabular}
\end{center}
\end{table}

\subsection{Photon polarization}

In the quantum theoretical description of Case I,
the whole system is described by the state
\begin{align}
\label{eq7}
| \Psi \rangle &=\frac{1}{\sqrt 2 }\left( {| H \rangle
_1 | V \rangle _2 -| V \rangle _1 | H
\rangle _2 } \right)
%\nonumber \\ &
=\frac{1}{\sqrt 2 }\left( {| {HV}
\rangle -| {VH} \rangle } \right),
\end{align}
where $H$ and $V$ denote the horizontal
and vertical polarization and the subscripts refer to photon 1 and 2, respectively.
The state $|\Psi\rangle$ cannot be written as a
product of single-photon states, hence it is an entangled state.

In Case II, the photons have a definite polarization $\eta_1$ and $\eta_2$ when
they enter the observation station.
The polarization of the two photons is described by
the product state
\begin{align}
\label{eq23}
|\Psi\rangle =&(\cos \eta_1|H\rangle_1 +\sin \eta_1|V\rangle_1)
%\nonumber \\ &\times
(\cos \eta_2 |H\rangle_2 +\sin \eta_2 |V\rangle _2).
\end{align}
Using the fact that the two-dimensional vector space with basis vectors $\{|H\rangle,|V\rangle\}$
is isomorphic to the vector space of spin-1/2 particles,
we may use the quantum theory of the latter to describe the EPRB experiments with
photons. The resulting expressions for the averages
are given in Table~\ref{tab2}. They are similar to those of the genuine
$S=1/2$ problem except for the restriction of ${\bf a}_1$ and ${\bf a}_2$ to lie in
planes orthogonal to the direction of propagation of the photons
and the factor of two that multiplies the angles.
The latter reflects the fact that the polarization is
defined modulo $\pi$, not $2\pi$ as in the case of $S=1/2$.

\begin{table}
\begin{center}
\caption{Quantum system of two photon polarizations:
The expectation values in the singlet state (Case I)
and in the product state (Case II) where
$\cos\theta_1={\bf a}_1\cdot{\bf S}_1$,
$\cos\theta_2={\bf a}_2\cdot{\bf S}_2$,
and
$\cos\theta_{1,2}={\bf S}_1\cdot{\bf S}_2$.
}
\label{tab2}       % Give a unique label
\begin{ruledtabular}
\begin{tabular}{lcc}
%\hline\noalign{\smallskip}
& Case I & Case II  \\
\noalign{\smallskip}\hline\noalign{\smallskip}
$\widehat E_1({\bf a}_1)$         & $0$ & $\cos2\theta_1$\\
$\widehat E_2({\bf a}_2)$         & $0$ & $\cos2\theta_2$\\
$\widehat E({\bf a}_1,{\bf a}_2)$ & $-\cos2\theta_{1,2}$ & $\cos 2\theta_1 \cos 2\theta_2$
%\noalign{\smallskip}\hline
\end{tabular}
\end{ruledtabular}
\end{center}
\end{table}

\section{Relating quantum theory and data}
\label{theorydata}

There is no doubt that quantum theory is very successful in describing a vast amount of phenomena
in which we observe the ensemble average of many measurements
that are repeated under the same external conditions~\cite{HOME97,BALL03}.
The EPRB experiments seem to be no exception:
The analysis of the experimental data according to the procedure discussed earlier, demonstrates that
$E(\mathbf{a}_1,\mathbf{a}_2)\approx \widehat E(\mathbf{a}_1,\mathbf{a}_2)$~\cite{FREE72,ASPE82a,ASPE82b,TAPS94,TITT98,WEIH98,ROWE01,FATA04}.

On the other hand, as is well known from the early days of quantum mechanics,
quantum theory itself has nothing to say about the individual events (quantum measurement paradox)~\cite{HOME97,BALL03}.
The very concept of an event cannot be reconciled with quantum theory~\cite{HOME97,BALL03}.

In this section, we elaborate on the relation between quantum theory and (experimental) data.

\subsection{From quantum theory to experimental data}
\label{quantumdata}

The fundamental problem of relating the object in the mathematical formalism of quantum theory
to experimental facts may be solved by
(1) interpreting the state of the system as the probability distribution for events to occur
and by
(2) supplementing quantum theory by a Bernouilli process~\cite{GRIM95,JAYN03} that generates logically
independent events according to the prescribed probability distribution, the so-called
measurement postulate.
Thus, we have
\begin{equation}
\label{logic1}
\hbox{Quantum theory} +
\hbox{Bernouilli process} \Rightarrow \hbox{Events}
.
\end{equation}
All treatments of quantum theory that we are aware of
turn the logical implication Eq.~(\ref{logic1}) around, without any justification and
declare all quantum events to be uncorrelated random.
Of course, it might be the case that the analysis of experimental data
supports the hypothesis that the events are generated as Bernoulli trials.
However, there is rather compelling experimental evidence that
successive events are correlated~\cite{JENNE00}.
Notwithstanding this, using Eq.~(\ref{logic1}) we are in the position to use
quantum theory and discuss events in a mathematically well-defined context.

For simplicity, in the example of the EPRB experiment,
we focus on the case where ${\bf a}_1$ and ${\bf a}_2$ are fixed in time.
Let us then inquire how we can simulate the quantum theoretical results
of the EPRB experiment (see Table I) using the procedure
laid out by Eq.~(\ref{logic1}).

According to the axioms of quantum theory, in each event
we observe only one of the eigenvalues of the dynamical variable that is being
measured~\cite{BALL03}.
For the case at hand, the eigenvalues of
$\mathbf{\sigma}_1\cdot\mathbf{a}_1$,
$\mathbf{\sigma}_2\cdot\mathbf{a}_2$,
and
$\mathbf{\sigma}_1\cdot\mathbf{a}_1\;\mathbf{\sigma}_2\cdot\mathbf{a}_2$
are $\pm1$.
Then, according to Eq.~(\ref{logic1}), what is left to do is
to imagine three Bernoulli processes that
generate sets of data ${\cal Q}=\{a_n=\pm1,b_n=\pm1,c_n=\pm1|n=1,\ldots,N\}$
such that for a sufficiently large number of events $N$,
\begin{eqnarray}
\frac{1}{N}\sum_{n=1}^N a_n
&\approx&
\widehat E_1({\bf a}_1)
,
\nonumber\\
\frac{1}{N}\sum_{n=1}^N b_n
&\approx&
\widehat E_2({\bf a}_2)
,
\nonumber\\
\frac{1}{N}\sum_{n=1}^N c_n
&\approx&
\widehat E({\bf a}_1,{\bf a}_2),
\label{abc}
\end{eqnarray}
for all ${\bf a}_1$ and ${\bf a}_2$, the expressions for
$\widehat E_1({\bf a}_1)$, $\widehat E_2({\bf a}_2)$ and
$\widehat E({\bf a}_1,{\bf a}_2)$ being given in Table~\ref{tab1}.
The fact that we use Bernouilli processes
in which every trial is drawn from the same probability
distribution guarantees, by the law of large numbers,
that the average over all events converges
with probability one to the ensemble average~\cite{GRIM95,JAYN03},
which in the present case is given by quantum theory.
Note that quantum theory does not impose
any relation (correlation) between the numbers $a_n$, $b_n$, and $c_n$,
other than that Eq.~(\ref{abc}) should hold.

In general, generating data $(x,y)$ according
to the probability distributions Eqs.~(\ref{Pxy}) and~(\ref{Pxyp})
is a nearly trivial exercise.
Once we have solved the quantum mechanical problem,
that is, once we have the explicit form of the wave function,
constructing a Bernoulli process that generates
events according to the explicit form is a simple task.
In practice, we assume that the pseudo-random
number generator that we employ produces Bernoulli trials,
a hypothesis that cannot be justified in a mathematically strict sense.

\subsection{Fundamental problem}
\label{FundamentalProblem}

Let us now try to relate the quantum theoretical expectation values that appear
in Eqs.~(\ref{Ei}) and (\ref{Eab}) to the actual data.
In general, the probability for observing a pair
of dichotomic variables $\{x,y\}$ can be written as
%
%\begin{eqnarray}
\begin{equation}
\label{Pxy1}
\widetilde P(x,y|\mathbf{a},\mathbf{b})=\frac{1+x \widetilde E_x(\mathbf{a},\mathbf{b})
+y \widetilde E_y(\mathbf{a},\mathbf{b})+xy \widetilde E_{xy}(\mathbf{a},\mathbf{b})}{4}
,
\end{equation}
%\end{eqnarray}
%
from which, by the standard rules of probability theory, it follows that
\begin{eqnarray}
\label{Px}
\widetilde P_y(x|\mathbf{a},\mathbf{b})&\equiv&\sum_{y=\pm1} \widetilde P(x,y|\mathbf{a},\mathbf{b})=
\frac{1+x \widetilde E_x(\mathbf{a},\mathbf{b})}{2}
,
\\
\label{Py}
\widetilde P_x(y|\mathbf{a},\mathbf{b})&\equiv&\sum_{x=\pm1} \widetilde P(x,y|\mathbf{a},\mathbf{b})=
\frac{1+y \widetilde E_y(\mathbf{a},\mathbf{b})}{2}
.
\end{eqnarray}
By definition, $x$ and $y$ are logically independent if and only if
$\widetilde P(x,y|\mathbf{a},\mathbf{b})=
\widetilde P_y(x|\mathbf{a},\mathbf{b})\widetilde P_x(y|\mathbf{a},\mathbf{b})$~\cite{GRIM95,BALL03,JAYN03}.
If $x$ and $y$ are logically independent it is easy to show that $\widetilde E_{xy}=\widetilde E_x\widetilde E_y$.
In general, the converse is not necessarily true~\cite{GRIM95,BALL03,JAYN03} but in this particular case it is.
Indeed, if $\widetilde E_{xy}=\widetilde E_x\widetilde E_y$, if follows directly from Eq.~(\ref{Pxy1}) that
$\widetilde P(x,y|\mathbf{a},\mathbf{b})=\widetilde P_y(x|\mathbf{a},\mathbf{b})\widetilde P_x(y|\mathbf{a},\mathbf{b})$.
Thus, for the case we are treating here, $\widetilde E_{xy}\not=\widetilde E_x\widetilde E_y$
if and only if $x$ and $y$ are logically dependent.

In quantum theory, we have two different cases also.
If the density matrix of the two spin-1/2 particle quantum system factorizes (Case II), we have
$\langle \mathbf{\sigma}_1\cdot\mathbf{a}\;\mathbf{\sigma}_2\cdot\mathbf{b} \rangle=
\langle \mathbf{\sigma}_1\cdot\mathbf{a} \rangle \langle \mathbf{\sigma}_2\cdot\mathbf{b} \rangle$
and the state of the system is completely characterized by
$\widehat E_1({\bf a})$ and $\widehat E_2({\bf b})$.
However, if the density matrix does not factorize (Case I), a complete
characterization of this entangled state requires the knowledge of
$\widehat E_1({\bf a})$, $\widehat E_2({\bf b})$, and
$\widehat E({\bf a},{\bf b})$.
Upto this point, it seems that there is full analogy with the probabilistic model
of the data, but we still have to relate the quantum theoretical
expressions to the observed data.

To this end, we invoke the postulate that states that the possible values of a dynamical
variable in quantum theory are the eigenvalues of the linear operator that corresponds to this variable~\cite{BALL03}.
For the case at hand, the operators are $\mathbf{\sigma}_1\cdot\mathbf{a}$,
$\mathbf{\sigma}_2\cdot\mathbf{b}$, and
$\mathbf{\sigma}_1\cdot\mathbf{a}\;\mathbf{\sigma}_2\cdot\mathbf{b}$,
with eigenvalues $\widehat x=\pm1$, $\widehat y=\pm1$ and $\widehat z=\pm1$, respectively.

It is evident that the triples $\{\widehat x,\widehat y,\widehat z\}$ cannot represent the
data Eq.~(\ref{Ups}) that is recorded and analyzed in real EPRB experiments~\cite{FREE72,ASPE82b,TAPS94,TITT98,WEIH98,ROWE01,FATA04}:
The quantum mechanical model is trivially incomplete in that it has no means to describe the time-tag data.
But, quantum theory is incomplete in a more fundamental sense~\cite{EPR35,BOHM51}.

First, let us consider an experiment that produces $z$ only.
In general, the probability to observe $z$ can be written as
%
%\begin{eqnarray}
\begin{equation}
\widetilde P(z|\mathbf{a},\mathbf{b})=\frac{1+ z \widetilde E_z(\mathbf{a},\mathbf{b})}{2}
.
\end{equation}
%\end{eqnarray}
%
A consistent application of the postulates of quantum theory yields
%
%\begin{eqnarray}
\begin{equation}
\label{Pz1}
P(\widehat z|\mathbf{a},\mathbf{b})=\frac{1+\widehat z \widehat E(\mathbf{a},\mathbf{b})}{2}
,
\end{equation}
%\end{eqnarray}
%
and we would use
$\widehat E(\mathbf{a},\mathbf{b}) = \widetilde E_{z}(\mathbf{a},\mathbf{b})$
to relate the theoretical result to the data.
Likewise, we could imagine an experiment that produces $x$ ($y$)
and use
$\widehat E_1(\mathbf{a}) = {E}_x(\mathbf{a},\mathbf{b})$
($\widehat E_2(\mathbf{b}) = {E}_y(\mathbf{a},\mathbf{b})$)
to relate the theoretical description to the data.

Second, we ask whether it is possible to describe by quantum theory, an experiment that yields the data $\{x,y\}$.
According to the postulates of quantum theory, the probabilities
for the eigenvalues to take the values $\{\widehat x,\widehat y\}$ are given by
\begin{eqnarray}
\label{Px1}
P(\widehat x|\mathbf{a},\mathbf{b})&=&\frac{1+\widehat x \widehat E_1(\mathbf{a})}{2}
,
\\
\label{Py1}
P(\widehat y|\mathbf{a},\mathbf{b})&=&\frac{1+\widehat y \widehat E_2(\mathbf{b})}{2}
,
\end{eqnarray}
where $\widehat x$, and $\widehat y$ are logically independent random variables,
that is each measurement of a dynamical variable constitutes a Bernouilli trial~\cite{BALL03}.
Then, we would use
$\widehat E_1(\mathbf{a}) = {E}_x(\mathbf{a},\mathbf{b})$
and
$\widehat E_2(\mathbf{b}) = {E}_y(\mathbf{a},\mathbf{b})$
to relate the theory to the data.

But the real data is $\{x,y\}$, not the
logically independent random variables $\{\widehat x,\widehat y\}$ of the mathematical model.
Therefore the quantum theoretical description of an experiment that yields $\{x,y\}$ is
necessarily incomplete if the data is such that $E_{xy}\not=E_xE_y$.

The fact that EPRB experiments show good
agreement with the quantum theory of two $S=1/2$ objects
is not in conflict with this reasoning:
In real EPRB experiments, the coincidences are computed according to
Eq.~(\ref{Cxy}), which includes the time-tag information,
about which quantum theory has nothing to say.
Hence there is no logical inconsistency.

\subsection{From experimental data to quantum theory}
\label{dataquantum}

Let us now turn things around and ask the much more interesting question how we,
as observers, relate the observed data of an EPRB experiment to quantum theory.
To simplify the discussion, we assume that the directions
${\bf a}_1$ and ${\bf a}_2$ are fixed.
Thus, we start from the data set
\begin{eqnarray}
\label{setE}
{\cal E}&=&\{ x_{n,1},x_{n,2},t_{n,1},t_{n,2} \vert n =1,\ldots ,N \},
\end{eqnarray}
and ask the question how to relate these numbers to the set of data
\begin{eqnarray}
\label{setQ}
{\cal Q}=\{a_n,b_n,c_n|n=1,\ldots,N\},
\end{eqnarray}
that we obtained by adopting the procedure Eq.~(\ref{logic1}).

It is not difficult to see that there are no a-priori rules.
How could there be rules?
In general, there is no guarantee that the data ${\cal E}$
that resides on the hard disk of the experimenter's
computer has been produced by a physical system and not by, for instance,
a bug in the operating system that is controlling the computer.
Moreover, for bonafide experimental data,
it should not matter who carries out the data analysis:
Once the data has been recorded and there is
agreement on the procedure to analyze this data,
the results (but not necessarily the subjective conclusions)
should not depend on whether or not the individual
that performs the data analysis ``knows'' about quantum theory.
The following example may be useful to understand the conceptual problem.

\subsubsection{Relating frequencies to probabilities}
\label{freq2prob}

Let us consider the experiment in which we toss a coin $N$ times.
The set of $N$ observations looks like $\{H,H,T,\ldots\}$
where $H$ and $T$ denote head and tails, respectively.
From the set of data, we find that the number of times that the coin ends up with
tails on the floor is $h$.
Thus, the frequency with which we observe head is then $f=h/N$,
which clearly is a well-defined number.
Little thought shows that without any further
knowledge/assumption about the experiment,
that is all we can say (of course, we could calculate
correlations between events and so on but this does not
change the essential point of the discussion).

Imagining that we can continue the experiment
forever does not help either because
$\lim_{N\rightarrow\infty} h/N$ is not well-defined~\cite{GRIM95,TRIB69,JAYN03}.
Indeed, it may happen that we never observe heads or always observe heads.
If, in our description of the experiment, we would like to go beyond
just giving the numbers $(h_i,N_i)$ for $i=1,\ldots,M$ repetitions
of the experiment, we have to make additional assumptions.

Implicit in the interpretation of most scientific experiments is the assumption
that there is some underlying process that generates the data.
In the simple case of the coin, assuming
Newton's law holds, solving the equations of motion
allows us to predict the outcome of each individual toss~\cite{JAYN03}.
This outcome depends on how well we know the initial
conditions, the precise form of the force field and so on.

If a description on the level of individual events
seems too complicated, or if we do not have enough knowledge to describe the whole
experimental situation (as in the case of the coin),
it is customary to postulate that there is some
underlying probabilistic process that determines
the frequency with which the events will be observed.

It is instructive to see how the process of reasoning works in the case of the coin
(the use of quantum theory to describe observed phenomena requires the same logic).
As usual, the simplest probabilistic model for the outcome of the experiment of tossing the coin,
assumes that (1) there is a probability $p$ to observe heads
and that (2) this probability is logically independent of what happens at other tosses.
Now, these are nice words but, in the absence of any experimental data, what do they mean?

The probability $p$ is a mathematical concept that we use to encode, by a real number in the interval $[0,1]$,
our state of knowledge about the problem~\cite{TRIB69,JAYN03}.
The statement that this probability is logically independent of what happens at other tosses
cannot be expressed in terms of frequencies~\cite{GRIM95,TRIB69,JAYN03}.
It is an hypothesis that we make without knowing what the frequencies and correlations
between the events will be.
Once we have collected the experimental data, we may compute the probability
for this hypothesis to be true or not and we may also use the
observed frequency to assign a value to the probability $p$~\cite{GRIM95,TRIB69,JAYN03}.

From a logical and conceptual point of view, it is extremely important
to realize that the first step is to define the concept of ''probability''
through the Kolmogorov set of axioms or through the more general inductive logic
approach (see also Section~\ref{Discussion})~\cite{JAYN03}.
Then, and only then, it may make sense to use the observed frequency
to assign a number to the probability for an event to occur.
We continue with the example of the coin to illustrate this point.

Now imagine a thought experiment (= a mental construct) in which we toss the coin $N$ times.
Note that in a strict mathematical sense,
the mathematical model cannot be simulated by an algorithm on a digital computer,
which by construction is a deterministic machine.
Of course, using pseudo-random numbers, we can simulate events that
are unpredictable to anyone who does not know the initial state
or the pseudo-random number generator algorithm.
The mathematical model can then be used to test
whether it describes the global features (but not the individual events) well.

A direct, constructive proof that probabilities are defined
through frequencies would be to invent a practical procedure (algorithm) that simulates
the tossing of the coin such that the probability for head is {\sl exactly} $p$ and
such that each toss is {\sl logically independent} from all others.
Such an algorithm does not exist:
The concept of probability is a mental construct and
has no meaning in the realm of algorithms that generate events
but, that does not imply that the concept of probability would be
useless for describing some of the features of the data generated by these algorithms.

In essence, we are repeating what has been said in the introduction.
Looking back at the diagram in Fig.~\ref{fig0},
the mathematical model is located at the right hand side (model space).
The mathematical model itself does not ``produce'' events.
This is done by some algorithm (data space).
We can test the various hypotheses that underpin the mathematical model by calculating
expectation values (ensemble averages in the case of the coin)
and, using the mathematical machinery of probability theory, compute the probability that these
hypotheses are correct.
Let us now see how this works in the case of the coin.

According to the assumed mathematical model,
the probability to observe $k$ heads and $N-k$ tails in a thought experiment involving $N$ tosses
is given by~\cite{GRIM95,TRIB69,JAYN03}
\begin{eqnarray}
\label{Pk}
P(k|N,Z)=\frac{N!}{k! (N-k)!} p^k (1-p)^{N-k},
\end{eqnarray}
where $Z$ represents all other knowledge about the experiment not contained in $k$ and $N$~\cite{TRIB69,JAYN03}.
If $m$ denotes the number of heads such that $P(m|N,Z)=\max_{k}P(k|N,Z)$, we have
\begin{eqnarray}
\label{Pk2}
\frac{P(m|N,Z)}{P(m+1|N,Z)}=\frac{(m+1)}{N-m} \frac{1-p}{p}\ge1,
\\ \noalign{\noindent and}%\nonumber\\
\frac{P(m|N,Z)}{P(m-1|N,Z)}=\frac{N-m+1}{m} \frac{p}{1-p}\ge1,
\end{eqnarray}
from which it follows that
\begin{eqnarray}
\frac{m}{N+1}&\le & p\le \frac{m+1}{N+1},
\\ \noalign{\noindent and}%\nonumber\\
(1+\frac{1}{N})p-\frac{1}{N}&\le & \frac{m}{N}\le(1+\frac{1}{N})p.
\label{fm}
\end{eqnarray}
According to our mathematical model, of all $k=0,\ldots,m$,
the value of $k$ that has the largest probability to occur is $m$
and from Eq.~(\ref{fm}) it follows that as $N$ increases, $m/N\rightarrow p$.
Of course, we can easily calculate other useful quantities
such as the ensemble average $\langle k/N \rangle=N^{-1}\sum_{k=0}^N k P(k|N,Z)=p$
and the variance $\langle (k/N)^2 \rangle-\langle k/N \rangle^2=N^{-1}p(1-p)$.

We now consider the real experiment in which we toss the coin and
assign the value $x_n=0,1$, if at the $n$th toss, we observe tail or head,
respectively. The frequency of heads is then $f=N^{-1}\sum_{n=1}^N x_n$.
The next logical step is to assume that the mathematical model, described above, is valid.
Then, for the most likely experiment (the one that occurs with the largest frequency)
we have $(1+N^{-1})p-N^{-1}\le f\le(1+N^{-1})p$.
Furthermore, the ensemble average of each event $x_n$ becomes a meaningful concept and,
if we compute the ensemble average of the frequency, we find $\langle f\rangle=p$ as naively expected.
Thus, it makes sense to use the observed frequency $f$ for assigning
a number to the symbol $p$ in the mathematical theory.
At this point, the mathematical theory has been ``connected'' to the observed phenomena.
Once this connection has been made, we can (and should) use the tools of probability theory
to compute the probability that the assumptions of the mathematical model are correct by confronting
the mathematical results for various ensemble averages to the corresponding averages of the
observed data.

From this simple example, we see that in order to
attach a meaning to the observed frequencies, we
first need to introduce a mathematical model, probability
theory in this case, and make the hypothesis
that the outcome of the toss is determined by
a Bernouilli process with probability $p$.
Only when this hypothesis has been made,
it can be proven that the observed frequency approaches $p$ as $N\rightarrow\infty$
with probability one~~\cite{GRIM95,TRIB69,JAYN03}.
Thus, the concept of probability and probability theory have to be introduced first.
Only then we can use probability theory to
relate the variables in the probabilistic model ($p$ in the example of the coin)
to the observed data.

This simple example clearly shows that frequencies and probabilities
have a different logical status~\cite{TRIB69,JAYN03,BALL03}.
Frequencies are the things that we observe (data space in Fig.~\ref{fig0})
and exhibit a causal dependence on the conditions under which the data is recorded.
Probability theory is a well-defined mathematical model (model space in Fig.~\ref{fig0})
that allows us to think in a rational, logical manner~\cite{TRIB69,JAYN03}.
Probabilities express logical relationships.
A problem with the conceptual difference is that in many instances, simply
using the frequency to assign a value to the probability works so well
that we may be inclined to forget that there is a fundamental
difference between the two.
Although it is generally recognized that logical implication is
not the same as physical causation, mixing up frequencies and probabilities leads
to bizarre conclusions~\cite{TRIB69,JAYN03,BALL03}.
As we discuss later, the mysteries surrounding
the EPR paradox and Bell's theorem dissolve if
one recognizes that physical cause and logical dependence
are fundamentally different concepts~\cite{JAYN03}.

\subsubsection{Relating EPRB data to quantum theory}

In the case of the EPRB experiment, we immediately see that we face the same fundamental problem
if we go beyond the description of merely giving the data collected in the experiment.
To make progress in understanding the behavior of the system as
it is revealed to us by our (experimental) method of questioning,
we have three options:

\begin{enumerate}
\item{Use the established mathematical framework
of probability theory to relate the
quantities that appear in this theory
(probabilities) to experimentally observed facts (expected frequencies).}
\item{Construct an event-based computer model that directly generates
the data set Eq.~(\ref{setE}), with expectation values that agree
with those of quantum theory.
\item{Without relying on concepts of quantum theory,
construct a probabilistic model that predicts
the expected frequencies as observed in the experiment.}
}

\end{enumerate}

As quantum theory has nothing to say about individual events~\cite{HOME97,BALL03},
logically speaking option (2) cannot make any reference to quantum theory.
Sections \ref{SimulationModel} and \ref{ProbabilisticModel} are devoted to options (2) and (3),
respectively.
For now, we continue with option (1).

If we measure a property of a single particle, from Eq.~(\ref{abc}) we naively expect that the assignment
\begin{eqnarray}
\label{E12}
\widehat E_1({\bf a}_1)
%&\approx&
&\leftarrow&
\frac{\sum_{x,y=\pm1} xC_{xy}}{\sum_{x,y=\pm1} C_{xy}}
,
\nonumber\\
\widehat E_2({\bf a}_2)
%&\approx&
&\leftarrow&
\frac{\sum_{x,y=\pm1} yC_{xy}}{\sum_{x,y=\pm1} C_{xy}}
,
\end{eqnarray}
holds with probability one.
Note that Eq.~(\ref{E12}) contains contributions from the events
that fall within the coincidence window only.
As explained earlier, for the assignments Eq.~(\ref{E12}) to make sense
mathematically, we have to assume that there is an underlying
probabilistic process that generates the data $\{x_{n,i}\}$.
The fact that quantum theory describes a very large variety of experimental data
strongly suggests that the assignment Eq.~(\ref{E12}) makes a lot of sense.

As explained earlier, for the quantum dynamical variable $\mathbf{\sigma}_1\cdot\mathbf{a}\; \mathbf{\sigma}_2\cdot\mathbf{b}$,
it is not clear at all how to relate its eigenvalue $c_n$ to the data set Eq.~(\ref{setE}).
What does it mean to measure a common property of a system of two particles?
Why is the time-tag data absent in the quantum theoretical description while it
is of vital importance for the experiment?
Evidently, we need a proper operational definition of ``a system of two particles''
in terms of the observed data.

As it is our aim to reproduce the experimental results as well as
the results of the quantum model for the experiment,
it would be logically inconsistent to adopt a definition
that is different from the one used in real EPRB experiments.
Therefore, we should consider the assignment
\begin{eqnarray}
\label{cn}
\widehat E({\bf a}_1,{\bf a}_2)
\leftarrow
\frac{\sum_{x,y} xyC_{xy}}{\sum_{x,y} C_{xy}}
,
\end{eqnarray}
where the frequency to observe systems of two particles is given by
\begin{eqnarray}
\label{fn}
f({\bf a}_1,{\bf a}_2,W)&=&\frac{1}{N}\sum_{x,y} C_{xy}
\nonumber \\
&=&\frac{1}{N}
\sumprime_{n=1}^N
\Theta(W-\vert t_{n,1} -t_{n ,2}\vert)
.
\end{eqnarray}
In Eq.~(\ref{fn}), the coincidence in time enters because it
is an essential ingredient in any EPRB experiment.
The expression for the coincidence is an operational procedure
to define precisely, in terms of the observed data, the meaning of
the statement that two particles constitute a two-particle system.

\section{Simulation model}
\label{SimulationModel}

In this section, we take up the main challenge, the
construction of locally causal (in Einstein's sense) processes
that generate the data sets Eq.~(\ref{Ups}) such that
they reproduce the results of quantum theory, summarized in Table~I.

A concrete simulation model of the EPRB experiment sketched in Fig.~\ref{fig1} requires
a specification of the information carried by the particles,
the algorithm that simulates the source and
the observation stations, the Stern-Gerlach magnets, and the procedure to analyze the data.
We now describe a computer simulation model that generates the data $\{\Upsilon_1,\Upsilon_2\}$, see Eq.~(\ref{Ups}).
From the specification of the algorithm, it will be clear that it complies with
Einstein's criterion of local causality on the ontological level: Once the particles
leave the source, an action at observation station 1 (2) can, in no way,
have a causal effect on the outcome of the measurement at observation
station 2 (1).

In this section, we limit the discussion to systems of two $S=1/2$ particles.
The algorithm that simulates the EPRB experiments with photons, as well as the results
of the simulations, are very similar to those presented here.
A detailed account of the simulations for the photon system can be found
elsewhere~\cite{RAED07a,RAED07b,ZHAO07a}.

\subsection{Algorithm}
\subsubsection{Source and particles}

As in the quantum theoretical treatment of the problem,
we will consider two different cases.
In Case I, the source emits particles that carry a unit vector
${\bf S}_{n,i}=(-1)^{i+1}(\cos\varphi_{n} \sin\theta_n,\sin\varphi_{n}\sin\theta_n,\cos\theta_n)$,
representing the magnetic moment (or spin) of the particles.
The spin of a particle is completely characterized by $\varphi_{n}$
and $\cos\theta_n$, which we assume to be distributed uniformly over the interval $[0,2\pi[$
and [-1,1], respectively.
In Case II, the source emits particles that carry fixed unit vectors ${\bf S}_{n,i}={\bf S}_{i}$.

\subsubsection{Observation station}
\label{ObservationStation}

Prior to the data collection, we fix the number $M$ of different directions of the Stern-Gerlach magnets.
We use $4M$ pseudo-random numbers to fill the arrays $({\bf b}_{i,1},...,{\bf b}_{i,M})$
for $i=1,2$ (in the photon experiments of Aspect \textit{et al.} and Weihs \textit{et al.}, $M=2$~\cite{ASPE82b,WEIH98}).

\subsubsection{Stern-Gerlach magnet}

The input-output relation of a Stern-Gerlach magnet is rather simple:
For a fixed direction $\mathbf{a}_i$ of the field,
the Stern-Gerlach magnet deflects a particle with magnetic moment ${\bf S}_{n,i}$
in a direction that we label by $x_{n,i}=\pm1$.
As the particle travels through the Stern-Gerlach magnet, the magnetic moment
of the particle changes from ${\bf S}_{n,i}$ to ${\bf S}_{n,i}=x_{n,i}\mathbf{a}_i$.

According to the simple quantum mechanical model
of the Stern-Gerlach experiment~\cite{BALL03},
for fixed ${\bf S}$ and fixed ${\bf a}_i$,
the probability to observe $x_{n,i}=\pm1$ is  $(1\pm{\bf S}\cdot\mathbf{a}_i)/2$.
Thus, in this case, the simulation algorithm should
generate the sequence $x_{n,i}=\pm1$ such that
\begin{equation}
\label{eq25}
\lim_{N\to \infty }
\frac{1}{N}\sum_{n=1}^N {x_{n,i}}=\langle x_{n,i}\rangle= {\bf S}\cdot{\bf a}_i,
\end{equation}
with probability one.
However, if the input consists of uniformly distributed ${\bf S}_{n,i}$,
the sequence of output bits $x_{n,i}=\pm1$ should satisfy
\begin{equation}
\label{eq26}
\lim_{N\to \infty }
\frac{1}{N}\sum_{n=1}^N {x_{n,i}}=\langle x_{n,i}\rangle=0,
\end{equation}
with probability one,
independent of the orientation ${\bf a}_i$ of the Stern-Gerlach magnet.
We now consider two algorithms, a deterministic and a pseudo-random one,
that simulate the operation of a Stern-Gerlach magnet.

\medskip
{\bf Deterministic model.}
Elsewhere, we have demonstrated that simple deterministic, local, causal and classical
processes that have a primitive form of learning capability
can be used to simulate quantum systems, not by solving a wave equation
but directly through event-by-event simulation~\cite{RAED05b,RAED05c,RAED05d,MICH05}.
The events are generated such that their frequencies of occurrence agree with the probabilities of
quantum theory.
In this simulation approach, the basic processing unit is called a
deterministic learning machine (DLM)~\cite{RAED05b,RAED05c,RAED05d,MICH05,download}.

A DLM is a device that
exchanges information with the particles that pass through it.
It learns by comparing the message carried by an event with predictions
based on the knowledge acquired by the DLM during the processing of previous events.
The DLM tries to do this in an efficient manner, effectively by minimizing the difference
of the data in the message and the DLM's internal representation of it~\cite{RAED05b,RAED05c,RAED05d,MICH05}.
A DLM learns by processing successive events but does not store the data contained in the individual events.

Connecting the input of a DLM to the output of another DLM yields
a locally connected network of DLMs.
A DLM within the network locally processes the data contained
in an event and responds by sending a message that may be used as input for another DLM.
Networks of DLMs process messages in a sequential manner and
only communicate with each other by message passing: They satisfy Einstein's criterion of local causality.

For the present purpose, we only need the simplest version of the DLM~\cite{RAED05b}.
The DLM that we use to simulate the operation of the Stern-Gerlach magnet is defined as follows.
The internal state of the $i$th DLM, after the $n$th event, is described by one real variable $u_{n,i}$.
Although irrelevant for what follows, this variable may be thought of as describing the fluctuations
of the applied field due to the passage of an uncharged particle that carries a magnetic moment.
As the particle with spin ${\bf S}_{n,i}$ communicates (interacts) with the DLM (applied field),
the latter updates its internal state according to
\begin{eqnarray}
\label{dlm1}
u_{n,i}=\left\{
\begin{array}{lllll}
l u_{n-1,i}+1-l & \mbox{if} & {\bf S}_{n,i}\cdot\mathbf{a}_i&\ge& l u_{n-1,i}\\
l u_{n-1,i}-1+l & \mbox{if} & {\bf S}_{n,i}\cdot\mathbf{a}_i&< &l u_{n-1,i}\\
\end{array}
\right.
,
\end{eqnarray}
and the spin changes according to
\begin{eqnarray}
\label{dlm2}
{\bf S}_{n,i}=\left\{
\begin{array}{lllll}
+\mathbf{a}_i& \mbox{if} & {\bf S}_{n,i}\cdot\mathbf{a}_i&\ge& l u_{n-1,i}\\
-\mathbf{a}_i& \mbox{if} & {\bf S}_{n,i}\cdot\mathbf{a}_i&< &l u_{n-1,i}\\
\end{array}
\right.
,
\end{eqnarray}
corresponding to spin up and spin down (relative to the direction
of the magnetic field $\mathbf{a}_i$), respectively.
If the DLM selects spin up (down), it generates a $x_{n,i}=+1$
($x_{n,i}=-1$) event.
In Eqs.~(\ref{dlm1}) and (\ref{dlm2}), $0<l<1$ is a parameter that controls the
speed with which the DLM learns (and forgets) about the incoming events.

The dynamic behavior of the DLM, defined by the rule Eq.~(\ref{dlm1})
is discussed in detail elsewhere~\cite{RAED05b} and may be summarized as follows:
\begin{enumerate}
\item{
If the DLM receives particles with fixed spin ${\bf S}_{n,i}={\bf S}$,
the sequence $\{x_{n,i}\}$ is periodic for all $n>n_0$,
$n_0$ depending on $u_{0,i}$ and $l$~\cite{RAED05b}.
For $n>n_0$, the frequency $N_{\pm}/(N_{+}+N_{-}$) of $x_{n,i}=\pm1$ events is given by
$(1+{\bf S}\cdot\mathbf{a}_i)/2$ and we have~\cite{RAED05b}
\begin{equation}
\label{dlm3}
\lim_{N\to \infty }
\frac{1}{N}\sum_{n=1}^N {x_{n,i}}={\bf S}\cdot{\bf a}_i,
\end{equation}
}%
exactly. Note that the limit $N\to \infty $ in
Eq.~(\ref{dlm3}) is well-defined because
the sequence $\{x_{n,i}\}$ is periodic with a finite periodicity~\cite{RAED05b}.
\item{If the DLM receives ${\bf S}_{n,i}$, statistically independent and uniformly distributed
over the unit sphere, then the DLM generates the sequence $x_{n,i}=\hbox{sign}({\bf S}_{n,i}\cdot\mathbf{a}_i)$
for all $n>n_0$, $n_0$ depending on $u_{0,i}$ and $l$~\cite{RAED05b}.
In this case we have
\begin{equation}
\label{dlm4}
\lim_{N\to \infty }
\frac{1}{N}\sum_{n=1}^N {x_{n,i}}=0
.
\end{equation}
}%
In this case, the $x_{n,i}$ are Bernoulli variables and the law
of large numbers then guarantees that Eq.~(\ref{dlm4}) holds
with probability one~\cite{GRIM95}.
\end{enumerate}
Thus, depending on the nature of the input sequence ${\bf S}_{n,i}$,
the DLM generates output sequences $\{x_{n,i}=\pm1\}$ and
particles with spin $S_{n,i}$ such that the time average
of these sequences agree with the experimental facts.

\medskip
{\bf Pseudo-random model.}
The simplest algorithm that performs the task of simulating a Stern-Gerlach magnet reads
\begin{eqnarray}
\label{sg1}
x_{n,i}=\left\{
\begin{array}{lll}
+1 & \mbox{if} & r_n\le {\bf S}_{n,i}\cdot\mathbf{a}_i\\
-1 & \mbox{if} & r_n > {\bf S}_{n,i}\cdot\mathbf{a}_i
\end{array}
\right.
,
\end{eqnarray}
where $-1\le r_n<1$ are uniform pseudo-random numbers
and the spin changes according to
\begin{eqnarray}
\label{sg2}
{\bf S}_{n,i}=\left\{
\begin{array}{lllll}
+\mathbf{a}_i& \mbox{if} & x_{n,i}=+1\\
-\mathbf{a}_i& \mbox{if} & x_{n,i}=-1
\end{array}
\right.
.
\end{eqnarray}
It is easy to check that on average,
the input-output behavior is the same as the one of the idealized Stern-Gerlach magnet.

\begin{figure*}[t]
\begin{center}
\mbox{
\includegraphics[width=9cm]{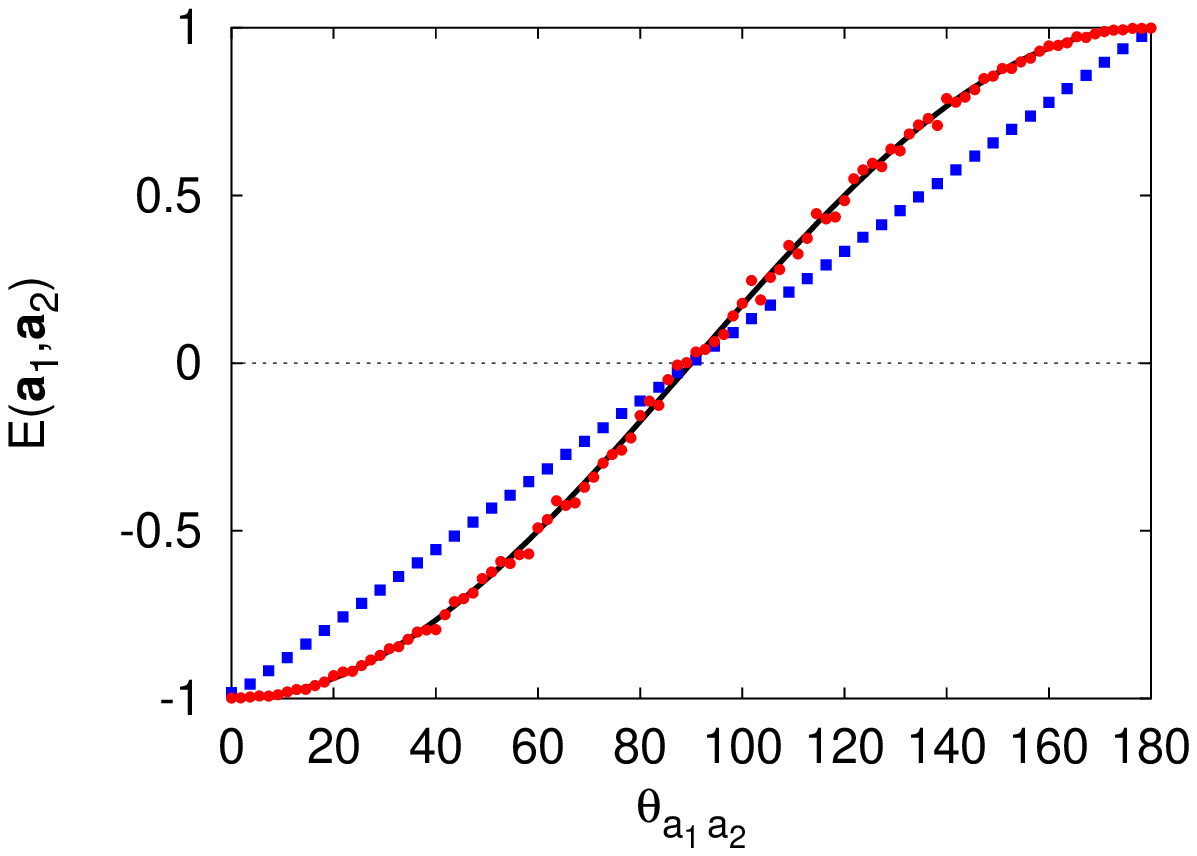}
\includegraphics[width=9cm]{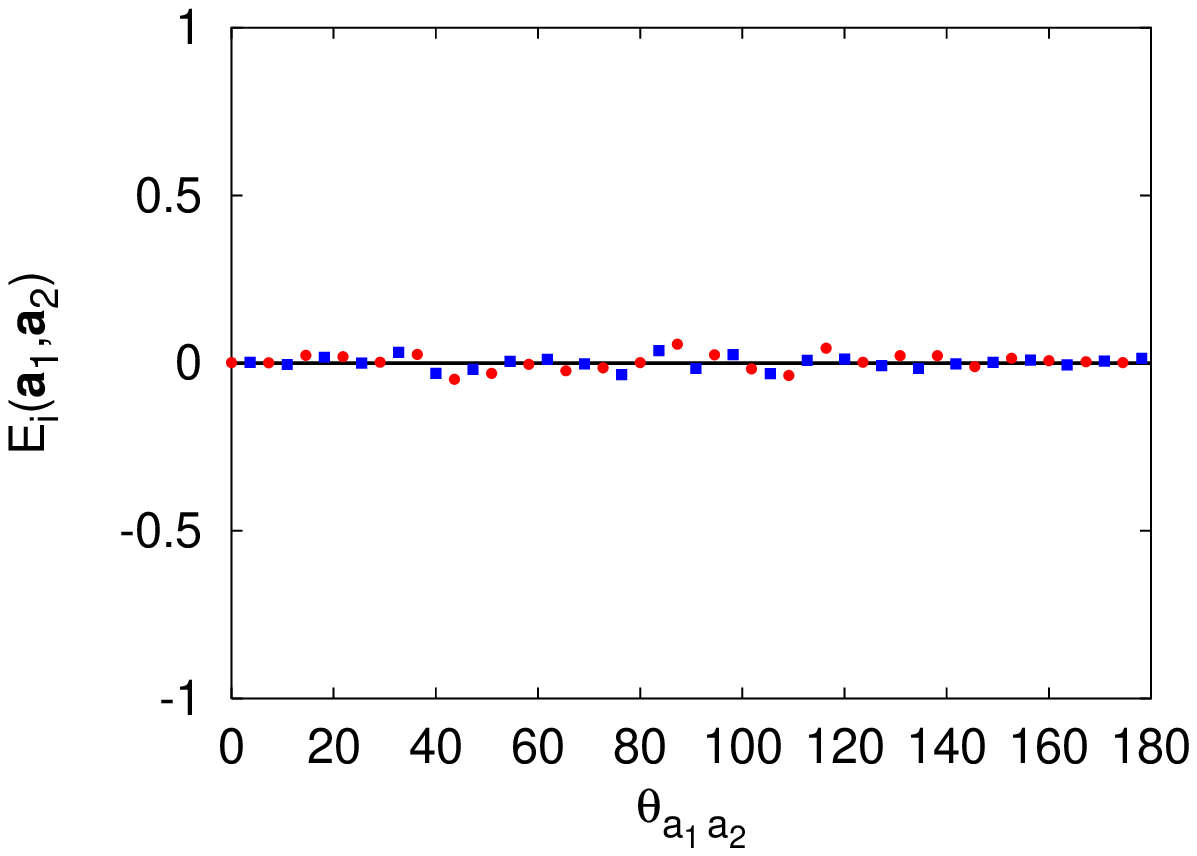}
}
\caption{
Comparison between the event-based simulation results obtained by using a deterministic model for the
Stern-Gerlach magnets and quantum theory for Case I.
Left: The two-particle correlation $E(\mathbf{a}_1,\mathbf{a}_2)$
as a function of $\theta_{\mathbf{a}_1\mathbf{a}_2}\equiv\arccos({\bf a}_1\cdot {\bf a}_2)$.
The simulation results are for $k=1$, $\tau =0.001$, $l=0.999$, $M=10$, % (see Section~\ref{ObservationStation}),
$N=10^6$, $d=3$ (red bullets) and $d=0$ (blue squares), the latter corresponding to
discarding the time-tag data (equivalent to $W>T_0$).
Solid line (black): $\widehat E(\mathbf{a}_1,\mathbf{a}_2)=-\cos\theta_{\mathbf{a}_1\mathbf{a}_2}$, as obtained from quantum theory.
Right: Single-particle expectation value
as a function of $\theta_i\equiv\arccos({\bf a}_i\cdot {\bf z})$, where ${\bf z}$ is the unit vector in the $z$-direction.
The simulation results are for $k=1$, $\tau =0.001$, $l=0.999$, $M=10$, % (see Section~\ref{ObservationStation}),
$N=10^6$, and $d=3$.
Bullets (red): $E_1(\mathbf{a}_1,\mathbf{a}_2)$;
Squares (blue): $E_2(\mathbf{a}_1,\mathbf{a}_2)$.
Solid line (black): $\widehat E_1(\mathbf{a}_1)=\widehat E_2(\mathbf{a}_2)=0$
for all orientations of the Stern-Gerlach magnets, as obtained from quantum theory.
}
\label{expI}
\end{center}
\end{figure*}

\begin{figure*}[t]
\begin{center}
\mbox{
\includegraphics[width=9cm]{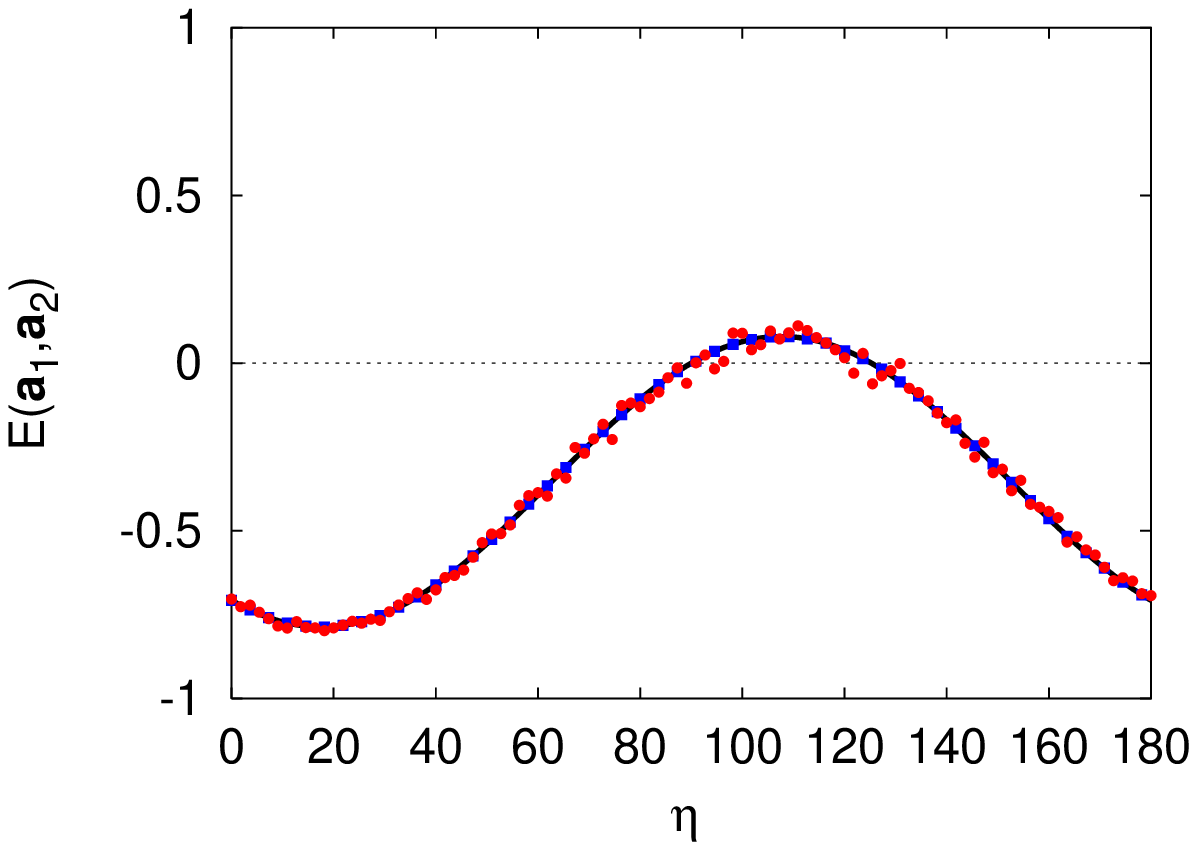}
\includegraphics[width=9cm]{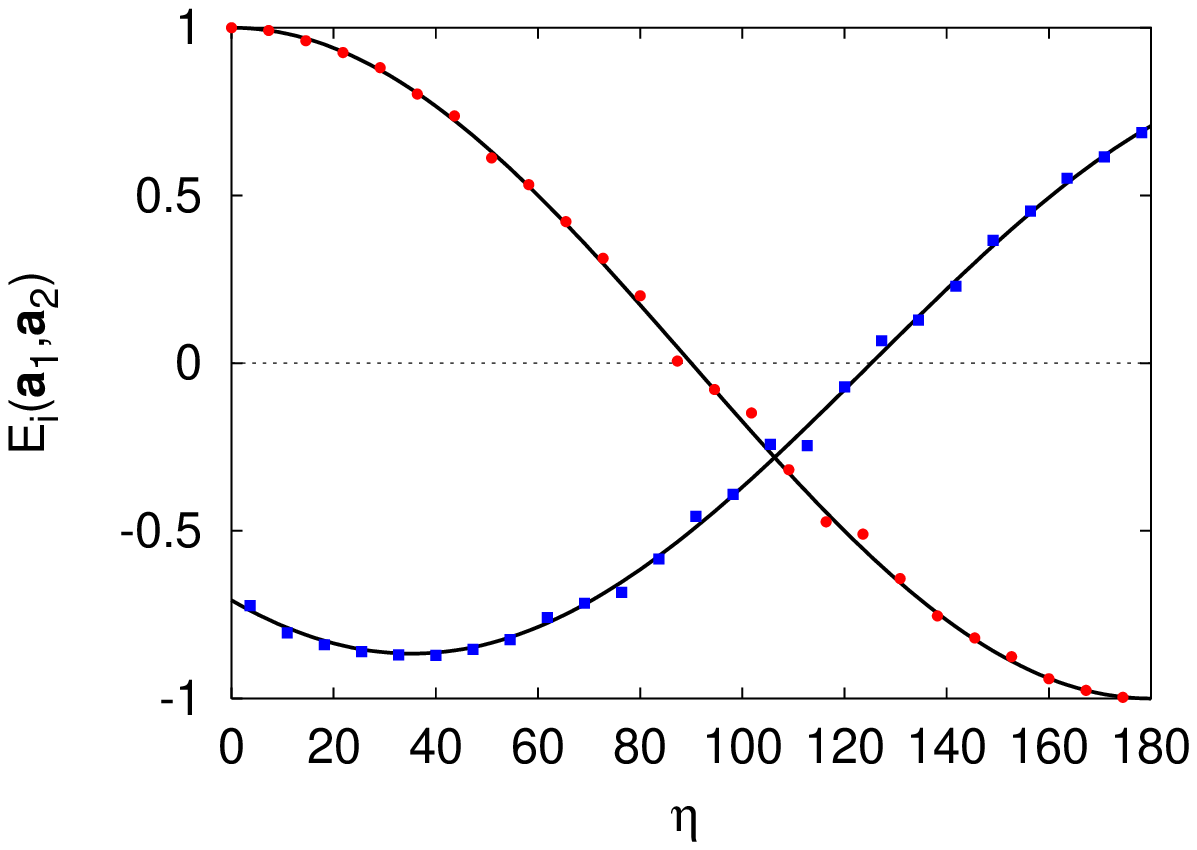}
}
\caption{
Same as in Fig.~\ref{expI} except that we simulate Case II with
${\bf a}_1=(0,0,1)$,  ${\bf a}_2=(1/2,1/2,1/\sqrt{2})$,
${\bf S}_{n,i}=(-1)^{i+1}(\sin\eta,0,\cos\eta)$ for $0\le\eta\le\pi$,
and that we plot the two-particle correlation
and the single-particle expectation value as a function of $\eta$.
}
\label{expII}
\end{center}
\end{figure*}

\medskip

\subsubsection{Time tags}

When a charge-neutral, magnetic particle passes through a Stern-Gerlach magnet
it experiences a time-delay that depends on the direction of its magnetic moment relative to the direction
of the field in the Stern-Gerlach magnet.
Experimentally, this time-delay is used to perform spectroscopy of atomic size
magnetic clusters~\cite{HEER89} and atomic interferometry~\cite{CHOR93}.
As a simple simulation model for this time delay mechanism,
we assume that the time delay ${t}_{n ,i}$ of a particle with spin ${\bf S}_{n,i}$
is distributed uniformly over the interval $[t_{0}, t_{0}+T_{n,i}]$.
Similarly, experimental evidence that the time-of-flight of single photons
passing through an electro-optic modulator fluctuates considerably
can be found in Ref.~\onlinecite{WEIH00}.
The idea that these fluctuations might be responsible for
the observed ``quantum correlations'' has been proposed
in our earlier work~\cite{RAED06c}.

From Eq.~(\ref{Cxy}), it follows that only differences of time delays matter.
Hence, we may put $t_0=0$.
The time-tag for the event $n$ is then $t_{n,i}\in[0,T_{n,i}]$.
We thus need an explicit expression for $T_{n,i}$.
The choice $T_{n,i}=constant$ is too simple: In this case we recover the model considered by Bell, which
is known not to reproduce the correct quantum correlation Eq.~(\ref{Eabs})~\cite{BELL93}.

Assuming that the particle only ``knows'' the direction
of its own spin relative to the direction of the magnetic field in the Stern-Gerlach magnet,
we can construct one number that is rotationally invariant, namely ${\bf S}_{n,i}\cdot{\bf a}_i$.
Thus, we assume $T_{n,i}=F({\bf S}_{n,i}\cdot{\mathbf a}_i)$.
As ${\bf S}_{n,i}\cdot \mathbf{a}_i=\cos\theta_{{\bf S}_{n,i}\mathbf{a}_i}$ determines
whether the particle generates a $+1$ or $-1$ signal,
it is not unreasonable to expect that $F$ is
a function of $\sin\theta_{{\bf S}_{n,i}\mathbf{a}_i}$.
After a few trials, we found that
$T_{n,i}=T_0|1-({\bf S}_{n,i} \cdot{\mathbf a}_i)^2|^{d/2}=T_0|{\bf S}_{n,i}\times{\mathbf a}_i|^d$,
yields interesting results.
Here, $T_0$ is the maximum time delay which defines the unit of time and
$d$ is a free parameter in our model.
In the sequel, we express $\tau$, $W$, $t_{n,i}$ and $T_{n,i}$ in units of $T_0$,
which for convenience we set equal to one.

\subsubsection{Data analysis}
The algorithm described earlier generates the
data sets $\Upsilon_i$ for spin-1/2 particles, just as experiment does for photons~\cite{WEIH98}.
In order to count the coincidences, we strictly follow
the procedure adopted in the EPRB experiment with photons~\cite{WEIH98}.
First, we choose a time-tag resolution $0<\tau<T_0 $ and a coincidence window $\tau\le W$.
We set the correlation counts $C_{xy} (\alpha _m ,\beta _{m'} )$ to zero for all $x,y=\pm 1$
and $m,m'=1,...,M$.
We compute the discretized time tags $k_{n ,i} =\lceil t_{n ,i}/ \tau\rceil $
for all events in both data sets.
Here $\lceil{x}\rceil$ denotes the smallest integer that is larger or equal to $x$, that is
$\lceil{x}\rceil-1<x\le\lceil{x}\rceil$.
According to the procedure adopted in the experiment~\cite{WEIH98},
an entangled pair is observed if and only if
$\left| {k_{n,1} -k_{n,2} } \right|<k=\lceil{W/\tau}\rceil$.
Thus, if $\left| {k_{n,1} -k_{n,2} } \right|<k$,
we increment the count $C_{x_{n,1},x_{n,2}} (\alpha _m ,\beta _{m'} )$.
After processing all the data for the $N$ events, we compute
the single-particle expectation values and the correlation according
to Eq.~(\ref{Ex}) and Eq.~(\ref{Exy}), respectively.

\subsection{Deterministic model: Results}

\subsubsection{Simulation of Case I and II}

We first demonstrate that the simulation model reproduces
the results of quantum theory in the case of the EPRB experiment (Case I).
In Fig.~\ref{expI} we show simulation data for $k=1$, $d=0,3$, $\tau =0.001$, $l=0.999$, and
$M=10$, $N=10^6$ for $100$ randomly chosen values of ${\bf a}_1\cdot {\bf a}_2$, covering the interval $[-1,+1]$.
At the $n$th event, two uniform pseudo-random numbers $1\le m,m'\le M$ are used
to select the rotation angles ${\bf a}_{n,i} ={\bf b}_{i,m}$.
Within the statistical errors, for the pseudo-random number generators that we use~\cite{PRES03},
the correlation between $m$ and $m'$ is zero.
The solid line is the prediction of quantum theory, see second column of Table \ref{tab1}.
It is clear that for $d=3$ there is excellent agreement between simulation and quantum
theory. This is not an accident.
Simulations for $d=3$ but with different values of the other parameters (results not shown)
confirm that for sufficiently small $\tau$ and sufficiently large $N$,
the simulation model reproduces the quantum theoretical results listed in the second column of Table \ref{tab1}.

Second, to simulate Case II, we let the source produce particles with fixed polarization but
we do not change the algorithm that simulates the observation stations.
In Fig.~\ref{expII}, we present simulation data for $k=1$, $d=0,3$, $\tau =0.001$, $l=0.999$, and
$N=10^6$, ${\bf a}_1=(0,0,1)$,  ${\bf a}_2=(1/2,1/2,1/\sqrt{2})$, and
${\bf S}_{n,i}=(-1)^{i+1}(\sin\eta,0,\cos\eta)$ for $0\le\eta\le\pi$.
For this choice of ${\bf a}_1$, ${\bf a}_2$ and ${\bf S}_{n,i}$, quantum theory predicts (see Table I)
\begin{eqnarray}
\label{caseIIdeterministic}
\widehat E_1({\bf a}_1) &=& \cos\eta,
\nonumber \\
\widehat E_2({\bf a}_2) &=& -\frac{\sin\eta+\sqrt{2}\cos\eta}{2},
\nonumber \\
\widehat E({\bf a}_1,{\bf a}_2) &=& -\frac{(\sin\eta+\sqrt{2}\cos\eta)\cos\eta}{2}
,
\end{eqnarray}
and for $d=3$, as shown in Fig.~\ref{expII}, the simulation model reproduces the quantum theoretical results
very well.

Extensive tests (data not shown) lead to the conclusion that for $d=3$ and to first order in $W$,
our simulation model reproduces the results of quantum theory of two $S=1/2$ objects,
for both Case I and Case II.

Also shown in the left panel of Figs.~\ref{expI} and \ref{expII} are the results for
$E({\bf a}_1,{\bf a}_2)$ if we ignore the time-delay data (equivalent to $d=0$ or $W>T_0$).
In Case I (see Fig.~\ref{expI}), we obtain simulation results that agree very well with the result
that is obtained by considering the class of models studied by Bell.
In Case II, $E({\bf a}_1,{\bf a}_2)$ is given by the
expression in Eq.~(\ref{caseIIdeterministic})
and up to the usual statistical fluctuations, the simulation data (see Fig.~\ref{expII}) do not depend on the
value of the time-tag parameter $d$ and the time window $W$.

\subsubsection{Case I: Numerical treatment}
\label{NumericalTreatment}

As a check on the simulation results for Case I,
we examine the limit $N\rightarrow\infty$
and show that to first order in $W$, the simulation model
yields the two-particle correlation that is characteristic for the singlet state~\cite{RAED06c,RAED07a}.

In the case of Case I we may replace
the DLM model for the Stern-Gerlach magnet by the more simple model that generates data
according to $x_{n,i}=\hbox{sign}({\bf S}_{n,i}\cdot\mathbf{a}_i)$.
For $N\rightarrow\infty$, Eq.~(\ref{Exy}) can be written as
\begin{equation}
\label{EabNinfinity}
E(\mathbf{a}_1,\mathbf{a}_2)=\frac{
 \int_0^\pi\int_0^{2\pi }x_1x_2D(T_1,T_2,W) \sin\theta d\theta d\varphi}{
 \int_0^\pi\int_0^{2\pi }D(T_1,T_2,W) \sin\theta d\theta d\varphi
},
\end{equation}
%
%\end{widetext}
where $D(T_1,T_2,W)$ is the density
of coincidences for fixed $\mathbf{a}_i$ and angles
$( {\varphi ,\theta })$ (within a small surface area $\sin\theta d\theta d\varphi$),
$T_i\equiv F(\mathbf{S}_i\cdot\mathbf{a}_i)$, $\mathbf{S}_i=\mathbf{S}_i (\varphi ,\theta )$
and $x_i=\mathop{\hbox{sign}}(\mathbf{S}_i\cdot\mathbf{a}_i)$.

\begin{figure}[t]
\begin{center}
\mbox{
\includegraphics[width=9cm]{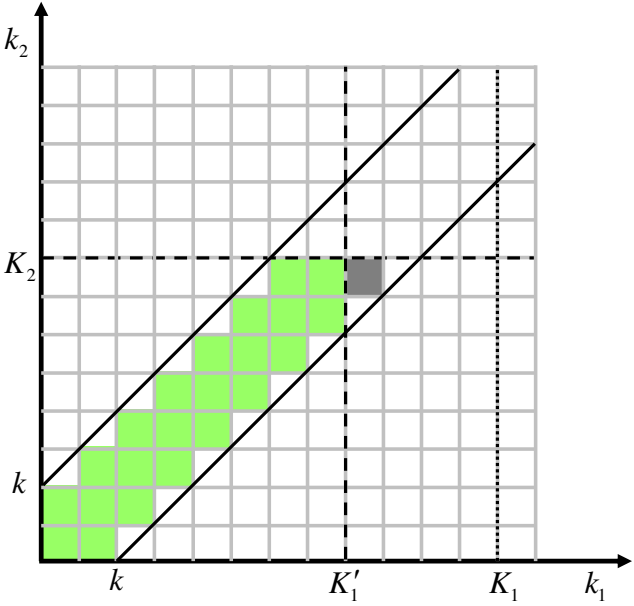}
}
\caption{
Graphical representation of the process of counting pairs. The time
interval is divided in bins of size \textit{$\tau $}, represented by the elementary squares.
The two parallel, 45$^{o}$ lines indicate the time window $W$, which was chosen
to be 2\textit{$\tau $} in this example. In the limit $N\rightarrow \infty $, the total number of
pairs for fixed $\mathbf{a}_i$ and $(\varphi,\theta)$
is given by the number of whole squares that fall within the time
window and satisfy $1\le k_i <K_i $ for $i=1,2$. For $K_1 >K_2 $, all filled
squares contribute while for ${K}'_1 =K_2 $, the dark gray square does not
contribute. For $K_1<K_2$ we interchange labels 1 and 2.
}
\label{math}
\end{center}
\end{figure}

\begin{figure}[t]
\begin{center}
\mbox{
\includegraphics[width=9cm]{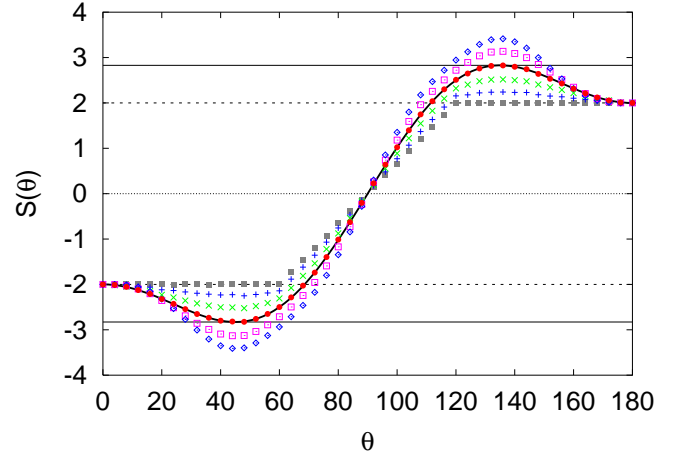}
}
\caption{Results for
$S(\theta)=S(\mathbf{a},\mathbf{b},\mathbf{c},\mathbf{d})$
where $\mathbf{a}\cdot\mathbf{c}=\mathbf{b}\cdot\mathbf{c}=\mathbf{b}\cdot\mathbf{d}=\cos\theta$ and
$\mathbf{a}\cdot\mathbf{d}=\cos3\theta$~\cite{BALL03},
for the case $k=W=1$ and $d=0,\ldots,5$,
as obtained by numerical evaluation of Eq.~(\ref{EabNinfinity}).
Solid squares (gray): $d=0$ (or $W>T_0$);
Plusses (blue): $d=1$;
Crosses (green): $d=2$;
Bullets (red): $d=3$;
Open squares (magenta): $d=4$;
Open diamonds (blue): $d=5$.
Dashed horizontal lines at $+2$ ($-2$): Maximum (minimum) value if the system is described by
a factorizable two-particle probability distribution.
Solid horizontal lines at $+2\sqrt{2}$ ($-2\sqrt{2}$): Maximum (minimum) value if the system is described by the quantum theory
for two spin-1/2 particles.
Solid line: $S(\theta)=\cos3\theta-3\cos\theta$, as obtained from quantum theory.
}
\label{Stheta}
\end{center}
\end{figure}

An analytical expression for $D(T_1,T_2,W)$ can be derived as follows.
For a fixed time-tag
resolution $0<\tau <1$, the discretized time-tag for the $n$th detection event is given by $k_{n,i}
=\left\lceil {t_{n,i} \tau ^{-1}} \right\rceil $ where $\left\lceil x
\right\rceil $ denotes the smallest integer that is larger or equal to $x$.
%that is $\left\lceil x \right\rceil -1<x\le \left\lceil x \right\rceil $.
The discretized time-tag $k_{n,i}$ takes integer values between 1 and $K_i
\equiv\lceil \tau ^{-1}T_i \rceil $, where $K_i$ is the
maximum, discretized time delay for a particle carrying angles $\left(
{\varphi ,\theta } \right)$ and passing through a Stern-Gerlach magnet with
orientation $\mathbf{a}_i$. If $\vert k_{n,1}
-k_{n,2} \vert <k=\left\lceil {\tau ^{-1}W} \right\rceil $, the two
spin-1/2 particles are defined to form a pair.
For fixed $\mathbf{a}_i$ and $\left( {\varphi ,\theta }
\right)$, we can count the total number of pairs, or coincidences $C(K_1 ,K_2 ,k)$, by
considering the graphical representation shown in Fig.~\ref{math}.
After a careful examination of all possibilities, we find that
the density can be written as $D(T_1,T_2,\tau)=C(K_1 ,K_2,1)/K_1 K_2 $
where
\begin{eqnarray}
 C(K_1 ,K_2 ,k)&=& (2k_0 -1)k_{12} -k_0 (k_0 -1)/2
\nonumber \\
 &-& \max (0,(K_{12}-1)\max (0,K_{12} )/2)
\nonumber \\
& +& \max (0,k-k_0 )k_0
\nonumber \\
& -&\max (0,kk_{12} -K_1 K_2 )
,
\label{CK}
\end{eqnarray}
and $k_0 =\min (K_1 ,K_2 ,k)$, $k_{12} =\min (K_1 ,K_2 )$, and $K_{12}
=k_{12} -\max (0,\max (K_1 ,K_2 )-k)$.
After substituting Eq.~(\ref{CK}) into Eq.~(\ref{EabNinfinity}),
the remaining integrals are easily calculated numerically.

In Fig.~\ref{Stheta} we present results for
$S(\theta)=S(\mathbf{a},\mathbf{b},\mathbf{c},\mathbf{d})$
for the case $k=W=1$ and $d=0,\ldots,5$ and the choice
$\mathbf{a}\cdot\mathbf{c}=\mathbf{b}\cdot\mathbf{c}=\mathbf{b}\cdot\mathbf{d}=\cos\theta$
and $\mathbf{a}\cdot\mathbf{d}=\cos3\theta$~\cite{BALL03}.
For $d=0$ (or $W>T_0$), we find that $S(\theta)\le2$.
Thus, we see that ignoring time-tag data automatically renders our model
incapable of producing data that violates the Bell inequalities~\cite{BELL93}.
For $1\le d<3$, $2<S_{max}<2\sqrt{2}$ and hence,
the model violates the Bell inequality but does not reproduce
the correlations of the singlet state.
As expected on the basis of our results for $E(\mathbf{a}_1,\mathbf{a}_2)$,
if $d=3$, the numerical results produced by our model are indistinguishable from the quantum theoretical
result $S(\theta)=\cos3\theta -3\cos\theta$.
For $d>3$, $2\sqrt{2}< S_{max}\le 4$, implying that our model
exhibits correlations that cannot be described by the quantum theory of two spin-1/2 particles,
even though it rigorously satisfies Einstein's criteria for local causality.

\begin{figure}[t]
\begin{center}
\mbox{
\includegraphics[width=9cm]{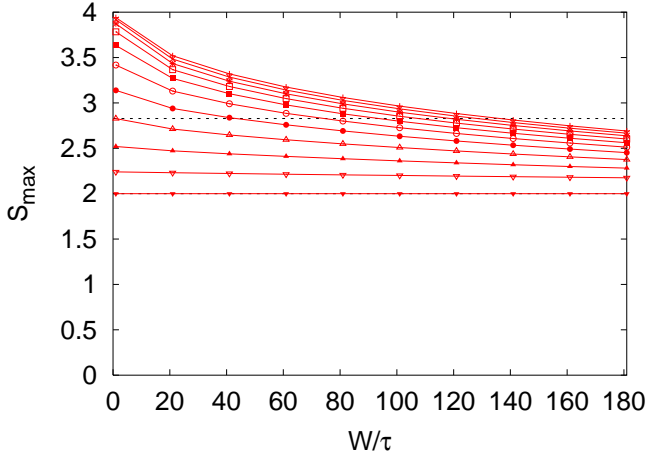}
}
\caption{
Maximum of $S(\mathbf{a},\mathbf{b},\mathbf{c},\mathbf{d})$
as a function of the time window $W$ relative to the time-tag resolution $\tau$
for $\mathbf{a}\cdot\mathbf{c}=\mathbf{b}\cdot\mathbf{c}=\mathbf{b}\cdot\mathbf{d}=\cos\theta$ and
$\mathbf{a}\cdot\mathbf{d}=\cos3\theta$.
Curves from bottom to top: Results obtained from Eq.~(\ref{EabNinfinity}) for $d=0, 1, \ldots , 10$.
Dashed line: Value of $S_{max}=2\sqrt{2}$ if the system is described by quantum theory.
}
\label{maxStheta}
\end{center}
\end{figure}

It is clear that the result for the coincidences depends on the
time-tag resolution $\tau$, the time window $W$ and the number of events $N$,
just as in real experiments~\cite{FREE72,ASPE82a,ASPE82b,TAPS94,TITT98,WEIH98,ROWE01,FATA04},
see Section~\ref{IIG}.
Expression Eq.~(\ref{EabNinfinity}) allows us to easily study the behavior of the model
as a function of the time window $W$, relative to the time-tag resolution $\tau$.
In Fig.~\ref{maxStheta} we plot $S_{max}$ as a function of $W/\tau$ for various values of $d$.
Note that the numerical results agree with the values of $S_{max}$ that can be obtained analytically
for the limiting cases $W=\tau\rightarrow 0$, $d=0,3$ and $W>T_0$ (see Sec.~\ref{ExactSolution}).
From Fig.~\ref{maxStheta}, it is clear that for $d=3$ and $W=0$,
the model reproduces the result of the quantum system in the fully entangled state.
Furthermore, Fig.~\ref{maxStheta} shows that, for sufficiently small time-tag resolution $\tau$,
increasing the time window changes the nature of the two-particle correlations.
Since $W$ is a parameter solely used in the data analysis procedure
and $S_{max}$ is a decreasing function of $W$, the value of $S_{max}$
and/or of the correlations are not sufficient to make
a definite statement about the nature of the source or even the nature of the
complete set-up.

\begin{figure}[t]
\begin{center}
\mbox{
\includegraphics[width=8.5cm]{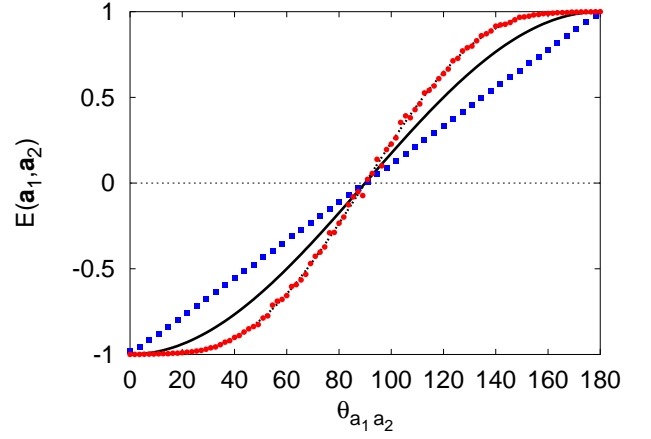}
}
\caption{
Comparison between the event-based simulation results obtained by using a deterministic model for the
Stern-Gerlach magnets, quantum theory and the exact solution for the analytical model in the limit $N\rightarrow\infty$.
The two-particle correlation $E(\mathbf{a}_1,\mathbf{a}_2)$ for Case I is shown
as a function of $\theta_{\mathbf{a}_1\mathbf{a}_2}\equiv\arccos({\bf a}_1\cdot {\bf a}_2)$.
Markers: Event-based simulation results obtained by using a deterministic model for the
Stern-Gerlach magnets. The simulation parameters are
$k=1$, $\tau =0.001$, $l=0.999$, $M=10$, $N=10^6$, $d=5$ (red bullets) and $d=0$ (blue squares),
the latter corresponding to discarding the time-tag data (equivalent to $W>T_0$).
Solid line (black): Quantum theory $\widehat E(\mathbf{a}_1,\mathbf{a}_2)=-\cos\theta_{\mathbf{a}_1\mathbf{a}_2}$.
Dashed line (black): Rigorous result Eq.~(\ref{EabW0d5}) for the simulation model for $d=5$ in the limit $W=\tau\rightarrow 0$.
}
\label{expId5}
\end{center}
\end{figure}
\begin{figure}[t]
\begin{center}
\mbox{
\includegraphics[width=8.5cm]{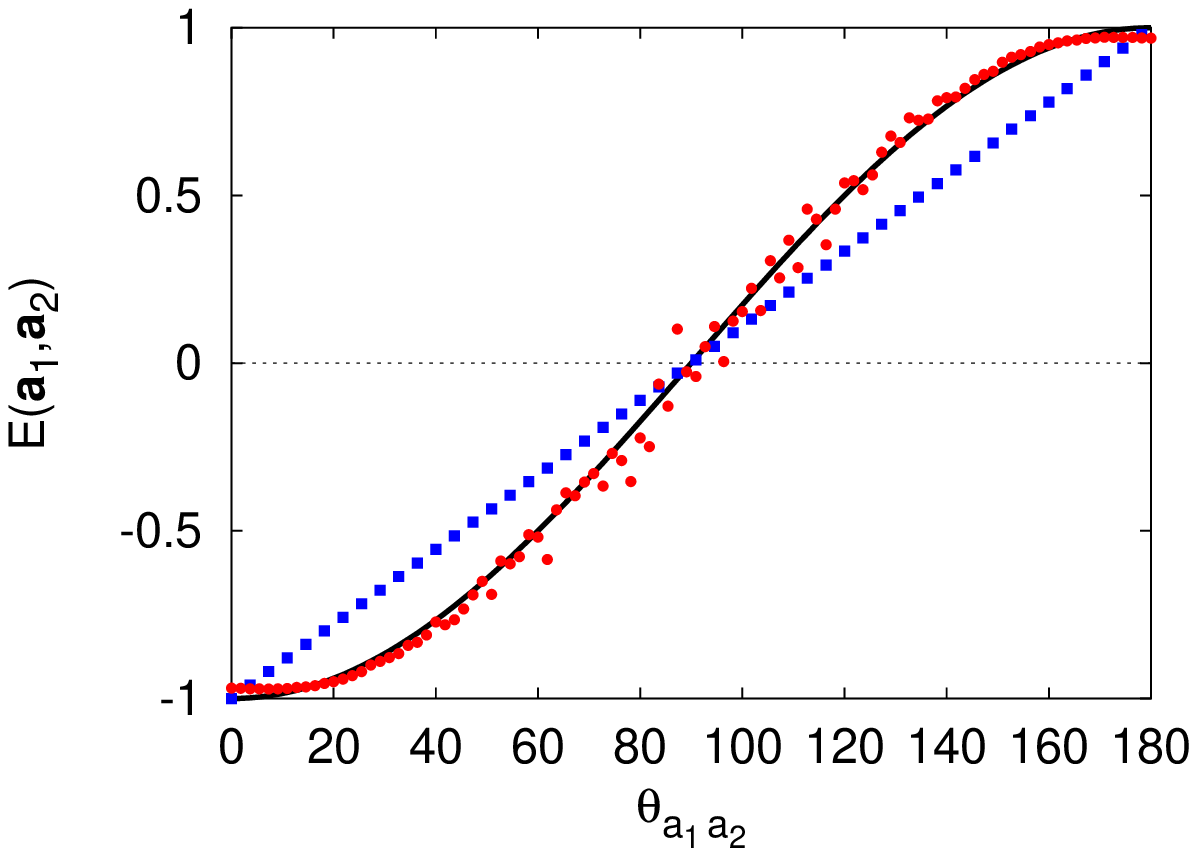}
}
\caption{
Comparison between the event-based simulation results obtained by using a pseudo-random model for the
Stern-Gerlach magnets, quantum theory and the exact solution for the analytical model in the limit $N\rightarrow\infty$.
The two-particle correlation $E(\mathbf{a}_1,\mathbf{a}_2)$ for Case I is shown
as a function of $\theta_{\mathbf{a}_1\mathbf{a}_2}\equiv\arccos({\bf a}_1\cdot {\bf a}_2)$.
Markers: Event-based simulation results obtained by using a pseudo-random model for the
Stern-Gerlach magnets. The simulation parameters are
$k=1$, $\tau =0.00001$, $M=10$, $N=10^9$, $d=7$ (red bullets) and $d=0$ (blue squares),
the latter corresponding to discarding the time-tag data (equivalent to $W>T_0$).
Solid line (black): Quantum theory $\widehat E(\mathbf{a}_1,\mathbf{a}_2)=-\cos\theta_{\mathbf{a}_1\mathbf{a}_2}$.
}
\label{stoc0}
\end{center}
\end{figure}
\begin{figure}[t]
\begin{center}
\mbox{
\includegraphics[width=8.5cm]{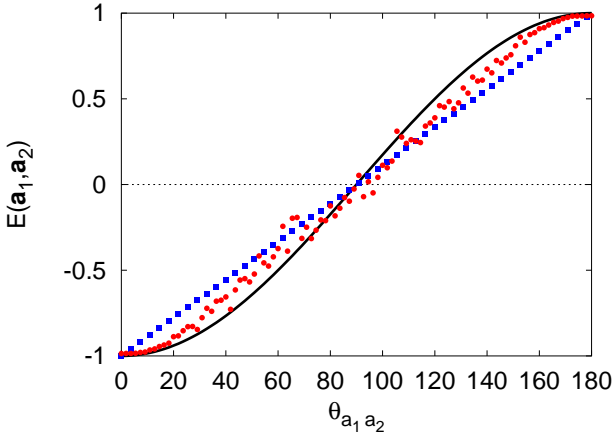}
}
\caption{
Same as Fig.~\ref{stoc0} except that bullets (red) are simulation results for $d=5$.
}
\label{stoc1}
\end{center}
\end{figure}

\subsubsection{Case I: Exact solution}
\label{ExactSolution}

For some choices of the parameters, Eq.~(\ref{EabNinfinity}) can be expressed in closed form.
We first examine the case $W>T_0$ for which $\Theta (W-\vert t_{n,1}-t_{n,2} \vert )=1$  and $D(T_1,T_2,W)=1$.
Without loss of generality, we may choose the coordinate system
such that $\mathbf{a}_1=(1,0,0)$ and $\mathbf{a}_2=(\cos \alpha ,\sin \alpha ,0)$.
Then, Eq.~(\ref{EabNinfinity}) reduces to~\cite{BELL93}
\begin{eqnarray}
\label{EabWinfinity}
E(\mathbf{a}_1,\mathbf{a}_2)&=&-\frac{1}{2\pi}\int_0^\pi \int_0^{2\pi}
x_1x_2{\sin \theta d\theta d\varphi }
\nonumber \\
&=& -1+\frac{2\vert \arccos(\mathbf{a}_1\cdot\mathbf{a}_2) \vert }{\pi }
.
\end{eqnarray}
Obviously, Eq.~(\ref{EabWinfinity}) does not agree with the quantum theoretical expression
Eq.~(\ref{Eabs}).

Second, we consider the case in which $W\rightarrow\tau$.
Formula Eq.~(\ref{CK}) greatly simplifies
if we consider the case $k=1$ ($W=\tau$), yielding
$C(K_1 ,K_2 ,1)=\min (K_1 ,K_2 )$
as is evident by looking at Fig.~\ref{math}.
For $W=\tau$ and fixed $\mathbf{a}_i$ and
$\left( {\varphi ,\theta } \right)$, the density
$D(T_1,T_2,\tau)=C(K_1 ,K_2,1)/K_1 K_2 $ that we register two particles with a
time-tag difference less than $\mathit{\tau }$ is bounded by
\begin{equation}
\tau \frac{\min (T_1+\tau ,T_2+\tau )}{( T_1+\tau)( T_2+\tau )}<D(T_1 ,T_2 ,\tau)\le\tau\frac{\min (T_1,T_2)}{T_1T_2}.
\label{CK1bound}
\end{equation}
For $W=\tau\rightarrow 0$ and $T_i=|\mathbf{S}_i\times\mathbf{a}_i|^3$,
the integrals in Eq.~(\ref{EabNinfinity}) can be evaluated in closed form.
Denoting $y_1=\mathop{\hbox{sign}}(\cos \varphi)$ and $y_2=\mathop{\hbox{sign}}(\cos (\varphi-\alpha))$
and using the same coordinate systems as above, we find
\begin{eqnarray}
\label{EabW0}
E(\mathbf{a}_1,\mathbf{a}_2)&=&-\frac{\int_0^{2\pi }{
%\mathop{\hbox{sign}}(\cos \varphi) \mathop{\hbox{sign}}(\cos (\varphi-\alpha))
y_1y_2\frac{\min (\sin ^2\varphi ,\sin ^2(\varphi -\alpha
))}{\sin ^2\varphi \sin ^2(\varphi -\alpha )}} d\varphi}{\int_0^{2\pi }
{\frac{\min (\sin ^2\varphi ,\sin ^2(\varphi -\alpha
))}{\sin ^2\varphi \sin ^2(\varphi -\alpha )}} d\varphi}
\nonumber \\
&=&-\mathbf{a}_1\cdot\mathbf{a}_2,
\end{eqnarray}
%
%\end{widetext}
which is exactly the same as the quantum theoretical result Eq.~(\ref{Eabs}).
In retrospect, it is remarkable that we obtain Eq.~(\ref{EabW0}) by
requiring that the results do not depend on $W$ and $\tau$, which
in this case is very much the same as hypothesis (3) of Sec.~\ref{ProbabilisticModel}, used
in the probabilistic modeling of the EPRB experiment.

For other integer values of $d$, the integrals can be worked out as well but the calculations
are rather tedious and the results are not very illuminating.
As an example, we give the expression for $d=5$:
\begin{eqnarray}
\label{EabW0d5}
E(\mathbf{a}_1,\mathbf{a}_2)&=&-\mathbf{a}_1\cdot\mathbf{a}_2\frac{15-7(\mathbf{a}_1\cdot\mathbf{a}_2)^2}{
11-3(\mathbf{a}_1\cdot\mathbf{a}_2)^2}.
\end{eqnarray}
In Fig.~\ref{expId5}, we demonstrate that the simulation data for $d=5$ agree very well with the
analytical result Eq.~(\ref{EabW0d5}). As shown in Fig.~\ref{Stheta}, for $d=5$, the data
not only violate the Bell inequality but also violate the rigorous upperbound $S_{max}\le2\sqrt{2}$ for
a quantum system of two $S=1/2$ particles.

\subsection{Pseudo-random model: Results}

Using the simple pseudo-random model for the Stern-Gerlach magnet yields results that
are qualitatively the same as those of the deterministic model.
Therefore, we present a few, representative simulation results only.
A detailed analytical treatment of the pseudo-random model is given in Section~\ref{Kolmogorov}
and fully supports the simulation results described next.

In Fig.~\ref{stoc0}, we demonstrate that the simulation results for $d=7$ are in excellent agreement
with the quantum theoretical expression for the correlation in the singlet state.
However, as we prove in Section~\ref{Kolmogorov}, if the number of events goes to infinity,
there is no exact agreement: There is a difference between the two curves of maximum 2\%.
Note that in the case of the deterministic model exact agreement is obtained for $d=3$.
Also notice that there is some weak but systematic deviation from the exact results for
$\theta_{\mathbf{a}_1\mathbf{a}_2}\approx0$ and $\theta_{\mathbf{a}_1\mathbf{a}_2}\approx\pi$.
This is due to the pseudo-random nature
of the model: It reproduces the perfect (anti) correlation at $\theta_{\mathbf{a}_1\mathbf{a}_2}=0,\pi$ in the limit
$N\rightarrow\infty$ only, as shown rigorously in Section~\ref{Kolmogorov}.

\begin{figure}[t]
\begin{center}
\mbox{
\includegraphics[width=8.5cm]{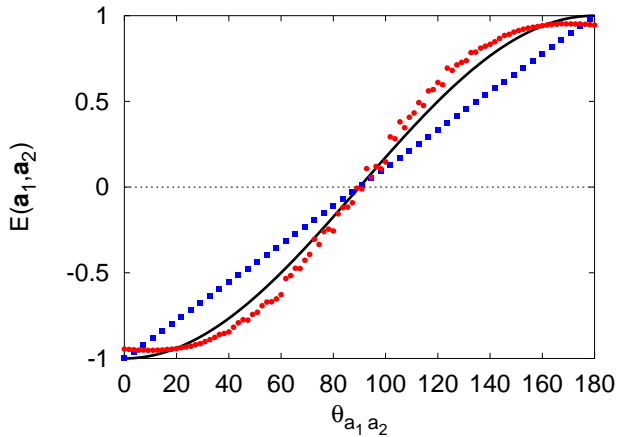}
}
\caption{
Same as Fig.~\ref{stoc0} except that bullets (red) are simulation results for $d=9$.
}
\label{stoc2}
\end{center}
\end{figure}

The results for $d=5$ and $d=9$, presented in Figs.~\ref{stoc1} and ~\ref{stoc2}, respectively,
show the same trend as we observed when using the
deterministic model for the Stern-Gerlach model: For $d=5$ the correlation is
less strong than for a quantum system in the singlet state but for $d\ge8$ it is definitely
stronger. Notice that for a fixed number of events $N$, the systematic deviation
from the perfect (anti) correlation at $\theta_{\mathbf{a}_1\mathbf{a}_2}=0,\pi$ increases with $d$.

\subsection{Summary}

Starting from the factual observation that experimental realizations of the EPRB experiment
produce the data $\{\Upsilon_1,\Upsilon_2\}$ (see Eq.~(\ref{Ups})) and that coincidence in time is
a key ingredient for the data analysis,
we have constructed computer simulation models that
satisfy Einstein's conditions of local causality
and, in the case that we employ a deterministic model for the Stern-Gerlach magnet,
exactly reproduce the correlation $\widehat E(\mathbf{a}_1,\mathbf{a}_2)=-\mathbf{a}_1\cdot\mathbf{a}_2$
that is characteristic for a quantum system in the singlet state.
In this case, both the simulation and a rigorous mathematical treatment of the model
lead to the conclusion that for $d=3$ and $W\rightarrow\tau\rightarrow0$,
the model reproduces the results (see Table I) of quantum theory for a
system of two $S=1/2$ particles.
The pseudo-random model for the Stern-Gerlach magnet yields data that are qualitatively similar
but, for integer values of $d$, do not exactly agree with quantum theory (see Section~\ref{Kolmogorov}).
It is of interest to mention here that if we simulate EPRB experiments that use the photon polarization
as a two-state system, both the deterministic and pseudo-random model exactly reproduce the quantum theoretical
results~\cite{RAED06c,ZHAO07a}.

Salient features of these models are that they
generate the data set Eq.~(\ref{Ups}) event-by-event,
use integer arithmetic and elementary mathematics to analyze the data,
do not rely on concepts of probability and quantum theory,
and provide a simple, rational and realistic picture of the mechanism that yields correlations such as Eq.~(\ref{Eabs}).

One may wonder why particles emitted by a source with definite spin orientations
that are exactly opposite to each other are not described by a density matrix that is a product state.
Of course, in this respect the description of our model may be deceptive.
In a naive picture one might think that the whole system is
described by a density matrix that is a product state.
The problem with this naive picture is that it often works extremely well but in
some cases leads to all kinds of logical inconsistencies
(see Ref.~\cite{BALL03} for an extensive discussion of this point)
and it should not come as a surprise that the EPR problem is the prime example
where the naive picture breaks down completely.
Quantum theory describes the system as a whole: It does not describe
a single pair of particles as they leave the source.

Another deceptive point may be that in our model,
one can compute the correlation of the particles right after they left the source.
This correlation is exactly minus one.
However, this correlation has no relevance to the experiment:
To measure the correlation of the particles, it is necessary to put in the Stern-Gerlach magnets,
detectors, timing logic and so on.
We emphasize that the simulation procedure counts
all events that, according to the same criterion as the one employed in experiment,
correspond to the detection of two-particle systems.

Our simulation results also suggest that we may have to reconsider
the commonly accepted point of view that the more certain we are about a measurement,
the more "classical" the system is.
Indeed, according to experiments and in concert with the prediction of our model,
this point of view is in conflict with the observation that
the more we reduce this uncertainty by letting $W\rightarrow0$,
the better the agreement with quantum theory becomes.
Both in experiments and in our model, the uncertainty is in the time-tag data
and it is this uncertainty that affects the coincidences and
yields the quantum correlations of the singlet state (if $W\rightarrow0$).
Isn't it then very remarkable that the agreement between experiment and quantum theory
improves by reducing (not increasing!) the uncertainty by making $W$ as small as technically feasible?

We have shown that whether or not these simulation models produce quantum correlations
depends on the data analysis procedure that is performed (long) after the data has been collected:
In order to observe the correlations of the singlet state,
the resolution $\tau$ of the devices that generate the
time-tags and the time window $W$ should be made as small as possible.
Disregarding the time-tag data ($d=0$ or $W>T_0$) yields results that disagree with quantum theory
but agree with the models considered by Bell~\cite{BELL93}.
Our results show that increasing the time window changes the nature of the two-particle correlations.
This prediction can easily be tested and is confirmed by re-analyzing available experimental data with different values of
the time window $W$, as we did in Section~\ref{IIG}.

In Case I, the two-particle correlation depends on the value of the time window $W$.
By reducing $W$ from infinity to zero, this correlation changes from typical Bell-like to singlet-like,
without changing the procedure by which the particles are emitted by the source.
Thus, the character of the correlation not only depends on the whole experimental setup
but also on the way the data analysis is carried out.
Hence, from the two-particle correlation itself, one cannot make any definite statement about the character of the source.
Thus, the two-particle correlation is a property of the whole system
(which is what quantum theory describes), not a property of the source itself.

In contrast, in Case II, the observation stations always receive particles with the same spin orientation and
although the number of coincidences decreases with $W$ (and the statistical fluctuations increase),
the functional form of the correlation does not depend on $W$: In Case II,
the single-particle and two-particle correlations do not depend on the value of the time window $W$.

\section{Einstein's locality versus Bell's locality}
\label{Discussion}

Starting from the data gathering and analysis
procedures used in EPRB (gedanken) experiments,
we have constructed an algorithm in which every
essential element in the experiment has a counterpart (see Section~\ref{EPRBexperiment}).
The algorithm generates the same type of data as recorded in the experiments.
The data is analyzed according to the experimental procedure to count coincidences.
The algorithm satisfies Einstein's criteria of local causality,
does not rely on any concept of quantum theory
but nevertheless reproduces the two-particle correlation of the singlet state
and all other properties of a quantum system consisting of two $S=1/2$ particles.

At first sight, our results may seem to be in contradiction with the folklore on the EPR paradox,
very often formulated in terms of Bell's theorem which
states that quantum theory cannot be described by a local hidden variable model.
In fact, there is no contradiction once one recognizes that
the concept of locality, as defined by Bell, is different from Einstein's definition of locality.
Bell made an attempt to incorporate Einstein's concept of locality
(defined on the level of individual events) to probabilistic theories,
apparently without realizing that probabilities express logical, not necessarily physical,
relationships between events.
However, the assumption that the absence of a causal influence implies logical independence
leads to absurd conclusions, even for very mundane problems~\cite{TRIB69,JAYN89}
and it is therefore not surprising that, when applied to the quantum problems, this
assumption can generate all kind of paradoxes~\cite{JAYN89}.

The simulation model that we describe in this paper, and
similar models that we described elsewhere~\cite{RAED05b,RAED05c,RAED05d,MICH05,RAED06a}
do not rely on concepts of probability theory: They operate on the ontological, event-by-event level.
Therefore it would be logically inconsistent to even attempt to apply Bell's notion of locality to these models.
However, the fact that we have proven that there exist event-based models that
satisfy Einstein's criterion of locality and causality and also
reproduce all properties of a quantum system consisting of two $S=1/2$ particles,
suggests that it may be of interest to revisit the relation between
locality \`a la Einstein and locality \`a la Bell.

Before we address this issue, we want to make clear
that we do not question the validity of the Bell-type inequalities.
These inequalities are mathematical identities that are useful
to characterize the amount of (quantum) correlation between two quantities.
In this section, we focus on the logic that is
used to address the meaning of ``locality'' in quantum physics.
In the discussion that follows, we assume that all processes are causal,
that is they should be physically realizable,
and we implicitly exclude all others.

\subsection{Einstein's locality criterion}

Einstein expressed the principle of locality as
\textit{the real factual situation of the system $S_2$
is independent of what is done with the system $S_1$,
which is spatially separated from the former}~\cite{BALL03}.
We formalize this by introducing

\medskip\noindent
{\bf Definition:} A theory is E-local if and only if it satisfies Einstein's principle of locality
for each individual event.

\medskip\noindent
Clearly, E-locality applies to each individual fact (ontological level).
Recall that quantum theory or probability theory have nothing to say about
individual events: They describe phenomena on the epistemological level.

The simulation model that we describe in this paper is a purely ontological model
of the EPRB experiment that can reproduce the results of quantum theory.
From the description of the simulation algorithm, % (see Methods),
it is evident that $x_{n ,i}$ and $t_{n,i}$ depend on the variables $(\varphi_n,\theta_{n})$
that represent the magnetic moment of a particle,
and on the orientation ${\bf a}_{n,i}$ of the Stern-Gerlach magnets,
which can be chosen at will for each $(n,i)$.
Furthermore, the event $n$ cannot affect the data recorded for all $n'<n$,
implying that the algorithm simulates a causal process.
In addition, it is obvious from the specification of the algorithm
that $x_{n,1}$, $t_{n,1}$, or ${\bf a}_{n,1}$ do not depend (in any mathematical sense)
on ${\bf a}_{n,2}$ nor do $x_{n,2} $, $t_{n,2} $, or ${\bf a}_{n,2}$ depend on ${\bf a}_{n,1}$.
This implies that for each event, the numbers $x_{n,1} $ and
$t_{n,1}$ ($x_{n,2} $ and $t_{n,2} )$ do not depend on whatever
action is taken at observation station 2 (1).
Summarizing: Our simulation model is E-local and causal.

\subsection{Bell's locality criterion}

To set the stage, we first recall the axioms of probability theory~\cite{GRIM95,JAYN03,BALL03}.
Let $A$, $B$, and $Z$ denote some propositions (events) that may be true (may occur) or false (may not occur).
The probability that $A$ is true, conditional on $Z$ being true, is denoted by $P(A\vert Z)$~\cite{GRIM95,JAYN03}.
The axioms of probability theory may be formulated as~\cite{JAYN03,BALL03}
\begin{enumerate}
\item{$0\le P(A|Z)\le1$.}
\item{$P(A|Z)+P(\bar A|Z)=1$, $\bar A$ denoting the logical negation of $A$.}
\item{$P(AB\vert Z)$=$P(A\vert BZ)P(B\vert Z)$=$P(B\vert AZ)P(A\vert Z)$.}
\end{enumerate}
These three axioms are necessary and sufficient to define a consistent mathematical
framework for probability theory.

By definition, two events $A$ and $B$ are logically independent if and only if
$P(A\vert BZ)=P(A\vert Z)$~\cite{GRIM95,JAYN03}.
If the events $A$ and $B$ are logically \textbf{dependent}, we have
\begin{eqnarray}
\label{eq28}
\frac{P(A\vert BZ)}{P(A\vert Z)}&=&\frac{P(B\vert AZ)}{P(B\vert Z)}
\nonumber \\
&=&\frac{P(B\vert AZ)P(A\vert Z)}{P(B\vert Z)P(A\vert Z)}
\nonumber \\
&=&\frac{P(AB\vert Z)}{P(A\vert Z)P(B\vert Z)}\ne 1,
\end{eqnarray}
showing that the assignment of the probability of the event $A$ ($B)$ depends on
the knowledge of the event $B$ ($A)$. From Eq.~(\ref{eq28}), we see that $P(A\vert BZ)\ne
P(A\vert Z)$ unless the events $A$ and $B$ are logically independent (we may
assume $P(A\vert Z)>0$ and $P(B\vert Z)>0$ because of the fact that we
actually registered $A$ and $B)$. As we shall see shortly, the
definition of logical independence is of extreme importance for understanding
the implications of Bell's definition of locality.

Bell considers theories (see Ref.~\onlinecite{BELL93} Chapt.7)
that assign a probability for an event $A$ to be registered,
given that the circumstances under which $A$ is registered are described by another event $Z$.
The events $A$ and $Z$ are propositions of the kind
``the values of the variables (as recorded by $m$ measurement devices)
are $A=\left\{ {A_1 ,\ldots ,A_m } \right\}$''
and ``the values of the variables (as recorded by $n$ measurement devices)
are $Z=\left\{ {Z_1 ,\ldots ,Z_n } \right\}$''.
Bell considers the case that the events $A$ and $B$ are localized in regions 1 and
2 respectively, and assumes that region 1 and 2 are separated in a
space-like way such that events in region 1 (2) have no causal influence on
events in region 2 (1)~\cite{BELL93}.

According to probability theory, we have~\cite{GRIM95,JAYN03}
\begin{equation}
\label{eq29}
P(\hat{A} \check{B}| \hat{a}\check{b}z)=P(\hat{A} |\check{B} \hat{a}\check{b} z)P(\check{B}| \hat{a} \check{b}  z),
\end{equation}
where we introduced the notation $\hat{X} $
and $\check{Y} $ to indicate that event $X$ ($Y$) can have no causal effect on event $Y$ ($X$).
We also made explicit that the condition $Z=\hat{a}\check{b}  z$
under which $A$ and $B$ have been registered may be written in terms of a common
condition $z$ and conditions $a$ and $b$ that may have a causal
effect on the outcome of $A$ and $B$, respectively.
Note that $a$, $b$ and $z$ are propositions too.

According to Bell, since the events $B$ and $b$ can have no causal effect on the
event $A$, in a local causal theory~\cite{BELL93}
\begin{equation}
\label{eq30}
P(\hat{A}| \check{B} \hat{a} \check{b} z)=P(\hat{A} |\hat{a}  z),
\end{equation}
and, similarly,
\begin{equation}
\label{eq31}
P(\check{B} | \hat{a} \check{b} z)=P(\check{B} |\check{b}  z),
\end{equation}
yielding
\begin{equation}
\label{eq32}
P(\hat{A}\check{B}| \hat{a} \check{b} z)=P(\hat{A} |\hat{a} z)P(\check{B}| \check{b}  z).
\end{equation}

The steps that take us from Eq.~(\ref{eq29}) to Eq.~(\ref{eq32}) clearly show that Bell
believes that the absence of a causal influence implies logical
independence. In fact, within probability theory,
Eq.~(\ref{eq32}) is the formal statement that
$A$ ($B$) is logically independent of $b$ ($a$) (see Eq.~(\ref{eq28})).

According to Bell, theories that do not satisfy Eq.~(\ref{eq32}),
such as quantum theory, are not locally causal~\cite{BELL93}.
Theories that satisfy Bell's criterion of locality,
as expressed by Eq.~(\ref{eq30}), will be called B-local.
We formalize this by introducing

\medskip\noindent
{\bf Definition:} A theory is B-local if and only if Eqs.~(\ref{eq30}) and (\ref{eq31})
are satisfied

\medskip\noindent
or equivalently,

\medskip\noindent
{\bf Definition:} A theory is B-nonlocal if and only if Eqs.~(\ref{eq30}) or (\ref{eq31})
are not satisfied.

\medskip\noindent
Clearly, B-locality is defined within the realm of probabilistic theories only.
Note that the folklore on the EPR paradox generally does not
distinguish between B-locality and E-locality, a remarkable logical leap because
E-locality is defined on the level of individual events whereas
B-locality is defined in terms of probabilities for events to occur.

A possible explanation for not noticing that this is a major
logical step to take is that it is quite common to mix up the meaning of frequencies and probabilities.
The former is a property that we measure by counting. It is a property of the whole system under study.
The latter is a mental, mathematical construct that allows us to reason about the former.
The reader who has difficulties to grasp this delicate but
fundamental point may find it useful to read Sec.~\ref{freq2prob} once more.

If the events $A$ and $B$ are represented by integer or real variables $A$ and $B$
(a minor abuse of notation), the expectation of the joint event $AB$ conditional on $ab$ is defined
by~\cite{GRIM95,JAYN03}
\begin{equation}
\label{eq32a}
\langle AB \rangle_{ab}=\sum_{A,B} AB P(AB|ab z).
\end{equation}
If Eq.~(\ref{eq32}) holds, we have
\begin{eqnarray}
\label{eq32b}
E_B(a,b)=\langle \hat{A}\check{B} \rangle_{\hat{a}\check{b}}=
\langle \hat{A} \rangle_{\hat{a}}
\langle \check{B} \rangle_{\check{b}},
\end{eqnarray}
where we used the subscript $B$ to indicate that
we have assumed that the theory is B-local.
Let us focus on the case that
$-1\le \hat{A}\le1$ and $-1\le \check{B}\le 1$.
Denoting
$a=\langle \hat{A} \rangle_{\hat{a}}$,
$b=\langle \hat{B} \rangle_{\hat{b}}$,
$c=\langle \hat{C} \rangle_{\hat{c}}$,
$a$, $b$, and $c$ all lie in the interval $[-1,1]$
and we have
\begin{eqnarray}
\label{eq32c}
|ab-ac|\le|b-c|\le 1- bc,
\end{eqnarray}
hence
\begin{eqnarray}
\label{eq32d}
|E_B(a,b)-E_B(a,c)|+E_B(b,c)\le 1,
\end{eqnarray}
which has the form of one of the Bell inequalities (other inequalities
can be derived in exactly the same manner) but
lacks the element of the hidden variables (see later).

A B-local theory can never violate the inequality Eq.~(\ref{eq32d}).
If, we find that inequality Eq.~(\ref{eq32d}) is violated
for some $E(a,b)$, the only conclusion that can be drawn
is that $E(a,b)$ cannot be obtained from a B-local probabilistic theory.

To appreciate the consequences of Bell's definition of a local theory,
it is very instructive to apply it to examples that do not
require concepts of quantum theory.
We first consider a very simple experiment that
shows that application of Bell's definition of locality leads
to the conclusion that an urn filled with balls of two different colors is
described by a theory that is B-nonlocal~\cite{JAYN89}.
Second, we show that Bell's assumption that the absence of causal influence implies logical
independence enforces very strong conditions
on the functional dependence of the probability distributions,
severely limiting the (classical) phenomena that a B-local theory can describe.

\subsubsection{Bernouilli's urn is B-nonlocal}

Let us take an urn filled with $M$ red and $N-M$ white balls (it is sufficient
to take $N=2$ and $M=1$ to see the consequences of Bell's definition of locality)~\cite{JAYN89}.
A blind monkey, having no knowledge about the position of the balls in the urn,
draws two balls without putting the first ball back into the urn.
We consider the events $R_{1 }$=``the result of the first draw is a red ball''
and $R_{2 }$=``the result of the second draw is a red ball''.
Denoting all other knowledge about this experiment by $Z$,
the probabilities for $R_{1}$ and $R_{2}$ are
\begin{equation}
\label{eq33}
P(R_1 | Z)=P(R_2 | Z)=\frac{M}{N}.
\end{equation}
If the result of the first draw is a red ball, the probability that the
result of the second draw is also a red ball is given by
\begin{equation}
\label{eq34}
P(R_2 | R_1 Z)=\frac{M-1}{N-1}.
\end{equation}
Let us now assume that the monkey hides the first ball from us but that it
shows us the second ball, which turns out to be red. As there can be no
causal effect of the second draw on the result of the first draw,
application of Bell's reasoning to this experiment yields
\begin{equation}
\label{eq35}
P(R_1 | R_2 Z)=P(\hat{R}_1 |\check{R}_2Z)=P(\hat{R} _1 | Z)=\frac{M}{N},
\end{equation}
which is obviously inconsistent with the basic rules of probability theory.
Indeed, from axiom 3, we have
\begin{equation}
\label{eq36}
P(R_1 | R_2 Z)P(R_2 | Z)=P(R_2 | R_1 Z)P(R_1 | Z),
\end{equation}
and using Eq.~(\ref{eq33}) we find
\begin{equation}
\label{eq37}
P(R_1 | R_2 Z)=P(R_2 | R_1 Z)=\frac{M-1}{N-1},
\end{equation}
which is definitely in conflict with Eq.~(\ref{eq35}).
Thus, Bell's assumption that the absence of a causal influence implies logical independence leads to
inconsistent results in probability theory when applied to the simple
physical system of an urn filled with red and white balls~\cite{JAYN89}.

\subsubsection{B-local hidden variable theories}

We now demonstrate that a consistent application of Bell's definition of
locality imposes severe constraints on the functional form of the probabilities.
Following Ref.~\onlinecite{JAYN03},
let us introduce a new set of $K$ exhaustive, mutually exclusive events $H_k$ ($k=1,\ldots,K$),
exhaustive implying that  $H_1+\ldots+H_K$ is always true.
Then, according to the rules of probability theory~\cite{JAYN03}
\begin{eqnarray}
\label{eq41a}
P(AB|ab z)&=&P(AB(H_1+\ldots+H_K)|ab z)
\nonumber \\
&=&
\sum_{k=1}^K P(ABH_k|ab z)
\nonumber \\
&=&\sum_{k=1}^K P(AB|H_kab z)P(H_k|ab z)
 .
\end{eqnarray}
To make contact to Bell's work, we write $\lambda$ instead of $H_k$, call them
hidden variables and replace the summation by an integration.
We have
\begin{equation}
\label{eq38}
P(\hat{A}\check{B}| \hat{a} \check{b}  z)=\int P(\hat{A}\check{B}| \hat{a} \check{b}  z\lambda )P(\lambda |\hat{a}\check{b}  z)
d\lambda.
\end{equation}
The variables \textit{$\lambda $} may have a causal influence on the events in regions 1 and 2,
hence they may affect the events $\hat{A} $ and/or $\check{B} $.
Invoking the product rule, we find~\cite{JAYN03}
\begin{equation}
\label{eq39}
P(\hat{A}\check{B}| \hat{a} \check{b}  z)=\int P(\hat{A}| \check{B} \hat{a} \check{b}  z\lambda)P(\check{B} |
\hat{a} \check{b}  z\lambda )P(\lambda |\hat{a}\check{b}  z)
d\lambda
.
\end{equation}
Following Bell~\cite{BELL93}, in his locally causal theory, Eqs.~(\ref{eq30}) and (\ref{eq31}) hold and
therefore Eq.~(\ref{eq39}) simplifies to
\begin{equation}
\label{eq40}
P(\hat{A}\check{B}| \hat{a} \check{b}  z)=\int P(\hat{A} | \hat{a}  z\lambda)P(\check{B}|
\check{b}  z\lambda )P(\lambda |\hat{a}\check{b}  z)
d\lambda
.
\end{equation}
Note that if there was a logical dependency between the events $\hat{A}$ and $\check{B}$,
we definitely destroyed it by dropping $\check{B}$ in $P(\hat{A}| \check{B} \hat{a} \check{b}  z\lambda)$.

Let us now make the (physically reasonable) assumption that the events
$\lambda$ are logically independent of $\hat{a} $ and $\check{b} $,
an assumption which is also implicit in the work of Bell
(because he ignored the difference between physical and logical independence).
In other words, it is assumed that
\begin{eqnarray}
\label{plambda}
P(\lambda |\hat{a}\check{b} z)=P(\lambda |\hat{a} z)=P(\lambda |\check{b} z)=P(\lambda |  z)
%\nonumber \\&= &
 .
\end{eqnarray}
Then, Eq.~(\ref{eq40}) simplifies to
\begin{equation}
\label{eq40a}
P(\hat{A}\check{B}| \hat{a} \check{b}  z)=\int P(\hat{A} | \hat{a}  z\lambda)P(\check{B}|
\check{b}  z\lambda )P(\lambda | z)
d\lambda
,
\end{equation}
which is the expression for the joint probability
$P(\hat{A}\check{B}| \hat{a} \check{b}  z)$ under the hypothesis of B-locality~\cite{BELL93}.

The famous Bell inequality~\cite{BELL93} follows from Eq.~(\ref{eq40a})
by repeating the steps that lead to Eq.~(\ref{eq32d}).
We denote the expectation value of $AB$ by
\begin{eqnarray}
\label{eq40b}
E_B^H(a,b)
%\langle AB \rangle_{ab}^{H}
&=&\int \sum_{\hat{A},\check{B}} \hat{A}\check{B}
P(\hat{A} | \hat{a}  z\lambda)P(\check{B}|
\check{b}  z\lambda )P(\lambda | z) d\lambda
\nonumber \\
&=&\int  \langle \hat{A} \rangle_{\hat{a},\lambda} \langle \check{B} \rangle_{\hat{b},\lambda}d\lambda
 ,
\end{eqnarray}
where the superscript $^H$ indicates that we compute the expectation
using the ``hidden variable'' probability distribution defined by Eq.~(\ref{eq40a}).
As before, we focus on the case that $-1\le \hat{A}\le1$ and $-1\le \check{B}\le 1$.
Then $a(\lambda)=\langle \hat{A} \rangle_{\hat{a},\lambda}$,
$b(\lambda)=\langle \hat{B} \rangle_{\hat{b},\lambda}$,
$c(\lambda)=\langle \hat{C} \rangle_{\hat{c},\lambda}$,
all lie in the interval $[-1,1]$.
We have
\begin{eqnarray}
\label{eq40c}
&&\left|
\int a(\lambda)b(\lambda)P(\lambda | z) d\lambda
-
\int a(\lambda)c(\lambda)P(\lambda | z) d\lambda
\right|
\nonumber \\
&&\le
\int |a(\lambda)b(\lambda)-a(\lambda)c(\lambda)|P(\lambda | z) d\lambda
\nonumber \\
&&\le
\int (1-b(\lambda)c(\lambda))P(\lambda | z) d\lambda
\nonumber \\
&&
=1-\int b(\lambda)c(\lambda)P(\lambda | z) d\lambda,
\end{eqnarray}
hence
\begin{eqnarray}
\label{eq40d}
|E_B^H(a,b)-E_B^H(a,c)|+E_B^H(b,c)\le 1.
\end{eqnarray}

Logical consistency of a B-local theory demands that we
may first apply Eqs.~(\ref{eq30}) and (\ref{eq31}) to
$P(\hat{A}\check{B}| \hat{a} \check{b}  z)$ and then insert the events $\lambda $.
This gives
\begin{eqnarray}
\label{eq41}
P(\hat{A}\check{B}| \hat{a} \check{b}  z)&= &P(\hat{A} | \hat{a}  z)P(\check{B} |\check{b}  z)
\nonumber \\
 &= &
 \left( \int P(\hat{A} | \hat{a}  z\lambda )P(\lambda |\hat{a}  z)d\lambda\right)
\nonumber \\
&&\times \left(\int  P(\check{B} | \check{b}  z\lambda )P(\lambda |\check{b}  z)d\lambda \right)
 .
\end{eqnarray}
Then, Eqs.~(\ref{eq40a}) and (\ref{eq41}) yield
\begin{eqnarray}
\label{eq42}
 P(\hat{A}\check{B}| \hat{a}\check{b}  z) &= & \int P(\hat{A}| \hat{a}  z\lambda)P(\check{B}| \check{b}  z\lambda )P(\lambda |  z)
 d\lambda
\nonumber \\
& = & \left( \int P(\hat{A} | \hat{a}  z\lambda )P(\lambda |  z) d\lambda\right)
\nonumber \\
&&\times\left( \int P(\check{B}| \check{b}  z\lambda )P(\lambda |  z)d\lambda\right)
,
\end{eqnarray}
and we see that in order for Bell's local probabilistic theory
to be mathematically consistent, the probabilities
$P(\hat{A} | \hat{a}  z\lambda )$ and $P(\check{B}| \check{b}  z\lambda )$
should satisfy Eq.~(\ref{eq42}),
for all $\hat{A}$, $\check{B}$, $\hat{a}$, and $\check{b}$, and for all
$P(\lambda |\hat{a}\check{b} z)$ satisfying Eq.~(\ref{plambda}).
Furthermore $P(\hat{A} \check{B} |\hat{a}\check{b} z)$ is completely
determined by $P(\hat{A} | \hat{a}  z\lambda )$ and $P(\check{B}| \check{b}  z\lambda )$.
Assuming, as is usually done, that the two measuring devices are the same, we
may write Eq.~(\ref{eq42}) as the functional equation
\begin{eqnarray}
\label{eq42a}
&&\int F({A},{a},\lambda)F({B},{b},\lambda)p(\lambda)d\lambda
\nonumber \\
&&=
\int F(A,a,\lambda)p(\lambda) d\lambda
\int F(B,b,\lambda)p(\lambda) d\lambda
,
\end{eqnarray}
where $0\le F(A,a,\lambda)\le1$, $0\le F(B,b,\lambda)\le1$ and $0\le p(\lambda)\le1$.

It may be of interest to note that the quantum theoretical
expression for the single-particle probability
\begin{eqnarray}
\label{eq42b}
F(A, a, {\bf S})&=&\frac{1+x{\bf a}\cdot{\bf S}}{2}
,
\end{eqnarray}
describing a Stern-Gerlach magnet (for which $A=\pm1$),
does not satisfy functional equation Eq.~(\ref{eq42a}),
assuming Eq.~(\ref{plambda}) holds here too.
Indeed, integrating over ${\bf S}$ over the unit sphere
yields $(1+AB{\bf a}\cdot{\bf b}/3)=1$
for the consistency condition Eq.~(\ref{eq42a}),
which obviously leads to a nonsensical conclusion (see also the Appendix).

Within probability theory, a mathematically consistent application
of B-locality severely limits the form of the probabilities and,
as in the case of the urn, leads to conclusions that defy common sense,
even in the realm of every-day experience.

\subsection{Reductio ad absurdum}
\label{sec:mylabel3
}
We now address the logic of the reasoning that
was used by EPR and then apply the same logic to the reasoning used by Bell.
We emphasize that we consider the logic of reasoning only. For instance,
whether or not quantum theory is a correct description of the experimental
data is not the issue here. We are concerned with logic only.

The argument put forward by EPR can be formalized as follows
\begin{enumerate}
\item $Q$ is true is equivalent to the statement that quantum theory is a correct description of the experimental data.
\item $C$ is true is equivalent to the statement quantum theory is complete.
Note that the precise definition of ``complete'' is irrelevant as far as the logic of reasoning is concerned.
\end{enumerate}
EPR use the formalism of quantum theory to prove that quantum theory is incomplete.
Thus, EPR show that if quantum theory is a correct description
of the experimental data and quantum theory is complete then quantum theory
is incomplete. This reasoning is an example of reductio ad absurdum: To
disprove a statement, we assume it is true and then prove that it leads to a
logical contradiction.

In formal language, EPR prove that
\begin{equation}
\label{eq43}
Q\wedge C\Rightarrow \overline{C},
\end{equation}
where $\wedge$, $\Rightarrow$ and $\overline{\phantom{C}}$ denote the logical ``and'' operation,
logical implication, and logical negation, respectively. Equivalently, we can write
\begin{equation}
\label{eq44}
C\Rightarrow \overline{Q}\vee \overline{C},
\end{equation}
where $\vee$ denotes the logical ``or'' operation. From Eqs.~(\ref{eq43}) or (\ref{eq44}),
it is clear that if we accept that statement $Q$ is true, statement $C$
must be false if we do not want to run into a contradiction.

We now apply the logic of the reasoning used by EPR to the reasoning used by Bell.
First, we introduce the symbol $E$:
\begin{enumerate}\setcounter{enumi}{2}
\item $E$ is true is equivalent to the statement that quantum theory obeys
Einstein's criterion of local causality (the precise meaning of this criterion is irrelevant for the logic of reasoning).
\end{enumerate}
Bell's extension of Einstein's criterion for a locally causal theory to probabilistic theories
can be formalized as follows:
\begin{enumerate}\setcounter{enumi}{3}
\item $B$ is true if and only if Einstein's criterion of local causality
is equivalent to the statement that if a variable $b$ has no causal effect on the variable $A$ then,
in a probabilistic theory, $P(A\vert bZ)=P(A\vert Z)$ must hold.
\end{enumerate}
Assuming $B$ is true, Bell derives
inequalities that are violated by quantum theory.
In formal language, Bell has shown that
\begin{equation}
\label{eq45}
Q\wedge E\wedge B\Rightarrow \overline{E},
\end{equation}
which is a logical contradiction.
Assuming that quantum theory gives a correct description of experimental
data, $Q$ is true. Then, from Eq.~(\ref{eq45}), if follows that
1) $B$ is false or 2) $E$ is false or 3) both $B$ and $E$ are false.
Bell apparently excluded the possibility that his probabilistic interpretation of Einstein's
criterion of local causality was wrong, hence he drew the conclusion
that quantum theory is E-nonlocal.
However, Bell's conclusion that quantum theory is E-nonlocal
has been drawn on the basis of a logically incorrect argument:
B-locality implicitly assumes
that the absence of a causal influence implies logical independence~\cite{JAYN89}
but, in probability theory, it is well-known that the assumption that the
absence of a causal influence implies logical independence
leads to logical inconsistencies~\cite{TRIB69,JAYN89}.
Hence, either $B$ is false or the mathematical framework of probability theory is logically inconsistent.

Excluding the hypothesis that probability theory is logically inconsistent,
it follows that $B$ is false but we cannot rule out that $E$ is false also.
However, Bell's general conclusion that an E-local, causal theory
cannot be a candidate for a more complete theory than quantum theory
is based on the wrong assumption that $B$ is true.
B-locality only looks deceptively similar to E-locality but is fundamentally different.

Thus, we are left with three options: (1) We adopt Bell's definition of locality,
keep insisting that causal indenpence implies logical independence and
learn to live with the fact that it leads to absurd conclusions such as an urn with two balls
being ``nonlocal'',
(2) we change the rules of probability theory~\cite{KHRE99} or
(3) we keep using probability theory as it is and reject Bell's definition of locality as a logically
consistent extension of Einstein's notion of locality to the domain of probabilistic theories.
We do not believe that option (1) is worth considering any further, nor that option (2) is a viable one,
in particular because quantum theory, being a very successful theory,
requires the established mathematical apparatus of
probability theory to make contact to experimental data.

\subsection{Alice on Earth and Bob on Mars}

For a logically local algorithm, such as the one described in Sec.~\ref{SimulationModel},
the condition that the two observation stations must be spatially separated is irrelevant.
To see this, imagine the following scenario.
We ask Bob to choose a set of directions ${\bf a}_{n,2}$ as he likes
and we also ask him to keep this set secret.
Then we send Bob to Mars.

After Bob has arrived on Mars, we (still on Earth) prepare data
sets $\{{\bf S}_{n,1}|n=1,\ldots,N\}$ and
$\{{\bf S}_{n,2}=-{\bf S}_{n,1}|n=1,\ldots,N\}$ for Case I
and send the second set by a radio link to an
observation station 2 that is located on Mars.
Once this data has been sent (which takes a few seconds at most), the link is destroyed.
Then, we give the first data set to Alice who is in charge of station 1 on planet Earth.
She processes her data for some set of directions ${\bf a}_{n,1}$ that she may choose as she likes
and obtains the data set $\Upsilon_1$.
This also takes a few seconds.

It takes at least five minutes before Bob, who controls station 2 on Mars,
starts to receive the data.
Bob processes this data, using a set of directions ${\bf a}_{n,2}$ he chose
before leaving for the mission to Mars and which he kept secret all the time,
and obtains the data set $\Upsilon_2$.
Then, Bob activates a radio link and sends the data set $\Upsilon_2$ to Alice (or a third person).
Alice analyses the data $\{\Upsilon_1,\Upsilon_2\}$ and computes the correlations
according to the procedure outlined in Sec.~\ref{SimulationModel}
and draws the unescapable conclusion that the data exhibit ``quantum correlations''.

If we assume that Alice and Bob never had the chance to communicate with
each other, there is no way, other than by telepathy, that
Bob could have influenced Alice's choice of ${\bf a}_{n,1}$.
Alice, not aware of the existence of Bob before Bob arrived on Mars
could not influence Bob's choice of ${\bf a}_{n,2}$ either.
In this hypothetical procedure, at the time that the data analysis was carried out,
the two systems were spatially and temporally separated and there
is no physical mechanism known to man by which Bob
could have influenced Alice's choice.

There is no point of sending Bob to Mars:
If Bob would have analyzed the data $\{{\bf S}_{n,2}\}$ on earth,
the data $\{\Upsilon_1,\Upsilon_2\}$ would be exactly the same
and so would be the conclusion that the data exhibit ``quantum correlations''.
This thought experiment (which can in fact be realized) is just another illustration
that correlations express logical but not necessarily physical dependencies.

\subsection{Summary}

In an attempt to extend Einstein's concept of a locally causal theory
to probabilistic theories, Bell implicitly assumed
that the absence of causal influence implies logical independence.
In general, this assumption prohibits the consistent application of probability theory and
leads to all kinds of paradoxes~\cite{TRIB69,JAYN03}.
However, if we limit our thinking to the domain of quantum physics,
the violation of the Bell inequalities by experimental data
should be taken as a strong signal that it is the correctness of this assumption
that one should question.
Instead of calling quantum mechanics (or an urn containing two balls)
a nonlocal theory, it would be more appropriate to reject the assumption
that the absence of causal influence implies logical independence.
This step is difficult to take unless one recognizes
that probabities are not {\sl defined} by frequencies:
Much of the recent controversies about the correctness and/or applicability
of Bell's theorem~\cite{ACCA05,HESS00,HESS01,HESS04,HESS05,KRAC05,HESS05b,GILL99,MERM05}
can be traced back to the failure of
keeping apart the concept of the frequency of events and the concept
of the probability to observe this frequency~\cite{TRIB69,JAYN03}.

Most importantly, it is simply logically incorrect to use probability theory
to even make a statement about the (non)existence of correlations in a set of experimental data.
At most, we can conclude that a probabilistic model
is compatible with the data, in which case we made a significant step
in describing the process that gave rise to the data.

The simulation models that we describe in this paper do not
rely on concepts of probability theory: They are purely ontological models of the EPRB experiment.
The expression for the coincidence Eq.~(\ref{Cxy}) cannot be written
in terms of a product of two single-particle probabilities,
an essential feature of the restricted class of local models examined by Bell~\cite{BELL93}.
Hence, the fact that we have discovered event-by-event simulation algorithms
that (1) generate the same type of data as recorded in the experiments,
(2) analyze data according to the experimental procedure to count coincidences,
(3) satisfy Einstein's criteria of local causality,
(4) do not rely on any concept of quantum theory or probability theory,
but nevertheless reproduce the two-particle correlation of the singlet state
and all other properties of a quantum system consisting of two $S=1/2$ particles
can never be in conflict with a theorem that has its roots in probability theory.

\medskip
\section{Probabilistic model of the simulation algorithm}
\label{Kolmogorov}

In this section, we use the probabilistic (Kolmogorov) approach to analyze the simulation model that we described
in Section~\ref{SimulationModel}.
This section serves three purposes.
First, it provides a rigorous proof that for to first order in $W$, the probabilistic description
of the simulation model can {\sl exactly} reproduce the results of quantum theory for
a system of two $S=1/2$ objects.
Second, it illustrates how the presence of the time-window introduces correlations
that cannot be described by a Bell-like ``hidden-variable'' model.
Third, it reveals a few hidden assumptions that are implicit in the
derivation of the specific, factorized form of the two-particle correlation
that is essential to Bell's work.

The first, fundamental step is to {\sl assume} that the simulation algorithm
can be replaced by an abstract mathematical model in which the quadruple $\{x_1,x_2,t_1,t_2\}$
is a random variable and
that the data occurs with probability $P(x_1,x_2,t_1,t_2|{\bf a}_1,{\bf a}_2)$.
We then use the standard rules of probability theory to write this probability
such that it can be evaluated analytically.

\medskip
\begin{widetext}

Using the product rule (see Eq.(~\ref{eq41a})),
we may always express the probability for observing the data $\{x_1,x_2,t_1,t_2\}$
as a sum over the mutual exclusive events.
Thus, we may write
\begin{equation}
\label{pepr0}
P(x_1,x_2,t_1,t_2|{\bf a}_1,{\bf a}_2)=\frac{1}{(4\pi^2)^2}\int d{\bf S}_1\,d{\bf S}_2\,
P(x_1,x_2,t_1,t_2|{\bf a}_1,{\bf a}_2,{\bf S}_1,{\bf S}_2) P({\bf S}_1,{\bf S}_2|{\bf a}_1,{\bf a}_2)
,
\end{equation}
where ${\bf S}_1$ and ${\bf S}_2$ denote the
three-dimensional unit vector representing the spin of the particles.
Representation Eq.~(\ref{pepr0}) is an exact expression for $P(x_1,x_2,t_1,t_2|{\bf a}_1,{\bf a}_2)$.
In the simulation model, $\{x_1,x_2,t_1,t_2\}$ are mutually independent
and $\{x_1,t_1\}$ ($\{x_2,t_2\}$) do not depend on $\{{\bf a}_2,{\bf S}_2\}$ ($\{{\bf a}_1,{\bf S}_1\}$).
This suggests that it is reasonable to {\sl assume} that
$\{x_1,x_2,t_1,t_2\}$ are mutually independent random variables and that
$\{x_1,t_1\}$ ($\{x_2,t_2\}$) are logically independent of $\{{\bf a}_2,{\bf S}_2\}$  ($\{{\bf a}_1,{\bf S}_1\}$).
Then, we have
\begin{eqnarray}
\label{pepr1}
P(x_1,x_2,t_1,t_2|{\bf a}_1,{\bf a}_2)
&=&
\frac{1}{(4\pi^2)^2}\int d{\bf S}_1\,d{\bf S}_2\,
P(x_1,t_1|x_2,t_2,{\bf a}_1,{\bf a}_2,{\bf S}_1,{\bf S}_2)
P(x_2,t_2|{\bf a}_1,{\bf a}_2,{\bf S}_1,{\bf S}_2)
P({\bf S}_1,{\bf S}_2|{\bf a}_1,{\bf a}_2)
\nonumber \\
&=&
\frac{1}{(4\pi^2)^2}\int d{\bf S}_1\,d{\bf S}_2\,
P(x_1,t_1|{\bf a}_1,{\bf S}_1)
P(x_2,t_2|{\bf a}_2,{\bf S}_2)
P({\bf S}_1,{\bf S}_2|{\bf a}_1,{\bf a}_2)
\nonumber \\
&=&
\frac{1}{(4\pi^2)^2}\int d{\bf S}_1\,d{\bf S}_2\,
P(x_1|{\bf a}_1,{\bf S}_1)
P(t_1|{\bf a}_1,{\bf S}_1)
P(x_2|{\bf a}_2,{\bf S}_2)
P(t_2|{\bf a}_2,{\bf S}_2)
P({\bf S}_1,{\bf S}_2|{\bf a}_1,{\bf a}_2)
\nonumber \\
&=&
\frac{1}{(4\pi^2)^2}\int d{\bf S}_1\,d{\bf S}_2\,
P(x_1|{\bf a}_1,{\bf S}_1)
P(t_1|{\bf a}_1,{\bf S}_1)
P(x_2|{\bf a}_2,{\bf S}_2)
P(t_2|{\bf a}_2,{\bf S}_2)
P({\bf S}_1,{\bf S}_2)
,
\end{eqnarray}
where, in the last step, we {\sl assumed} that
${\bf S}_1$ and ${\bf S}_2$ are logically independent of ${\bf a}_1$ and ${\bf a}_2$,
which is reasonable because in the simulation algorithm
${\bf S}_1$ and ${\bf S}_2$ are independent of ${\bf a}_1$ and ${\bf a}_2$.
Note that Eq.~(\ref{pepr1}) gives the exact probabilistic
description of our simulation model.

The reader may wonder why in the present case it is allowed to go from Eq.~(\ref{pepr0}) to Eq.~(\ref{pepr1})
while in Section~\ref{Discussion}, we demonstrated that making these steps may lead to logical inconsistencies.
The difference is in the fact that in Section~\ref{Discussion}, we use probability theory to make {\sl inferences}
about logical dependencies whereas in the present case we know for certain (by assumption)
which variables are logically dependent on others and which variables are not.
Thus, in the present case it is mathematically correct to describe our simulation model
by the probability Eq.~(\ref{pepr1}). However, if we analyze data for logical dependencies,
it is logically inconsistent to draw conclusions from an analysis based on Eq.~(\ref{pepr1}).
In essence, we are repeating ourselves: We can cross the line in Fig.~\ref{fig0}, separating model space
from data space from right to left because we know the properties of our simulation model but
crossing the line in the opposite direction is impossible without making additional assumptions.

Up to this point, Eq.~(\ref{pepr1}) has the same structure as the expression
that is used in the derivation of Bell's results and if we would go ahead in the same way,
our model also cannot produce the correlation of the singlet state.
However, the real factual situation in the experiment is different:
The events are selected using a time window $W$ that the experimenters try to make as small as possible~\cite{WEIH00}.
Accounting for the time window, that is multiplying Eq.~(\ref{pepr1}) by the step function,
and integrating over all $t_1$ and $t_2$, the expression for the probability for observing
the event $(x_1,x_2)$ reads
\begin{eqnarray}
\label{pepr2}
P(x_1,x_2|{\bf a}_1,{\bf a}_2)
&=&
\frac{
\int d{\bf S}_1\,d{\bf S}_2\,
P(x_1|{\bf a}_1,{\bf S}_1)P(x_2|{\bf a}_2,{\bf S}_2)
w({\bf a}_1,{\bf a}_2,{\bf S}_1,{\bf S}_2,W)
P({\bf S}_1,{\bf S}_2)
}{
\sum_{x_1,x_2=\pm1}
\int d{\bf S}_1\,d{\bf S}_2\,
P(x_1|{\bf a}_1,{\bf S}_1)
P(x_2|{\bf a}_2,{\bf S}_2)
w({\bf a}_1,{\bf a}_2,{\bf S}_1,{\bf S}_2,W)
P({\bf S}_1,{\bf S}_2)
}
\nonumber \\
&=&
\frac{
\int d{\bf S}_1\,d{\bf S}_2\,
P(x_1|{\bf a}_1,{\bf S}_1)P(x_2|{\bf a}_2,{\bf S}_2)
w({\bf a}_1,{\bf a}_2,{\bf S}_1,{\bf S}_2,W)
P({\bf S}_1,{\bf S}_2)
}{
\int d{\bf S}_1\,d{\bf S}_2\,
w({\bf a}_1,{\bf a}_2,{\bf S}_1,{\bf S}_2,W)
P({\bf S}_1,{\bf S}_2)
}
,
\end{eqnarray}
where, in general, the weight function
\begin{eqnarray}
\label{pepr3}
w({\bf a}_1,{\bf a}_2,{\bf S}_1,{\bf S}_2,W)
&=&
\int_{-\infty}^{+\infty}dt_1 \int_{-\infty}^{+\infty} dt_2\,
P(t_1|{\bf a}_1,{\bf S}_1)
P(t_2|{\bf a}_2,{\bf S}_2)
\Theta(W-|t_1 -t_2|)
,
\end{eqnarray}
will be less than one (because
$\int_{-\infty}^{+\infty}dt_1 \int_{-\infty}^{+\infty} dt_2\,P(t_1|{\bf a}_1,{\bf S})P(t_2|{\bf a}_2,{\bf S})=1$)
unless $W$ is larger than the range of $(t_1,t_2)$ for which $P(t_1|{\bf a}_1,{\bf S}_1)$ and
$P(t_2|{\bf a}_2,{\bf S}_2)$ are nonzero.
Unless $w({\bf a}_1,{\bf a}_2,{\bf S}_1,{\bf S}_2,W)=w_1({\bf a}_1,{\bf S}_1,W)w_2({\bf a}_2,{\bf S}_2,W)$,
Eq.~(\ref{pepr2}) cannot be written in the factorized form
$P(x_1,x_2|\alpha,\beta)=\int P(x_1|\alpha,\lambda)P(x_2|\beta,\lambda)\rho(\lambda)d\lambda$
that is essential to derive the Bell inequalities (see Section~\ref{Discussion}).

In the light of the discussion in Sections~\ref{Introduction} and \ref{dataquantum}, it is not without importance
to note that Eq.~(\ref{pepr2}) can be written down directly (as we did in Section~\ref{ExactSolution}), without
reference to concepts of probability theory.
Indeed, it suffices to replace the sums over the pseudo-random numbers by
discrete sums over equally spaced intervals and let these intervals go to zero.
Then the total number of events goes to infinity and we recover Eq.~(\ref{pepr2}), except that
the $P$'s that appear in Eq.~(\ref{pepr2}) do not have the meaning of probabilities.
Again, we see that the use of probabilistic models requires additional assumptions, the
correctness of which can be established a-posteriori only.

First, let us consider Case II, that is we assume that the source
emits pairs of particles with fixed, known directions ${\cal S}_1$ and ${\cal S}_2$.
Then, $P({\bf S}_1,{\bf S}_2)=\delta({\bf S}_1-{\cal S}_1)\delta({\bf S}_2-{\bf{\cal S}}_2)$,
the weight function $w({\bf a}_1,{\bf a}_2,{\cal S}_1,{\cal S}_2,W)$ drops out
and Eq.(\ref{pepr2}) reduces to
\begin{eqnarray}
\label{pepr2b}
P(x_1,x_2|{\bf a}_1,{\bf a}_2)
&=&
P(x_1|{\bf a}_1,{\cal S}_1)P(x_2|{\bf a}_2,{\cal S}_2)
,
\end{eqnarray}
which agrees with the expression for the quantum system of two $S=1/2$ particles in the product state.

Second, we put $P({\bf S}_1,{\bf S}_2)=\delta({\bf S}_1+{\bf S}_2)$.
Then, ${\bf S}_1=-{\bf S}_2$ is a random variable that covers
the unit sphere in a uniform manner, that is we are treating Case I.
In our simulation model, the time delays $t_i$ are distributed uniformly
over the interval $[0,T_i]$ where $T_i\equiv T_0|1-({\bf a_i}\cdot{\bf S}_i)^2|^{d/2}$ for $i=1,2$.
Thus, $P(t_i|{\bf a_i},{\bf S}_i)= \Theta(t_i)\Theta(T_i - t_i)/T_i$ and
\begin{eqnarray}
\label{pepr4}
w({\bf a}_1,{\bf a}_2,{\bf S}_1,{\bf S}_2,W,d)
&=&\frac{1}{T_1T_2}
\int_{0}^{T_1}dt_1 \int_{0}^{T_2} dt_2\,
\Theta(W-|t_1 -t_2|)
,
\end{eqnarray}
where we added the parameter $d$ to the list of variables to make explicit that we
adopted the time-tag model that we employ in the simulation.
The integrals in Eq.(\ref{pepr4}) can be worked out analytically, yielding
\begin{eqnarray}
\label{pepr5}
w({\bf a}_1,{\bf a}_2,{\bf S}_1,{\bf S}_2,W,d)
=\frac{1}{4T_1T_2}&[&T_1^2+T_2^2+2(T_1+T_2)W+(W-T_1)|W-T_1|+(W-T_2)|W-T_2|
\nonumber \\
&&-(W-T_1+T_2)|W-T_1+T_2|-(W+T_1-T_2)|W+T_1-T_2|\;\;]
.
\end{eqnarray}
Clearly, Eq.~(\ref{pepr5}) cannot be written in the factorized form $w_1({\bf a}_1,{\bf S}_1,W)w_2({\bf a}_2,{\bf S}_2,W)$.
Hence, it should not come as a surprise that as soon as we want to model the real experiment in which the time window is essential,
we may obtain correlations that cannot be described by Bell-like models.

We now consider the relevant limiting cases for which we can easily
derive closed-form expressions for the expectation values.
From Eq.~(\ref{pepr5}), it follows that
\begin{eqnarray}
\label{pepr7a}
w({\bf a}_1,{\bf a}_2,{\bf S}_1,{\bf S}_2,W\rightarrow\infty,d)&=&w({\bf a}_1,{\bf a}_2,{\bf S}_1,{\bf S}_2,W\ge T_0,d)=1,
\\
\label{pepr7b}
w({\bf a}_1,{\bf a}_2,{\bf S}_1,{\bf S}_2,W<T_0,d=0)&=&W(2W-T_0)/T_0^2,
\\
\label{pepr7c}
w({\bf a}_1,{\bf a}_2,{\bf S}_1,{\bf S}_2,W\rightarrow0,d)&=&\frac{2W}{\max(T_1,T_2)}+{\cal O}(W^2)
.
\end{eqnarray}
If the weight function is a constant, as in Eqs.~(\ref{pepr7a}) and (\ref{pepr7b}),
Eq.~(\ref{pepr2}) reduces to
\begin{eqnarray}
\label{pepr2a}
P(x_1,x_2|{\bf a}_1,{\bf a}_2)
&=&
\int d{\bf S}_1\,d{\bf S}_2\,
P(x_1|{\bf a}_1,{\bf S}_1)P(x_2|{\bf a}_2,{\bf S}_2)
P({\bf S}_1,{\bf S}_2)\quad;\quad d=0\;\;\hbox{or}\;\;W\ge T_0
,
\end{eqnarray}
and takes the factorized form that is characteristic for the probabilistic models considered by Bell~\cite{BELL93}.
Hence, we know that we cannot recover the results of quantum theory in the limiting cases
$d=0$ or $W\ge T_0$ in which the time-tag information plays no role.

From now on, we focus on the experimentally relevant case of small $W$, that is we neglect contributions of ${\cal O}(W^2)$.
We insert in Eq.(\ref{pepr2}), the probability distributions $P(x|{\bf a},{\bf S})=\Theta(x{\bf a}\cdot{\bf S})$
or $P(x|{\bf a},{\bf S})=(1+x{\bf a}\cdot{\bf S})/2$, corresponding to the deterministic
and pseudo-random model for the Stern-Gerlach magnet respectively.
By symmetry we have $E_1({\bf a}_1,{\bf a}_2,W\rightarrow0)=E_2({\bf a}_1,{\bf a}_2,W\rightarrow0)=0$
for all values of $d$, in agreement with quantum theory (see the second column of Table~\ref{tab1}).
The two-particle correlations are given by
\begin{eqnarray}
\label{pepr8}
E({\bf a}_1,{\bf a}_2,W\rightarrow0)
&=&
-\frac{\int d{\bf S}\;\hbox{sign}({\bf a}_1\cdot{\bf S})\hbox{sign}({\bf a}_2\cdot{\bf S})
\max^{-1}(T_1,T_2)}{\int d{\bf S} \max^{-1}(T_1,T_2)
},
\nonumber \\
\noalign{\noindent and}
E({\bf a}_1,{\bf a}_2,W\rightarrow0)
&=&
-\frac{\int d{\bf S}\;{\bf a}_1\cdot{\bf S}\;{\bf a}_2\cdot{\bf S}\max^{-1}(T_1,T_2)}{\int d{\bf S}\,\max^{-1}(T_1,T_2)}
,
\end{eqnarray}
for the deterministic and the pseudo-random  model, respectively.

Without loss of generality, we may choose the coordinate system
such that $\mathbf{a}_1=(1,0,0)$ and $\mathbf{a}_2=(\cos \alpha ,\sin \alpha ,0)$.
Then, substitution of $T_i=|\mathbf{S}_i\times\mathbf{a}_i|^d$ into Eq.~(\ref{pepr8}) yields
\begin{eqnarray}
\label{pepr8a}
E({\bf a}_1,{\bf a}_2,W\rightarrow0)
&=&
-\frac{\int_{-1}^{+1} dx\,\int_{\theta/2}^{\pi/2+\theta/2} d\phi\, g(\phi,\theta,x) (\sin^2\phi+x^2\cos^2\phi)^{-d/2}
}{
\int_{-1}^{+1} dx\, \int_{\theta/2}^{\pi/2+\theta/2} d\phi\,(\sin^2\phi+x^2\cos^2\phi)^{-d/2}
},
\end{eqnarray}
where $g(\phi,\theta,x)=\hbox{sign}(\cos\phi\cos(\phi-\theta))$
or $g(\phi,\theta,x)=(1-x^2)\cos\phi\cos(\phi-\theta)$ for the deterministic or pseudo-random
model for the Stern-Gerlach magnet, respectively.
Here and in the remainder of this section, we define $\cos\theta\equiv{\bf a}_1\cdot{\bf a}_2$.

For specific values of $d$, Eq.~(\ref{pepr8a}) can be written in terms of elementary functions.
In the case of the deterministic model for the Stern-Gerlach magnet, we find
\begin{eqnarray}
\label{pepr8b}
E({\bf a}_1,{\bf a}_2,W\rightarrow0)
&=&
\left\{
\begin{array}{ll}
-{\displaystyle1 + \frac{2}{\pi}\arccos(\cos\theta)}&, d=0\\ \\
-\cos\theta &, d=3\\ \\
-{\displaystyle\frac{15\cos\theta-7\cos^3\theta}{11-3\cos^2\theta}}&, d=5\\ \\
-{\displaystyle\frac{6890\cos\theta - 895\cos3\theta + 149\cos5\theta}{ 5774 + 280\cos2\theta + 90\cos4\theta}}&, d=7
\end{array}
\right.
.
\end{eqnarray}
In the case of the pseudo-random model for the Stern-Gerlach magnet, we obtain
\begin{eqnarray}
\label{pepr8c}
E({\bf a}_1,{\bf a}_2,W\rightarrow0)
&=&
\left\{
\begin{array}{ll}
-{\displaystyle\frac{1}{3}\cos \theta}&, d=0\\ \\
-{\displaystyle\frac{8\cos\theta}{8+3\sin\theta}}&, d=5\\ \\
-{\displaystyle\frac{2992\cos\theta+80\cos3\theta}{2887+140\cos2\theta+45\cos4\theta}}&, d=7\\ \\
-{\displaystyle\frac{84026\cos\theta + 8169\cos3\theta  - 35\cos5\theta}{45666+16254\cos2\theta-1680\cos^4\theta+3150\cos^6\theta}}&, d=9
\end{array}
\right.
.
\end{eqnarray}

All the $d>0$ results in Eqs.~(\ref{pepr8b}) and (\ref{pepr8c}) violate the Bell inequalities but, as we already explained, this finding
has no significant consequences.
From Eq.~(\ref{pepr8b}), we conclude that for $d=3$ and the deterministic model
of the Stern-Gerlach magnet, the expression is identical to the correlation of a system of two $S=1/2$ particles in the singlet state.
The result for the pseudo-random model of the Stern-Gerlach magnet and $d=7$ (see Eq.~(\ref{pepr8c})) is very close (with a maximum error of less that 2\%)
to the singlet correlation. Of course, there is no fundamental reason why $d$ should be an integer.
Finally, we note the almost trivial fact that for $W\rightarrow0$, the results are insensitive to small variations
in $W$, in agreement with the general idea, explored in the next section, that quantum theory is one out of the many probabilistic theories
that has the special feature that its predictions are insensitive to small changes of the parameters of the model.

For completeness, we list the analytical results for the case of the photon polarization.
For the deterministic model of a polarizer (which does not reproduce Malus law),
the probabilistic treatment yields~\cite{RAED07a}
\begin{eqnarray}
\label{shua1}
E({\bf a}_1,{\bf a}_2,W\rightarrow0)
&=&
\left\{
\begin{array}{ll}
-{\displaystyle1 + \frac{2}{\pi}\arccos(\cos2\theta)}&, d=0\\ \\
-{\displaystyle
-\frac{
\ln \frac{1+|\cos \theta|}{1-|\cos \theta|}\frac{1-|\sin \theta|}{1+|\sin \theta|}
}{
\ln \frac{1+|\cos \theta|}{1-|\cos \theta|}\frac{1+|\sin \theta|}{1-|\sin \theta|}}
}&,d=1\\ \\
-\cos2\theta &, d=2\\ \\
-{\displaystyle\frac{3\cos 2\theta-\cos^32\theta}{2}}&, d=4
\end{array}
\right.
.
\end{eqnarray}
In the case that we adopt the pseudo-random model for the polarizer that can reproduce Malus law,
the probabilistic model yields~\cite{ZHAO07a}
\begin{eqnarray}
\label{shua2}
E({\bf a}_1,{\bf a}_2,W\rightarrow0)
&=&
\left\{
\begin{array}{ll}
-{\displaystyle\frac{1}{2}\cos 2\theta}&, d=0\\ \\
{\displaystyle\frac{\pi}{4}\sin2\theta\cos2\theta - \cos2\theta +\ln[|\tan \theta|^{\sin^22\theta/2}]}&, d=2\\ \\
-\cos 2\theta&, d=4\\ \\
-{\displaystyle\frac{43\cos 2\theta+5\cos 2\theta\cos4\theta}{38+10\cos4\theta}}&, d=6\\ \\
-{\displaystyle\frac{53\cos2\theta+7\cos6\theta}{39+21\cos4\theta}}&, d=8
\end{array}
\right.
,
\end{eqnarray}
where we have omitted the expressions for odd $d$ because they cannot be written
in terms of elementary functions.
In passing, we note that the mathematically rigorous result for $d=4$ disposes of
the widespread believe~\cite{BELL93} that perfect correlation of the singlet state requires some form
of determinism.

\end{widetext}
\medskip

\section{Derivation of the Quantum Theory of the EPRB experiment}
\label{ProbabilisticModel}

In the quantum theoretical model, the choice of the state that describes
the EPRB experiment is an educated guess.
There is no underlying principle that guides us to this choice other than that
the particular averages (of time series) that we compute from the experimental data agree
with the expectation values (ensemble averages) obtained from the theory.

From the work of Cox and Jaynes in the early 60's, we know that once
we have agreed to represent the degree of the plausibility of a proposition
by a real number, then there is a unique set of rules,
identical to the standard rules of probability theory,
that we must adhere to in order that the logical inferences
we make do not violate elementary desiderata
of rationality and consistency~\cite{COX61,TRIB69,JAYN03}.
In this case, the rules of probability theory are used
as a vehicle for carrying out probable inference~\cite{COX61,TRIB69,JAYN03}
and have a much broader range of applications than the Kolmogorov theory of probability.
The latter is incorporated in the former.
As mentioned earlier, and as is most evident in Section~\ref{Discussion},
we mainly use probability theory as a vehicle to make
statements about propositions, that is we use it
in its extended logic mode.

An intriguing question now arises: Would it be possible to derive
the quantum theoretical description of the EPRB experiment from the
general principles of logical inference and empirical knowledge about
the results of the experiment, not involving concepts from quantum theory at all?
Elsewhere, we have shown that Malus law can be derived in this manner~\cite{RAED06a}.
In this section, we show that the same approach yields the probability
distributions Eqs.~(\ref{Pxy}) and (\ref{Pxyp})
without making the detour via quantum theory.

The approach that we take here is very much inspired by the work of Frieden~\cite{FRIE04}.
Frieden has shown that one can recover all the fundamental equations of
physics by finding the extremes of the Fisher information plus the ``bound'' information~\cite{FRIE04}.
According to Frieden, the act of measurement elicits a physical law
and quantum mechanics appears as the result of what Frieden calls ``a smart measurement'',
a measurement that tries to make the best estimate~\cite{FRIE04}.
Although our approach is similar to Frieden's, our line of reasoning is different.
We do not invoke concepts from estimation theory, such as the
estimators and the Cram\'er-Rao inequality, nor do we require the concept of random noise.
%Furthermore, in Frieden's treatment of quantum systems, the parameters to be
%estimated (such as the position) are of the same kind as the measured quantities.
%This is not the case for the systems that we treat here.

The probabilistic model that we will develop is based on the following four hypotheses:
\begin{enumerate}
\item{
Each detection event constitutes a Bernouilli trial,
that is we assume that the events are logically independent~\cite{TRIB69,GRIM95,JAYN03}.
Note that the absence of statistical correlation in the data recorded
in an experiment is an indication but definitely not a proof that the events
are logically independent~\cite{TRIB69,GRIM95,JAYN03}.
On the other hand, if the data would exhibit correlations,
the events would be logically dependent~\cite{TRIB69,GRIM95,JAYN03}.
}
\item{
The time series recorded during
an experiment suggest that the averages of the data are rotational invariant.
This observation we formalize by making the hypothesis that
the expectation values (not necessarily the probabilities)
are invariant for rotations of the conditions
under which the experiments are carried out.
}
\item{
The model operates according to the principle of efficient data processing~\cite{RAED06a}:
It generates the events such that
the probability distribution is least sensitive to small variations
in the conditions under which the experiment is carried out.
In other words, the probability distribution
should be as smooth as possible, for all values of the parameters
that determine these conditions.
}
\item{
The time series that we observe is the one which is most likely to be observed, that
is its probability is maximum.
}
\end{enumerate}
In the remainder of this section, we will simplify the notation a little
by omitting from the conditions that appear in the probabilities,
the proposition that expresses the knowledge about the problem
that we do not need to specify explicitly.

We begin by demonstrating that these four assumptions suffice to
derive the probability distribution
\begin{equation}
\label{prob1}
P(x|{\bf a},{\bf S})=\frac{1+x{\bf a}\cdot{\bf S}}{2}
,
\end{equation}
of a single Stern-Gerlach magnet.
Then, using the same four assumptions, we derive the probability distribution
\begin{eqnarray}
\label{prob2}
P(x,y|{\bf a}_1,{\bf a}_2)&=&
\frac{1-xy{\bf a}_1\cdot{\bf a}_2}{4},
\end{eqnarray}
for the EPRB experiment.

\begin{widetext}
\subsection{Stern-Gerlach magnet}

We consider the case that
the direction ${\bf a}$ of the applied field in the Stern-Gerlach magnet
and the magnetic moment ${\bf S}$ of the particles do not change with time.
The measuring apparatus (Stern-Gerlach magnet + particle detector)
transforms the input, $N$ particles with magnetic moment ${\bf S}$,
into a time series $\{x_n|n=1,\ldots,N\}$ of signals $x_n=\pm1$.
By hypothesis (1), the probability $P(x_1,\ldots,x_N|{\bf a},{\bf S},N)$
to observe the data record $\{x_n|n=1,\ldots,N\}$
can be written as $\prod_n^NP(x_n|{\bf a},{\bf S})$.
As $x=\pm1$, $P(x|{\bf a},{\bf S})$ is completely
determined by its first moment, that is we can write
\begin{equation}
\label{prob3a}
P(x|{\bf a},{\bf S})=\frac{1+xE({\bf a},{\bf S})}{2}
,
\end{equation}
where $E({\bf a},{\bf S})=\sum_{x=\pm1}xP(x|{\bf a},{\bf S})$.
By hypothesis (2), $E({\bf a},{\bf S})=E({\bf a}\cdot{\bf S})$
and hence the probability for a single event $x$ is given by
\begin{equation}
\label{prob3}
P(x|{\bf a},{\bf S})=P(x|{\bf a}\cdot{\bf S})=P(x|\theta)
,
\end{equation}
and is conditional on the relative angle $\theta$
between the magnetic moment ${\bf S}$ of the
particle and the direction ${\bf a}$ of the applied field.
Denoting $p(\theta)=P(x=+1|\theta)$,
the probability for observing a time series
$\{x_n|n=1,\ldots,N\}$ in which $m$ of the events $x_n$
take the value $+1$, that is $\sum_{n=1}^N x_n=2m-N$,
is given by~\cite{TRIB69,GRIM95,JAYN03}
\begin{equation}
P(m|\theta,N)=\frac{N!}{m! (N-m)!} p^m(\theta) [1-p(\theta)]^{N-m}
.
\label{prob4}
\end{equation}

We now consider the likelihood that the observed sequence of $\{x_n\}$
was generated by $p(x|\theta+\epsilon)$ instead of $p(x|\theta)$,
$\epsilon$ being a small positive number.
The log-likelihood $L$ that the data was generated
by $p(x|\theta+\epsilon)$ instead of by $p(x|\theta)$
is given by~\cite{TRIB69,JAYN03}
\begin{eqnarray}
\frac{L}{N}&=&
\frac{1}{N}\ln \frac{P(m|\theta+\epsilon,N)}{P(m|\theta,N)}
%,\nonumber \\ &=&
=\frac{m}{N}\ln \frac{p(\theta+\epsilon)}{p(\theta)}
+(1-\frac{m}{N})\ln \frac{1-p(\theta+\epsilon)}{1-p(\theta)}
.
\label{prob5}
\end{eqnarray}
According to hypothesis (3), the variation of $L$ with $\epsilon$ should be minimal.
Then, the results (averages over the time-series)  will be least sensitive to
small variations of the conditions under which the experiment is carried out.

We bring the problem of determining the function $p(\theta)$ in a mathematically trackable form by
using the Taylor expansion with respect to $\epsilon$.
We find
\begin{eqnarray}
\frac{L}{N}&=&
-
\frac{\epsilon^2}{2}
\frac{(p'(\theta))^2}{(1-p(\theta))p(\theta)}
+(\frac{m}{N}-p(\theta))\left(
\epsilon\frac{p'(\theta)}{(1-p(\theta))p(\theta)}
-
\frac{\epsilon^2}{2}
\frac{(1-2p(\theta))(p'(\theta))^2+(1-p(\theta))p''(\theta)}{(1-p(\theta))^2p^2(\theta)}
\right)
.
\label{prob6}
\end{eqnarray}
Invoking hypothesis (4), $m$ is the value that maximizes $P(x_1,\ldots,x_N|{\bf a},{\bf S},N)$.
A simple calculation (see Section~\ref{freq2prob}) shows that
\begin{equation}
(1+\frac{1}{N})p(\theta)-\frac{1}{N}\le\frac{m}{N}\le (1+\frac{1}{N})p(\theta)
.
\label{prob7a}
\end{equation}
Hence, for large $N$ we may set $m/N=p(\theta)$ in Eq.~(\ref{prob6}) and then
the second term of the right hand side vanishes.
Then, $L$ will be least sensitive to changes in $\epsilon$ if
\begin{equation}
I_F=\frac{1}{p(\theta)(1-p(\theta))}
\left(\frac{\partial p(\theta)}{\partial \theta}\right)^2
,
\label{prob7}
\end{equation}
is minimal.
The quantity $I_F$ is the Fisher information~\cite{TREE68,FRIE04,JAYN03}
\begin{eqnarray}
I_F
&=&\frac{1}{N}\sum_{x_1,\ldots,x_N=\pm1}
\frac{1}{P(x_1,\ldots,x_N|\theta,N)}\left(\frac{\partial P(x_1,\ldots,x_N|\theta,N)}{\partial \theta}\right)^2
%\nonumber \\
%&=&
=\sum_{x=\pm1}
\frac{1}{p(x|\theta)}\left(\frac{\partial p(x|\theta)}{\partial\theta}\right)^2
,
\label{prob8}
\end{eqnarray}
for this particular problem.
Hypothesis (1) was used to obtain the right hand side of Eq.~(\ref{prob8}), which
upon substitution of $p(x=+1|\theta)=p(\theta)$ and $p(x=-1|\theta)=1-p(\theta)$
turns into Eq.~(\ref{prob7}).

We find the minimum of the Fisher information $I_F$ by
substituting $p(\theta)=\cos^2 g(\theta)$ and obtain
\begin{equation}
I_F=4\left[\frac{\partial g(\theta)}{\partial \theta}\right]^2
.
\label{prob9}
\end{equation}
Rotational invariance requires that $I_F$ is independent of $\theta$,
hence $g(\theta)=a\theta+b$, where $a$ and $b$ are constants still to be determined.
Rotational invariance further requires that $p(\theta)=\cos^2(a\theta+b)=p(\theta+2\pi)$,
hence $a=k/2$ and $I_F=k^2$, with $k$ an integer number.
We may exclude the case $k=0$ because then $p(\theta)$ does not depend
on $\theta$ and a Stern-Gerlach magnet that operates according to this $k=0$ model would not be a useful device.
Thus, $I_F$ is minimal if $k=1$ and we may set the irrelevant phase factor $b$ to zero.
Therefore, using the four hypotheses given earlier, we have found that the probabilistic
model for the Stern-Gerlach magnet generates events with probabilities
\begin{eqnarray}
P(x=+1|{\bf a},{\bf S})&=&\cos^2\frac{\theta}{2}=\frac{1+{\bf a}\cdot{\bf S}}{2},
\nonumber \\
%&=&
P(x=-1|{\bf a},{\bf S})&=&\sin^2\frac{\theta}{2}=\frac{1-{\bf a}\cdot{\bf S}}{2}
,
\label{prob10}
\end{eqnarray}
which is in exact agreement with Eq.~(\ref{prob1}).

\medskip
\subsection{EPRB gedanken experiment}

We consider the case that the directions ${\bf a}_1$ and ${\bf a}_2$
of the applied fields in the Stern-Gerlach magnets do not change with time
(as in the quantum model)
and that we have no knowledge about the direction of the magnetic moments ${\bf S}_1$ and
${\bf S}_2$ of the particles.

The measuring equipment (Stern-Gerlach magnets + particle detectors + time-coincidence logic)
transforms the input, $N$ pairs of particles with unknown magnetic moments
into a time series $\{x_n,y_n|n=1,\ldots,N\}$ of signals $x_n=\pm1$ and $y_n=\pm1$.
By hypothesis (1), the probability $P((x_1,y_1),\ldots,(x_N,y_N)|{\bf a}_1,{\bf a}_2,N)$
to observe the data $\{x_n,y_n|n=1,\ldots,N\}$
can be written as $\prod_n^NP(x_n,y_n|{\bf a}_1,{\bf a}_2)$.
As $x,y=\pm1$, $P(x,y|{\bf a}_1,{\bf a}_2)$ can be written as
\begin{equation}
\label{prob20a}
P(x,y|{\bf a}_1,{\bf a}_2)=\frac{1+xE_1({\bf a}_1,{\bf a}_2)+yE_2({\bf a}_1,{\bf a}_2)+xyE({\bf a}_1,{\bf a}_2)}{4}
,
\end{equation}
where $E_1({\bf a}_1,{\bf a}_2)=\sum_{x,y=\pm1}xP(x,y|{\bf a}_1,{\bf a}_2)$,
$E_2({\bf a}_1,{\bf a}_2)=\sum_{x,y=\pm1}yP(x,y|{\bf a}_1,{\bf a}_2)$,
and $E({\bf a}_1,{\bf a}_2)=\sum_{x,y=\pm1}xyP(x,y|{\bf a}_1,{\bf a}_2)$.
Using the empirical (experimental) knowledge that
the averages are rotational invariant (hypothesis (2)), we have
$E_1({\bf a}_1,{\bf a}_2)=E_1({\bf a}_1\cdot{\bf a}_2)$,
$E_2({\bf a}_1,{\bf a}_2)=E_2({\bf a}_1\cdot{\bf a}_2)$, and
$E({\bf a}_1,{\bf a}_2)=E({\bf a}_1\cdot{\bf a}_2)$.
Furthermore, experiments indicate that frequencies
of the $x=\pm1$ ($y=\pm1$) events (not of the correlated events!)
are the same.%  and logically independent of $y$ ($x$).
We formalize this knowledge by the hypothesis that
\begin{eqnarray}
E_1({\bf a}_1\cdot{\bf a}_2)=E_2({\bf a}_1\cdot{\bf a}_2)=0
,
\label{prob24}
\end{eqnarray}
from which it immediately follows that
\begin{eqnarray}
\label{prob26}
\label{prob20}
P(x,y|{\bf a}_1,{\bf a}_2)=\frac{1+xyE({\bf a}_1\cdot{\bf a}_2)}{4}\equiv p(x,y|\theta)=\frac{1+xyE(\theta)}{4}.
\end{eqnarray}
Thus, the probability $p(x,y|\theta)$ for a single-event $(x,y)$
is conditional on the relative angle $\theta$ between the two unit vectors ${\bf a}_1$
and ${\bf a}_2$.

The probability that an experiment of $N$ events yields
$n(x,y)$ events of the type $(x,y)$ is given by
\begin{equation}
P(n(+1,+1),n(-1,-1),n(+1,-1),n(-1,+1)|\theta,N)=
%N!\prod_{x,y=\pm1} \frac{\left[p(x,y|\theta)\right]^{n(x,y)}}{n(x,y)!}
N!\prod_{x,y=\pm1} \frac{p(x,y|\theta)^{n(x,y)}}{n(x,y)!}
,
\label{prob21}
\end{equation}
where $n(+1,+1)+ n(-1,-1) + n(+1,-1) + n(-1,+1)=N$.
Adopting the same strategy as in the case of the single Stern-Gerlach magnet,
we consider the log-likelihood
\begin{eqnarray}
\frac{L}{N}&=&
\frac{1}{N}\ln \frac{
P(n(+1,+1),n(-1,-1),n(+1,-1),n(-1,+1)|\theta+\epsilon,N)}
{P(n(+1,+1),n(-1,-1),n(+1,-1),n(-1,+1)|\theta,N)}
%,\nonumber \\&=&
=
\sum_{x,y=\pm1}
\frac{n(x,y)}{N}\ln \frac{p(x,y|\theta+\epsilon)}{p(x,y|\theta)}
,
\label{prob22}
\end{eqnarray}
that the data was generated
by $p(x,y|\theta+\epsilon)$ instead of $p(x,y|\theta)$.
Repeating the steps that lead from Eq.~(\ref{prob5}) to Eq.~(\ref{prob7}),
we find that for small $\epsilon$ minimization of $L$
is tantamount to finding the probability $p(x,y|\theta)$
that minimizes the Fisher information
\begin{equation}
I_{F}= \sum_{x,y=\pm1}
\frac{1}{p(x,y|\theta)}
\left(\frac{\partial p(x,y|\theta)}{\partial \theta}\right)^2
.
\label{prob23}
\end{equation}
Using Eq.~(\ref{prob20}), we can write Eq.~(\ref{prob23}) as
\begin{equation}
\label{prob27}
I_F=\frac{1}{1-E^2(\theta)}
\left(\frac{\partial E(\theta)}{\partial \theta}\right)^2
,
\end{equation}
which, in essence, is the same expression as the one that
we obtained for the case of the Stern-Gerlach magnet.
Of course, the solution of the minimization problem is also the same.
Solving Eq.~(\ref{prob27}) for $E(\theta)$, we find
\begin{equation}
\label{IFE}
E(\theta)=\sin(\theta\sqrt{I_F}+b).
\end{equation}
In the case that one uses the magnetic moment of the particles,
the experimental data indicates that $E(\theta)$ is periodic in $\theta$ with a period
of $2\pi$ ($\pi$ if the experiment measures the polarization, as in EPRB experiments with photons).
This implies that $I_F$ should be an integer number.
The solution $I_F=0$ can be discarded because then $E(\theta)$ would not depend on $\theta$,
which would contradict the experimental observations.
Therefore, the nontrivial solution with minimum Fisher information is $I_F=1$.
Using the fact that the solution of the minimization problem
is determined up to an arbitrary phase $b$,
the two-particle correlation can be written as
\begin{eqnarray}
E({\bf a}_1,{\bf a}_2) %&=&-\sum_{x,y=\pm1} xy p(x,y|\theta)\nonumber \\
&=&-\cos\theta=-{\bf a}_1\cdot{\bf a}_2,
\label{prob29}
\end{eqnarray}
in agreement with the expression of the correlation of two $S=1/2$
particles in the singlet state.
Thus, we may conclude that we can derive the results of quantum theory
for the singlet state from a straightforward application
of probability theory, without making reference to
concepts of quantum theory.
%
%%
%\begin{eqnarray}
%p(+1,+1|\theta)&=&p(-1,-1|\theta)=\frac{1}{2}\sin^2\frac{\theta}{2}
%,\nonumber \\
%p(+1,-1|\theta)&=&p(-1,+1|\theta)=\frac{1}{2}\cos^2\frac{\theta}{2}
%,
%\label{prob28}
%\end{eqnarray}
%%
%which agrees with Eq.~(\ref{prob2}).

\subsection{Real EPRB experiment}

As explained in Section~\ref{EPRBexperiment}, real EPRB experiments produce the data sets
\begin{eqnarray}
\label{Ups2}
\Upsilon_i=\left\{ {x_{n,i} =\pm 1,t_{n,i},{\bf a}_{n,i} \vert n =1,\ldots ,N } \right\}
.
\end{eqnarray}
We assume that this data set can be described by a probabilistic model that satisfies hypothesis (1).
Let $P(x_1,x_2,t_1,t_2|{\bf a}_1,{\bf a}_2)$ denote the probability for
observing the data $\{x_1,t_1\}$ and $\{x_2,t_2\}$ at station 1 and 2, respectively.
Without loss of generality, we can use the exact representation
\begin{equation}
\label{pxx0}
P(x_1,x_2,t_1,t_2|{\bf a}_1,{\bf a}_2)=
\frac{f_0(t_1,t_2|{\bf a}_1,{\bf a}_2)+x_1f_1(t_1,t_2|{\bf a}_1,{\bf a}_2)
+x_2f_2(t_1,t_2|{\bf a}_1,{\bf a}_2)+x_1x_2f_3(t_1,t_2|{\bf a}_1,{\bf a}_2)}{4}
,
\end{equation}
to express the single- and two-particle correlations in terms of the
functions $f_i(t_1,t_2|{\bf a}_1,{\bf a}_2)$ for $i=0,\ldots,3$.
Because $0\le P(x_1,x_2,t_1,t_2|{\bf a}_1,{\bf a}_2)\le1$,
the functions $f_i(t_1,t_2|{\bf a}_1,{\bf a}_2)$ must satisfy the inequalities
$0\le f_0(t_1,t_2|{\bf a}_1,{\bf a}_2)\pm f_1(t_1,t_2|{\bf a}_1,{\bf a}_2)+f_2(t_1,t_2|{\bf a}_1,{\bf a}_2) \pm f_3(t_1,t_2|{\bf a}_1,{\bf a}_2)\le4$
and
$0\le f_0(t_1,t_2|{\bf a}_1,{\bf a}_2)\pm f_1(t_1,t_2|{\bf a}_1,{\bf a}_2)-f_2(t_1,t_2|{\bf a}_1,{\bf a}_2) \mp f_3(t_1,t_2|{\bf a}_1,{\bf a}_2)\le4$.
The mathematical expectation of the coincidences $C_{xy}$ (see Eq.~(\ref{Cxy})), that is the average computed with
$P(x_1,x_2,t_1,t_2|{\bf a}_1,{\bf a}_2)$, is given by
\begin{equation}
\label{pxx1}
\langle C_{xy}\rangle \equiv N
\int_{-\infty}^{+\infty}dt_1 \int_{-\infty}^{+\infty} dt_2\,P(x,y,t_1,t_2|{\bf a}_1,{\bf a}_2)\Theta(W-|t_1 -t_2|)
.
\end{equation}
We find
\begin{eqnarray}
\label{pxx2}
E_1({\bf a}_1,{\bf a}_2,W)=
\frac{\sum_{x,y=\pm1} x\langle C_{xy}\rangle}{\sum_{x,y=\pm1} \langle C_{xy}\rangle}
&=&
\frac{\int_{-\infty}^{+\infty}dt_1 \int_{-\infty}^{\infty} dt_2\,\Theta(W-|t_1 -t_2|) f_1(t_1,t_2|{\bf a}_1,{\bf a}_2)}{
\int_{-\infty}^{+\infty}dt_1 \int_{-\infty}^{+\infty} dt_2\,\Theta(W-|t_1 -t_2|) f_0(t_1,t_2|{\bf a}_1,{\bf a}_2)},
\nonumber \\
E_2({\bf a}_1,{\bf a}_2,W)=
\frac{\sum_{x,y=\pm1} y\langle C_{xy}\rangle}{\sum_{x,y=\pm1} \langle C_{xy}\rangle}
&=&
\frac{\int_{-\infty}^{+\infty}dt_1 \int_{-\infty}^{\infty} dt_2\,\Theta(W-|t_1 -t_2|) f_2(t_1,t_2|{\bf a}_1,{\bf a}_2)}{
\int_{-\infty}^{+\infty}dt_1 \int_{-\infty}^{+\infty} dt_2\,\Theta(W-|t_1 -t_2|) f_0(t_1,t_2|{\bf a}_1,{\bf a}_2)},
\nonumber \\
E({\bf a}_1,{\bf a}_2,W)=
\frac{\sum_{x,y=\pm1} xy\langle C_{xy}\rangle}{\sum_{x,y=\pm1} \langle C_{xy}\rangle}
&=&
\frac{\int_{-\infty}^{+\infty}dt_1 \int_{-\infty}^{\infty} dt_2\,\Theta(W-|t_1 -t_2|) f_3(t_1,t_2|{\bf a}_1,{\bf a}_2)}{
\int_{-\infty}^{+\infty}dt_1 \int_{-\infty}^{+\infty} dt_2\,\Theta(W-|t_1 -t_2|) f_0(t_1,t_2|{\bf a}_1,{\bf a}_2)}
.
\end{eqnarray}
At this point, we feel that we lack the necessary mathematical tools for carrying out
the procedure that we successfully applied to the simpler cases treated earlier.
First, it is difficult to see how the empirical knowledge that
single-particle averages are zero and that the two-particle average is rotational invariant
leads to useful conditions on the form of the $f_i(t_1,t_2|{\bf a}_1,{\bf a}_2)$.
Second, the presence in Eq.~(\ref{pxx2}) of the step functions introduces nontrivial correlations
and prevents us from making further progress in the
mathematical treatment of this problem.
Third, the description now contains a new parameter ($W$ to which we should also apply hypothesis (3))
as well as extra variables ($t_1$ and $t_2$).
We leave the problem of the analytical treatment of the general case for future research.

\end{widetext}
\medskip

\subsection{Summary}

The assumption that there is an underlying probabilistic process
that gives rise to the observation of the data as obtained in
Stern-Gerlach and EPRB experiments, together with the very simple,
plausible hypotheses (1)-(4) are sufficient to derive
the probability distributions of quantum theory for the EPRB experiment,
without using a single concept of quantum theory.
In addition, this derivation suggests that quantum theory
is the probabilistic model for the set of data that is most likely
to be observed.

From a more general perspective, this section demonstrates,
by way of a successful application to specific problems,
how to formalize the process of inductive inference
and derive useful results (those of quantum theory) from it.
This derivation builds on prior, empirical knowledge that we have
acquired through experiments, the application of probability
theory as mathematical vehicle for rational reasoning,
and the metaphysical principle that we, human observers,
have great difficulties to interpret experimental data
that is not robust with respect to small changes in the
conditions under which the experiments are carried out~\cite{SCHOM98}.

\section{Conclusion}
\label{Conclusions}

Starting from nothing more than the observation that an EPRB experiment produces pairs of triples of data
$\{\Upsilon_1,\Upsilon_2\}$,
we have constructed computer simulation models that reproduce
the results of all single-particle and two-particle correlations
of a quantum system of two $S=1/2$ particles.
Salient features of these models are that they

\begin{itemize}
\item{Generate, event-by-event, the same kind of data set $\{\Upsilon_1,\Upsilon_2\}$
as the one recorded in real EPRB experiments}
\item{Satisfy Einstein's criteria of local causality}
\item{Count all events in which systems of two particles have been detected, using the same
time-coincidence criterion as used in real EPRB experiments}
\item{Provide a simple, rational and realistic picture of a mechanism that yields the correlations
of an ``entangled state''}
\item{Do not rely on any concept of quantum theory or probability theory}
\end{itemize}

A key ingredient of these models, not present in the textbook treatments
of the EPRB {\sl gedanken} experiment, is the time window $W$ that
is used to detect coincidences.
We have demonstrated (see Section~\ref{IIG})
the importance of the choice of the time window
by analyzing a data set of a real EPRB experiment with photons~\cite{WEIH98}.

The mathematical treatment of the models yields results that are in exact
agreement with quantum theory. The condition under which an EPRB
experiment yields results that agree with quantum theory is evident: The
resolution $\tau$ of the devices that generate the time-tags and the time window
$W$ should be much smaller than the time delays, the range which is determined
by $T_0$.
Disregarding the timing data yields a result that disagrees with
quantum theory and with experiment. The EPR paradox reappears when the
experiments are analyzed in terms of an incomplete set of data.

We have demonstrated that the event-by-event simulation of
EPRB experiments allows us to reproduce not only the results of quantum
theory but also allows us to consider cases that are not described by
quantum theory. Therefore, for this type of experiments, the
two-particle ``world'' that we can simulate contains the two-particle
``world'' described by quantum theory as a special case.

As our work shows that it is possible to construct event-based simulation models
that satisfy Einstein's criteria of local causality
and reproduce the expectation values calculated by quantum theory
it opens new routes to ontological descriptions of
microscopic phenomena~\cite{RAED05b,RAED05c,RAED05d,MICH05,RAED06c,RAED07a,RAED07b,ZHAO07a}.

We have resolved the apparent conflict between
the fact that there exist event-based simulation models
that satisfy Einstein's criteria of local causality and reproduce
the results of the quantum theory of two $S=1/2$ particles
and the folklore about Bell's theorem, stating
that such models are not supposed to exist.
The origin of this conflict has been traced back to
Bell's extension of Einstein's concept of locality
to the domain of probabilistic theories, the fundamental assumption being
that the absence of a causal influence implies logical independence~\cite{JAYN89}.
This leaves two options:
\begin{itemize}
\item{One accepts the assumption that the absence of a causal influence implies logical independence
and lives with the logical paradoxes that this assumption creates}
\item{One recognizes that logical independence and the absence of a causal influence are
different concepts~\cite{COX61,TRIB69,JAYN03} and one searches for rational explanations
of experimental facts that are logically consistent, as we did in this paper}
\end{itemize}

Finally, we have demonstrated that it is possible to derive,
without resorting to concepts of quantum theory,
the quantum theoretical description of the EPRB experiment
from the general principles of logical inference, developed
by Cox and Jaynes,~\cite{COX61,TRIB69,JAYN03}
and empirical knowledge about the results of the experiment.

The computer models we have invented can be built with
macroscopic, say mechanical parts (in principle a digital computer can
be built from mechanical parts).
To the experimenter who has no knowledge of what is
going on inside the building where the mechanical machine is operating,
there is no way of telling whether the data he/she receives is generated by
a quantum system or not. In a sense, this supports Bohr's point of view that
``There is no quantum world. There is only an abstract quantum theoretical
description''~\cite{JAMM74}.

~
\section*{Appendix}

For the singlet state, the probability $P(x,y|{\bf a},{\bf b})$ reads
\begin{equation}
\label{Pxy-a}
P(x,y|{\bf a},{\bf b})=\frac{1-xy{\bf a}\cdot{\bf b}}{4},
\end{equation}
where $x,y=\pm1$, ${\bf a}$ and ${\bf b}$ are unit vectors.
Let us now try to write Eq.~(\ref{Pxy-a}) in the form
\begin{equation}
\label{Fxy-a}
P(x,y|{\bf a},{\bf b})=\int F(x,{\bf a},\lambda)F(y,{\bf b},\lambda) d\lambda,
\end{equation}
where $\lambda$ denotes a set of auxiliary variables that may be chosen at will.
%and we impose no restrictions of $F(x,{\bf a},\lambda)$ other than that is should
%be positive.

A simple solution to this problem is given by
\begin{equation}
\label{Sxy-a}
P(x,y|{\bf a},{\bf b})=\int
\frac{1+\sqrt{3}x{\bf a}\cdot{\bf S}}{2}
\frac{1-\sqrt{3}y{\bf b}\cdot{\bf S}}{2}  d{\bf S},
\end{equation}
where ${\bf S}=(\sin\theta\cos\phi,\sin\theta\sin\phi,\cos\theta)$ and the integral
is over the unit sphere.
In this case, the function $F(x,{\bf a},{\bf S})=(1+\sqrt{3}x{\bf a}\cdot{\bf S})/2$
can take negative values and therefore it does not qualify as a probability distribution.

\bibliography{../../epr}

\end{document}